\documentclass[11pt,a4paper]{article}

\def\bea#1\eea{\begin{eqnarray}#1\end{eqnarray}}
\def\be#1\ee{\begin{equation}#1\end{equation}}
\def\ba#1\ea{\begin{align}#1\end{align}}

\usepackage{tikz-cd}
\usepackage{graphicx}

\usepackage{subcaption}

\usepackage{pgfplots}
\usepackage[utf8]{inputenc}
\usepackage{jheppub}
\usepackage{amsmath}
\usepackage{multicol}
\usepackage{bbm}
\usepackage{enumerate}
\usepackage[inline]{enumitem}
\usepackage{bibentry}
\usepackage{comment}
\usepackage{amsthm}
\usepackage{mathrsfs}
\usepackage{upgreek}
\usepackage{amssymb}
\usepackage{bm}
\usepackage{setspace}
\usepackage{array,multirow,arydshln}
\usepackage{bigdelim}
\usepackage{scalerel}
\usepackage{diagbox}
\usepackage{caption}
\usepackage{tabularx}
\usepackage{empheq}
\usepackage{lipsum}

\usepackage{adjustbox}

\usepackage{tikz}

\usetikzlibrary{shapes.geometric,arrows,arrows.meta,decorations.pathmorphing,decorations.markings,patterns}

\newcommand*{\halfway}{0.5*\pgfdecoratedpathlength+4pt}
\tikzset{vertex/.style={inner sep=0,minimum size=3pt,circle,fill}}

\def\<{\langle}
\def\>{\rangle}

\newcommand{\minorangsq}[1]{ \langle#1\rangle }

\pgfplotsset{compat=1.17}

\preprint{LCTP-21-04}

\title{Algebraic branch points at all loop orders from positive kinematics and wall crossing}

\author{Aidan Herderschee}
\emailAdd{aidanh@umich.edu}

\affiliation{Leinweber Center for Theoretical Physics, \\
Randall Laboratory of Physics, Department of Physics, \\
University of Michigan, Ann Arbor, MI 48109, USA}

\date{\today}

\abstract{There is a remarkable connection between the boundary structure of the positive kinematic region and branch points of integrated amplitudes in planar $\mathcal{N}=4$ SYM. A long-standing question has been precisely how algebraic branch points emerge from this picture. We use wall crossing and scattering diagrams to systematically study the boundary structure of the positive kinematic regions associated with MHV amplitudes. The notion of asymptotic chambers in the scattering diagram naturally explains the appearance of algebraic branch points. Furthermore, the scattering diagram construction also motivates a new coordinate system for kinematic space that rationalizes the relations between algebraic letters in the symbol alphabet. As a direct application, we conjecture a complete list of all algebraic letters that could appear in the symbol alphabet of the 8-point MHV amplitude. }

\begin{document}

\maketitle
\flushbottom

\section{Introduction}

Scattering amplitudes are one of the most fundamental observables in modern high-energy physics. Significant strides have been made in both understanding the underlying structure of amplitudes and developing new computation methods. Many techniques for computing amplitudes are under active research, such as the double copy \cite{Bern:2010ue,Bern:2019prr,Bern:2019crd}, positive geometry \cite{Arkani-Hamed:2017mur,Arkani-Hamed:2017tmz,Arkani-Hamed:2019mrd}, intersection theory \cite{Mizera:2017rqa, Mastrolia:2018uzb,Mizera:2020wdt} and much more \cite{Elvang:2013cua,Arkani-Hamed:2017jhn,Chung:2018kqs,Chicherin:2020umh,Elvang:2020lue,Elvang:2018dco,Bern:2020buy,Trnka:2020dxl}. In particular, a number of techniques have emerged that leverage non-perturbative properties of the S-matrix, such as unitarity and cluster decomposition. For example, cutting rules for integrands that hold at all orders can be derived from unitarity \cite{Bern:2011qt,Bourjaily:2017wjl,Bourjaily:2019iqr}, and bounds on couplings can be derived by testing whether the theory allows for macroscopic superluminal signal transmission \cite{Adams:2006sv}. Recently, significant focus in this program has been placed on studying the analytic structure of amplitudes, which encode the causal dynamics of the underlying theory. Imposing that the amplitude has the correct analytic structure imposes many nontrivial properties, such as bounds on higher dimension operators \cite{Camanho:2014apa,Arkani-Hamed:2020blm,Caron-Huot:2020cmc,Trott:2020ebl} and Steinmann relations \cite{2018PhRvL.120p1601D,Drummond:2018dfd,Bourjaily:2020wvq}. 

\indent In this paper, we study what singularities and branch cuts can appear in integrated Maximal Helicity Violating (MHV) amplitudes at all loop orders in $\mathcal{N}=4$ planar super Yang-Mills (pSYM).\footnote{``Integrated amplitude'' refers to the BDS-like normalized amplitude \cite{Alday:2009dv,Caron-Huot:2020bkp}.} We restrict ourselves to MHV amplitudes due to their simplicity and plethora of computational data. Remarkably, all evidence to date suggests that integrated MHV amplitudes do not have any singularities or branch cuts inside the positive kinematic region \cite{Arkani-Hamed:2019rds}. Furthermore, by studying the boundary structure of the positive kinematic region, one can make predictions for what branch points can appear at any loop order \cite{Golden:2013xva}. However, the boundary structure of the positive kinematic region is very difficult to study due to the subtleties involved in choosing particular compactifications of the positive kinematic region. Cluster algebras provide a precise understanding of the boundary structure of the positive kinematic region at 6-point and 7-point \cite{Golden:2013xva,Golden:2013lha,Golden:2014xqa}. At 8-point and beyond though, many new features appear that are still under active investigation and not well understood from the cluster algebra perspective. Foremost among these features is the appearance of algebraic letters in the symbol alphabet \cite{Prlina:2017azl,Prlina:2017tvx,Zhang:2019vnm}.  

\indent To approach these questions, we use scattering diagrams\footnote{The term ``scattering diagrams'' in this context has nothing to do with Feynman diagrams.} \cite{Kontsevich:2013rda,2007math......3822G,2004math......6564K,2018arXiv180605094R,2017arXiv171206968R,2014arXiv1411.1394G}, a natural generalization of the cluster algebra framework, to study different compactifications of the positive kinematic region of the 8-point MHV amplitude. We show how the boundary structure of the positive kinematic region can be systematically studied using scattering diagrams and find that algebraic letters naturally emerge from the notion of asymptotic chambers in the scattering diagram. We ultimately found a list of 72 multiplicatively independent letters, of which at most 52 are algebraic, associated with the asymptotic chambers apparently relevant for the 8-point MHV amplitude. We argue this alphabet includes all algebraic letters that could appear in the 8-point MHV symbol alphabet. Furthermore, we also discuss how scattering diagrams provide a new approach to studying rational letters in the symbol alphabet. 

\subsection{\texorpdfstring{$\mathcal{N}=4$}{N=4} super Yang-Mills }
\noindent Amplitudes in $\mathcal{N}=4$ pSYM are an ideal testing ground for exploring the analytic structure of planar scattering amplitudes. For instance, amplitudes in $\mathcal{N}=4$ pSYM have a finite number of branch points associated with solutions to the Landau equations and are expected to have a finite radius of convergence in perturbation theory \cite{Prlina:2018ukf}. Significant progress has been made in understanding the structure of $\mathcal{N}=4$ pSYM amplitudes beyond Feynman diagrams. At weak coupling, deep geometric structures, such as the amplituhedron, have emerged that provide both powerful computational techniques for computing integrands at any loop order and a radically different perspective on the nature of locality and unitarity \cite{Arkani-Hamed:2013jha,Arkani-Hamed:2017vfh,Herrmann:2020qlt,ArkaniHamed:2012nw,Kojima:2020tjf}. At strong coupling, holographic calculations provide non-trivial predictions for the behavior of $\mathcal{N}=4$ pSYM amplitudes in the form of the BDS-ansatz \cite{Anastasiou:2003kj,Bern:2005iz,Alday:2007hr} and its generalizations \cite{Alday:2009dv,Yang:2010as}. Other formalisms motivated by the duality between Wilson loops and scattering amplitudes have also emerged \cite{Brandhuber:2007yx,Alday:2010ku,CaronHuot:2010ek,Eden:2011yp,Basso:2013vsa,He:2020lcu, He:2020uxy,He:2021esx}. 

\indent MHV amplitudes in $\mathcal{N}=4$ pSYM are particularly simple and have been useful litmus tests for conjectures. MHV $n$-point amplitudes are transcendental functions of fixed weight at each loop order that can be expressed in terms of multi-polylogarithms (MPLs) at all orders calculated to date \cite{Goncharov:2010jf,Gaiotto:2011dt,2011arXiv1105.2076G,ArkaniHamed:2012nw,Golden:2013xva,Duhr:2014woa,GONCHAROV1995197}.\footnote{Using the Grassmannian form of $\mathcal{N}=4$ pSYM loop integrands, one can directly show all integrals in the MHV (and NMHV) sector can be written as iterated integrals of $d\log$-forms \cite{ArkaniHamed:2012nw}. Unfortunately, this does not necessarily mean they integrate to a function that can be written in terms of MPLs \cite{Brown:2020rda}.} These transcendental functions of weight $W$, $F_{W}$, are the generalizations of logarithms that obey extremely nice properties. Primarily, the \textit{symbol} provides a map from the amplitude to a sum of $W$-fold tensor products:
\begin{equation}\label{symboldecompositionsss}
F \rightarrow \sum F^{\phi_{\alpha_{1}},\phi_{\alpha_{2}},\ldots,\phi_{\alpha_{W}}}_{0}[\log(\phi_{\alpha_{1}})\otimes \log(\phi_{\alpha_{2}})\otimes \ldots \otimes \log(\phi_{\alpha_{W}}) ]    \ ,
\end{equation}
\noindent where $F_{0}$ are rational numbers. Each factor in the tensor product behaves similarly to a logarithm, leading to properties like
\begin{equation}\label{productstructuess}
[\ldots \otimes\log(\phi_{1}\phi_{2})\otimes \ldots]=[\ldots \otimes \log(\phi_{2})\otimes \ldots]+[\ldots \otimes \log(\phi_{1})\otimes \ldots] \ .  
\end{equation}
\noindent The $\phi_{i}$ in eq. (\ref{symboldecompositionsss}) are functions of external kinematic data and correspond to branch points of $F_{W}$. The set of all $\phi$ that can appear in eq. (\ref{symboldecompositionsss}) is called the symbol alphabet of $F_{W}$. 

\indent The symbol provides a very transparent understanding of the analytic structure of $F_{W}$. We will focus on finding a minimal symbol alphabet, a set of multiplicatively independent letters that all letters in the original symbol alphabet can be written as monomials of. For example, consider the initial symbol alphabet $\{ \phi_{1}, \ \phi_{2}, \ \phi_{1}\phi_{2} \}$. One minimal symbol alphabet is $\{ \phi_{1}, \ \phi_{2}\}$ as $\phi_{1}\phi_{2}$ factors into $\phi_{1}$ and $\phi_{2}$. An alternative minimal symbol alphabet is $\{ \phi_{2}, \ \phi_{1}\phi_{2} \}$, as $\phi_{1}=(\phi_{1}\phi_{2})/\phi_{2}$. Given a minimal symbol alphabet, one can use eq. (\ref{productstructuess}) to construct a complete basis of possible tensors. Finding a minimal symbol alphabet of the 8-point MHV amplitude would be a major achievement and open up the possibility of bootstrapping 8-point MHV higher loop amplitudes. We take an important step towards this goal by proposing a minimal symbol alphabet for algebraic letters. \newline 

\subsection{The positive kinematic region}
\noindent Scattering amplitudes are functions on kinematic space. The positive kinematic region is a region of kinematic space where planar gauge theory amplitudes are conjectured to have no poles or branch cuts. More precisely, in all examples studied to date, the Landau equations admit no solutions when the external data is taken to be in the positive kinematic region. The positive kinematic region for a given ordering of externals, $\alpha\in \textrm{Perm}[1,2,\ldots,n]$, is associated with the region where all planar variables are positive definite,
\begin{equation}\label{positiveregion4d}
X_{i,j}=\left ( \sum_{a=i}^{j-1}p_{\alpha(a)} \right )^{2}>0 \ ,   
\end{equation}
\noindent along with additional constraints \cite{Golden:2013xva,Arkani-Hamed:2017vfh,Arkani-Hamed:2019rds}. Although not well understood, the positive kinematic region implicitly appears in many computations. For example, the integrands that appear in open superstring scattering amplitudes are generically divergent unless evaluated in the positive kinematic region \cite{Arkani-Hamed:2019mrd,Sen:2019jpm}. Therefore, when evaluating superstring integrands without using string field theory techniques or taking sophisticated Pochhammer contours in the string moduli space \cite{Witten:2013pra}, one must implicitly work in the positive kinematic region, only taking an analytic continuation to generic momentum configurations at the end of the calculation. Again, we emphasize that the physical significance of the positive kinematic region remains mysterious and its importance has been found through direct computation. 

\indent Since we are studying the positive kinematic region of massless planar gauge theory amplitudes in 4 dimensions, we parameterize our external kinematic data using momentum twistors \cite{Golden:2013xva}; $Z_{i}^{A}$ is the momentum twistor of state $i$ and the $A$ index transforms in the fundamental representation of the dual conformal algebra, $SU(2,2)$. Individual momentum twistors are projective:
\begin{equation}
Z_{i}^{A} \sim t_{i}Z_{i}^{A} \ . 
\end{equation}
\noindent Therefore, the kinematic space of the $n$-point amplitude can be interpreted as a quotient of the Grassmannian, $Gr(4,n)/T$, where $T$ acts on columns by a re-scaling. The positive kinematic region is then a quotient of the positive Grassmannian, $Gr_{+}(4,n)/T$, cut out by the inequalities
\begin{equation}
0<\langle i,j,k,l\rangle \textrm{ when } i<j<k<l \ ,
\end{equation}
\noindent where $\langle \ldots \rangle$ corresponds to a minor of columns ``$\ldots$''.\footnote{This is only true in the MHV sector. Beyond MHV, the kinematic region is most naturally interpreted as bundles over $Gr(4,n)/T$ \cite{Arkani-Hamed:2017vfh,Arkani-Hamed:2019rds}.} 

\begin{figure}
\centering
  \includegraphics[scale=0.6]{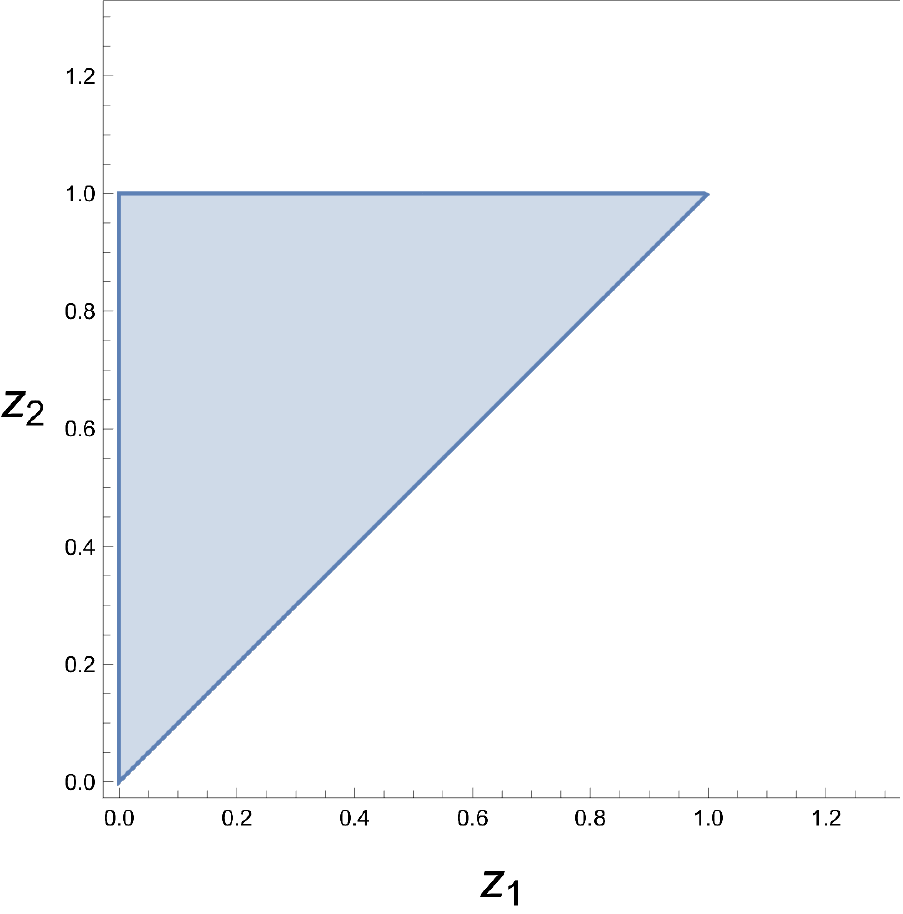} \quad \quad 
   \includegraphics[scale=0.6]{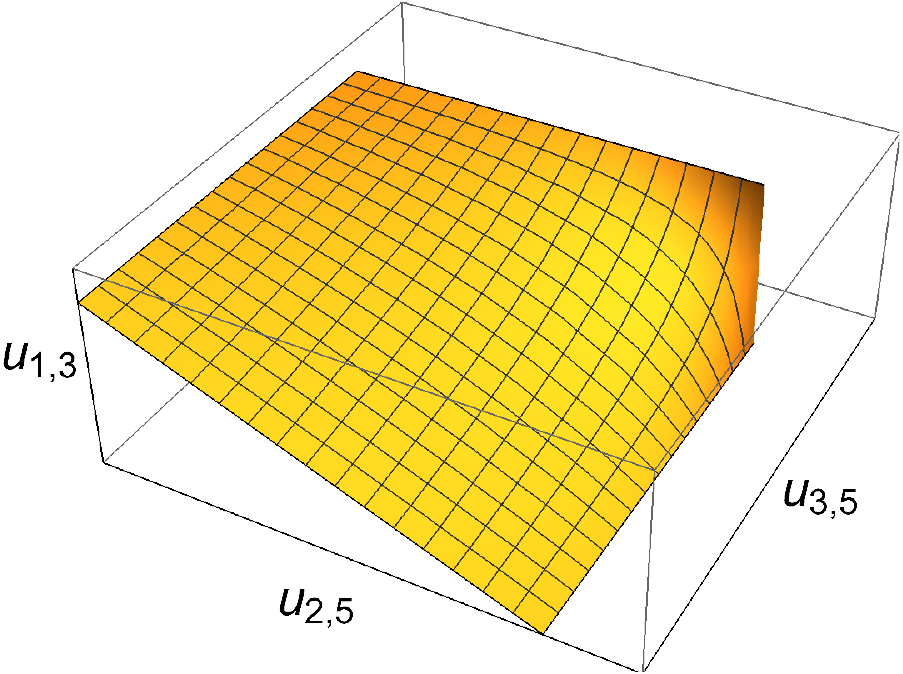} 
  \caption{Different parameterizations of $Gr_{+}(2,5)/T$. In the $z$-variable parameterization, only three boundaries are manifest. However, the $u$-variable parameterization makes all five boundaries manifest.}
  \label{fig:exampleofcompactificationsA2}
\end{figure}

\indent We are particularly interested in the boundary structure of the positive kinematic region since the boundary is where planar gauge theory amplitudes can have singularities \cite{Prlina:2018ukf,Arkani-Hamed:2019rds}. This is most easily seen at tree level, where poles of the form $X_{i,j}\rightarrow 0$ manifestly correspond to boundaries of the region defined by eq. (\ref{positiveregion4d}). Remarkably, not only singularities, but also possible branch cuts are encoded in the boundary structure of the positive kinematic region. However, the problem of finding distinct boundaries is more subtle than one might naively expect. Instead of investigating $Gr_{+}(4,n)/T$, let us consider a simpler version of the problem by studying the boundaries of $Gr_{+}(2,5)/T$. A naive parameterization of this space is 
\begin{equation}\label{eq:naiveparameterizatioss}
C_{i}^{\alpha}\sim \begin{pmatrix}
1 & 1 & 1 & 1 & 0 \\ 
0 & z_{1} & z_{2}  & 1 & 1
\end{pmatrix} \ ,   
\end{equation}
\noindent where $0<z_{1}<z_{2}<1$ defines the positive region. We graph the positive region explicitly in fig. \ref{fig:exampleofcompactificationsA2}, where 3 boundaries are clearly manifest. However, let us now consider an alternative parameterization of the space with $u$-variables \cite{Arkani-Hamed:2019plo,Arkani-Hamed:2017mur,Arkani-Hamed:2020tuz}
\begin{equation}
u_{i,j}=\frac{\langle i,j-1\rangle\langle i-1,j\rangle}{\langle i,j\rangle\langle i-1,j-1\rangle} \ ,
\end{equation}
\noindent where $\langle a,b\rangle$ denotes a minor of $C_{i}^{\alpha}$. These $u_{i,j}$ variables obey the non-linear relations:
\begin{equation}
u_{1,3}=1-u_{2,4}u_{2,5}, \quad (\textrm{Cyclic permutations}) ,
\end{equation}
\noindent and the bounds, $0<u_{i,j}<1$, in the positive region. The second plot of the positive region in fig. \ref{fig:exampleofcompactificationsA2} using $u$-variables shows that there are 5 boundaries, not 3. The underlying problem with eq. (\ref{eq:naiveparameterizatioss}) is that a single set of coordinates, unless chosen very carefully, will not manifest all possible boundaries of the positive region. In other words, to study the boundary structure of the positive kinematic region, we need to study the compactified positive kinematic region, $\overline{Gr(4,n)/T}$. \newline 

\subsection{Critically positive coordinates and cluster algebras}
\noindent Previous research into the connection between the positive kinematic region and $\mathcal{N}=4$ pSYM amplitudes has generally focused on the cluster algebra structure of the positive kinematic region \cite{Dixon:2011nj,Golden:2013xva,Golden:2013lha,Dixon:2013eka,Golden:2014pua,Golden:2014xqa,Drummond:2014ffa,Dixon:2014iba,Dixon:2014voa,Dixon:2015iva,Caron-Huot:2016owq,Drummond:2018caf,Caron-Huot:2019vjl,Caron-Huot:2019bsq}.\footnote{Note that alternate approaches have also been very successful without directly referencing the cluster algebra structure of the positive kinematic region. The $\bar{\mathcal{Q}}$ approach in particular has been extremely useful in probing $n\geq 8$ higher loop amplitudes \cite{CaronHuot:2011kk,Zhang:2019vnm,He:2020vob}. Computations with irrational Yangian invariants provide a very clever probe of the algebraic letters \cite{He:2020uhb,Mago:2020kmp,Mago:2020nuv}. Finally, studying branch points using Landau-equations and the amplitudhedron have allowed direct computations of the singularity structure at high loop order \cite{Dennen:2016mdk,Prlina:2017azl,Prlina:2017tvx}.} More concretely, the positive kinematic region of MHV $\mathcal{N}=4$ pSYM amplitudes corresponds to a $\mathcal{X}$-type cluster algebra\footnote{A quick introduction to cluster algebras is provided in appendix \ref{sec:introductionclusters}.} which associates to the positive kinematic region a set of \textit{critically positive} coordinates called $\hat{y}$-variables. A coordinate that is critically positive vanishes on at least one boundary of the positive region. Although each cluster parameterization makes only a sub-set of boundaries manifest, considering all cluster parameterizations together allows one to study all the possible boundaries. At 6-point and 7-point, the symbol alphabet, the $\phi_{i}$ in eq. (\ref{symboldecompositionsss}), consists solely of the $\hat{y}$-variables, implying that $\hat{y}$-variables correspond to logarithmic branch cuts! Calculations are further simplified by considering a minimal multiplicative basis of $\hat{y}$-variables instead of the set of $\hat{y}$-variables themselves. Given an initial cluster, one minimal multiplicative basis consists of the $\hat{y}$-variables of an initial cluster, $\hat{y}_{i}$, along with some non-factorable Laurent polynomials of $\hat{y}_{i}$. We denote this set of non-factorable Laurent polynomials as $\mathcal{O}(\mathcal{X})$. 

\indent Starting at 8-point, two problematic features appear in the cluster algebra approach:
\begin{itemize}
    \item There are an infinite number of $\hat{y}$-variables in the cluster algebra. 
    \item Algebraic letters start to appear in the symbol alphabet.
\end{itemize}
\noindent Several approaches to tackling these problems have appeared in the literature and significant progress has been made. 

\indent The first problem is troublesome because a key restriction for calculations at 6-point and 7-point is that the symbol alphabet is finite. Upon finding that the cluster algebra is infinite at 8-point, one might be tempted to assume that the symbol alphabet at 8-point is also infinite. However, it has been proven that the $n$-point amplitude in $\mathcal{N}=4$ pSYM has a finite number of branch points associated with solutions to the Landau equations \cite{Prlina:2018ukf}, implying that the symbol alphabet could also be finite. Following this train of thought, several truncation procedures have been proposed, motivated by connections between stringy canonical forms and compactifications of configuration spaces \cite{Arkani-Hamed:2019rds,He:2020ray,Arkani-Hamed:2020tuz,Arkani-Hamed:2020cig}. 

\indent The second problem has proven a major obstacle for interpreting letters as cluster variables because cluster variables are rational by construction. Multiple methods have been developed to extract algebraic functions from cluster algebras and then match these functions with algebraic letters that appear in direct calculations \cite{Mago:2020kmp,Arkani-Hamed:2019rds,Henke:2019hve,Drummond:2019cxm,He:2020uhb}. We use the term cluster algebraic letters as an umbrella term for all such cluster-like variables that are algebraic.\footnote{Notably,
the initial definition of cluster algebraic functions in ref. \cite{Arkani-Hamed:2019rds} only included 2 algebraic letters for each limiting ray in the $\overline{Gr(4,8)/T}$ $g$-vector fan. However, at least 18 algebraic letters seem to appear in the NMHV 2-loop amplitude at 8-point \cite{Zhang:2019vnm}.} However, no unified picture has emerged that provides a systematic understanding of these cluster algebraic functions. 
\begin{figure}
\centering
  \begin{tikzpicture}[every node/.style={font=\footnotesize}]
	\draw [thick,-stealth] (0,0) -- (-3,3);
	\draw [thick,-stealth] (0,0) -- (3,0);
	\draw [thick,-stealth] (0,0) -- (-3,0);
	\draw [thick,-stealth] (0,0) -- (0,3);
	\draw [thick,-stealth] (0,0) -- (0,-3);
	\filldraw (0,0) circle (1pt) ;
	\end{tikzpicture} \quad \includegraphics[scale=1]{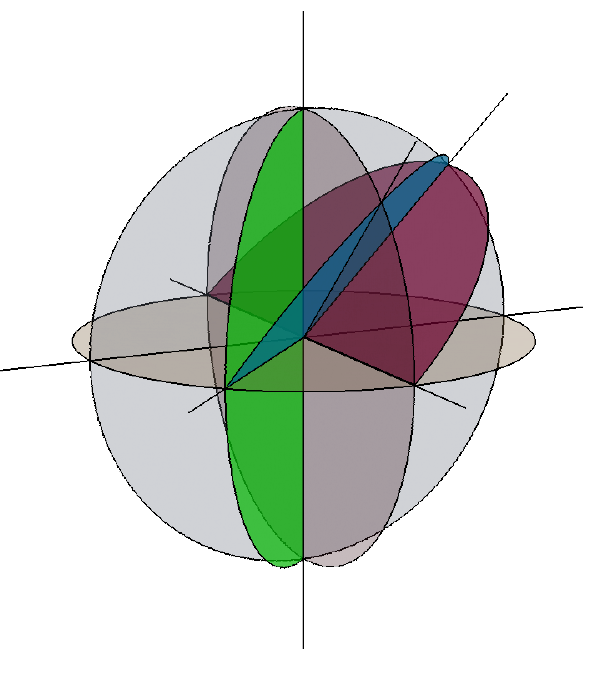}
  \caption{Pictures of the scattering diagrams corresponding to $\overline{Gr(2,5)/T}$ (left) and $\overline{Gr(2,6)/T}$ (right). }
  \label{fig:exameplA2}
\end{figure}
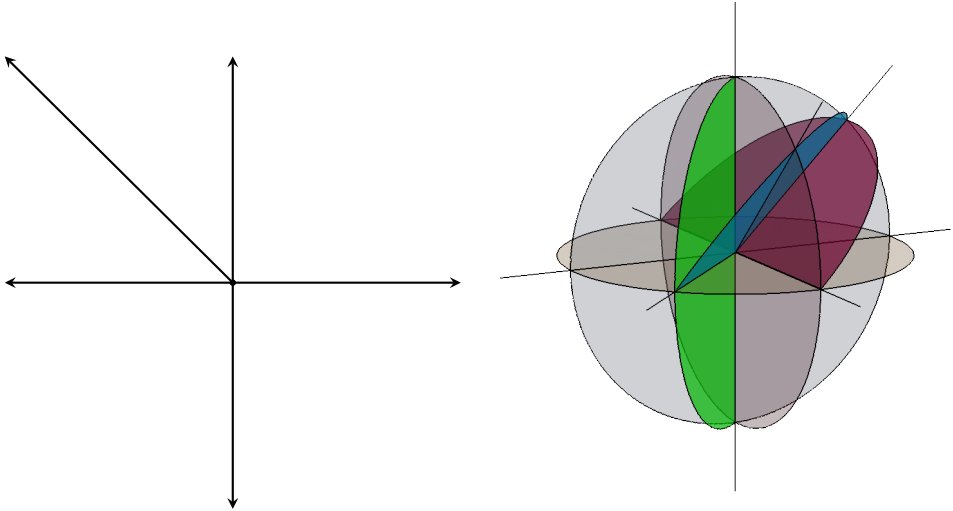\newline

\subsection{Scattering diagrams and asymptotic chambers}
\noindent In this paper, we propose wall crossing, and scattering diagrams more specifically, as a useful framework to address these issues \cite{Kontsevich:2013rda,2007math......3822G,2004math......6564K,2018arXiv180605094R,2017arXiv171206968R,2014arXiv1411.1394G}. Wall crossing has found applications in a number of research areas, such as the moduli spaces of $\mathcal{N}=2$ gauge theories and black hole entropy formulas \cite{Gaiotto:2009hg,Gaiotto:2010be,Andriyash:2010qv,Manschot:2010qz,Cordova:2013bza,Gaiotto:2015aoa}. However, we are not studying any kind of entropy formula or moduli space, but instead compactifications of the positive kinematic region, $\overline{Gr(4,n)/T}$. The application of wall crossing and scattering diagrams to partial compactifications is best understood in the context of mirror symmetry \cite{2011arXiv1106.4977G,Kontsevich:2013rda,2014arXiv1411.1394G}, but such a discussion is unfortunately beyond the scope of this paper.\footnote{The schematic connection between mirror symmetry and cluster algebras is as follows. We can interpret $\mathcal{X}$ as the blow-up of an associated toric geometry. Cluster transformations correspond to changing the blow up description by an elementary transformation. Scattering diagrams provide a framework to systematically ``sew'' these different parameterizations together using a fan defined by tropical points of the dual mirror manifold, $\mathcal{A}^{\vee}$. This framework is famous for giving a geometric interpretation of the connection between tropical points of $\mathcal{A}^{\vee}$ and regular functions on $\mathcal{X}$ using mirror symmetry.} Instead, we take a more practical approach, giving a computational definition of a scattering diagram with examples and then making the connection to cluster algebras. We argue that scattering diagrams, which represent a more general mathematical framework than cluster algebras, are useful for studying cluster algebraic functions that appear in the symbol alphabet of $\mathcal{N}=4$ pSYM.

\indent The scattering diagram of a rank $N$ cluster algebra corresponds to a fan in $\mathbb{Z}^{N}$, where each cone in the fan corresponds to a different coordinate system for $\mathcal{X}$. Cones of the scattering diagram correspond to clusters of the cluster algebra. In the case of finite cluster algebras, crossing between adjacent cones in the scattering diagram always corresponds to a cluster mutation. For example, the scattering diagrams of $\overline{Gr(2,5)/T}$ and $\overline{Gr(2,6)/T}$ are provided in fig. \ref{fig:exameplA2}. Crucially, the scattering diagram perspective motivates an alternate set of coordinates for $\mathcal{X}$, denoted as $\hat{y}_{\gamma}$-variables. For a given cone/cluster, the $\hat{y}_{\gamma}$-variables can be written as monomials of the $\hat{y}$-variables and vice-versa. Therefore, the $\hat{y}_{\gamma}$-variables and $\hat{y}$-variables have the same multiplicative basis. 

\begin{figure}
\centering
  \begin{tikzpicture}[every node/.style={font=\footnotesize}]
	%\node at (1.875,4.3) {$\mathcal{A}_5[(1,2),3,(4,5)]$};
	%\fill [fill=red!50!white] (1,0) ++(2,2) -- (1,0) -- (0,0) -- (0,1) -- ++(2,2);0.5*
	\filldraw (0,0) circle (1pt) ;
    \coordinate (TL) at (-0.5*4,0.5*4);
    \coordinate (TC) at (0,0.5*4);
    \coordinate (TR) at (0.5*4,0.5*4);
    \coordinate (CL) at (-0.5*4,0);
    \coordinate (CC) at (0,0);
    \coordinate (CR) at (0.5*4,0);
    \coordinate (LL) at (-0.5*4,-0.5*4);
    \coordinate (LC) at (0,-0.5*4);
    \coordinate (LR) at (0.5*4,-0.5*4);
    \draw [thick] (TL.center) -- (LR.center);
    \draw [thick] (LC.center) -- (TC.center);
    \draw [thick] (CL.center) -- (CR.center);
    \coordinate (yT1) at (0,0.5*2);
    \coordinate (yT2) at (0,0.5*1);
    \coordinate (yT3) at (0,0.5*0.6666);
    \coordinate (yT4) at (0,0.5*0.5);
    \draw (yT1.center) -- (LR.center);
    \draw (yT2.center) -- (LR.center);
    \draw (yT3.center) -- (LR.center);
    \draw (yT4.center) -- (LR.center);
    
    \coordinate (yR1) at (0.5*4,0);
    \coordinate (yR2) at (0.5*1.3333,0);
    \coordinate (yR3) at (0.5*0.8,0);
    \coordinate (yR4) at (0.5*0.5714,0);
    \draw (yR1.center) -- (TL.center);
    \draw (yR2.center) -- (TL.center);
    \draw (yR3.center) -- (TL.center);
    \draw (yR4.center) -- (TL.center);
    
    \coordinate (yL1) at (-0.5*2,0);
    \coordinate (yL2) at (-0.5*1,0);
    \coordinate (yL3) at (-0.5*0.6666,0);
    \coordinate (yL4) at (-0.5*0.5,0);
    \draw (yL1.center) -- (LR.center);
    \draw (yL2.center) -- (LR.center);
    \draw (yL3.center) -- (LR.center);
    \draw (yL4.center) -- (LR.center);
    
    \coordinate (yB1) at (0,-0.5*4);
    \coordinate (yB2) at (0,-0.5*1.3333);
    \coordinate (yB3) at (0,-0.5*0.8);
    \coordinate (yB4) at (0,-0.5*0.5714);
    \draw (yB1.center) -- (TL.center);
    \draw (yB2.center) -- (TL.center);
    \draw (yB3.center) -- (TL.center);
    \draw (yB4.center) -- (TL.center);
    \node at (0,0) [circle,green,fill,inner sep=2pt]{};
	\end{tikzpicture}
  \caption{A schematic representation of the cone structure near the limiting ray in some 3-dimensional scattering diagram. We are looking down on the limiting ray, which corresponds to the green dot.}
  \label{fig:sketchofwallA21pre}
\end{figure}
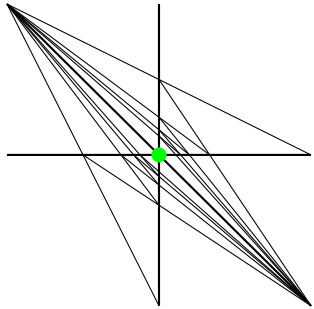

\indent In the finite case, the walls corresponding to cluster mutations define a complete scattering diagram. In some sense, the finite scattering diagram is simply a rewriting of the cluster algebra and contains no new information. In the infinite case, where there are an infinite number of cones, scattering diagrams are a genuine generalization of the cluster algebra framework. In particular, infinite sequences of cones appear in the scattering diagrams that asymptotically approach limiting rays, as schematically drawn in fig. \ref{fig:sketchofwallA21pre}. We use these infinite sequences of cones to define the notion of asymptotic chambers: cones that are asymptotically close to the limiting ray. Although there are always an \textit{infinite} number of walls as you approach the limiting ray, we argue that walls not intersecting the limiting ray can be ignored when calculating relations between the $\hat{y}_{\gamma}$-variables in this asymptotic limit. For example, there are 6 asymptotic chambers in fig. \ref{fig:sketchofwallA21pre} as only three walls intersect the limiting ray. We can calculate relations between the $\hat{y}_{\gamma}$-variables of distinct asymptotic chambers using the wall crossing framework. 

\indent The initial motivation for asymptotic chambers actually came from $\mathcal{N}=2$ super-symmetric gauge theories. For specific $\mathcal{N}=2$ gauge theories on $\mathbb{R}^{3}\times S^{1}$, the moduli space corresponds to a $\mathcal{X}$-type cluster algebra \cite{Gaiotto:2010be}. This connection between cluster algebras and $\mathcal{N}=2$ gauge theories led to a number of interesting results, such as a connection between canonical bases of the cluster algebra and the set of simple line defects in the theory \cite{Cordova:2013bza}. The concept of an asymptotic chamber was proposed in ref. \cite{Gaiotto:2010be}, although initial calculations were first performed in Section 5.9 of ref. \cite{Gaiotto:2009hg} using different terminology. Later generalizations made connections between asymptotic chambers and Fenchel-Nielsen coordinates of (higher) Teichmuller spaces \cite{Gaiotto:2012rg,Hollands:2013qza,Hollands:2017ahy}. However, to our knowledge, the notion of asymptotic chambers in the context of higher dimension scattering diagrams has been largely unstudied for general $\mathcal{X}$ spaces. 

\indent Crucially, although the $\hat{y}$-variables often diverge in the asymptotic limit, the $\hat{y}_{\gamma}$-variables themselves remain finite. These ``asymptotic'' $\hat{y}_{\gamma}$-variables correspond to the algebraic letters that appear in the 8-point symbol alphabet! Using scattering diagrams and the notion of asymptotic chambers, we conjecture a complete multiplicative basis for all algebraic letters that could appear in the $\mathcal{N}=4$ pSYM symbol alphabet at 8-point. Remarkably, we found at most 52 multiplicatively independent algebraic letters associated with the asymptotic chambers. This result systematizes the techniques in refs. \cite{Henke:2019hve,Drummond:2019cxm}, which effectively analyzed a particular subset of asymptotic chambers and did not study the relations between the algebraic letters of different asymptotic chambers. 

\indent The scattering diagram approach also offers a new perspective on proposed truncation procedures for $\hat{y}$-variables. We take a similar philosophy to refs. \cite{Arkani-Hamed:2019rds,Henke:2019hve,Drummond:2019cxm}, arguing that the positive kinematic region is not maximally compactified, so not all boundaries appear. However, in contrast to refs. \cite{Arkani-Hamed:2019rds,Henke:2019hve,Drummond:2019cxm}, which argue for a truncation of the $x$-variables, we instead argue for a truncation of clusters in the cluster algebra, or equivalently cones in the scattering diagram. We further argue that such a truncation naturally leads to the notion of asymptotic chambers and algebraic critical coordinates. 

\subsection{Outline of this paper}
\indent The paper is structured as follows:

\begin{itemize}
    \item Section \ref{sec:wallcrossingasymptotquiver}: We first introduce the notion of scattering diagrams and wall crossing for finite cluster algebras before defining the notion of an asymptotic chamber. We show that the cluster algebraic functions follow naturally from the notion of asymptotic chambers. The core result of this section is the conjectured bound in eq. (\ref{boundonclusteralgebraicfunctionss}), which is a necessary condition for asymptotic chambers to be well defined.  
    \item Section \ref{sec:toymodelclusteralgebrass}: We study asymptotic chambers in several examples, eventually studying the asymptotic chambers associated with $\overline{Gr(4,8)/T}$. The core results of this section are eqs. (\ref{eq:initialparameterizyhat}), (\ref{eq:explciitmapfrominitialtoprinc}), (\ref{eq:explicitsolutiongr48princ}) and (\ref{eq:explicitalphaGr48}), which together give an explicit alphabet for the algebraic letters at 8-point in terms of momentum twistors. We conclude with some comments about possible obstructions to applying the same techniques beyond 8-point.
    \item Section \ref{sec:degscatteringtrop}: We introduce the notion of degenerate scattering diagrams, motivating their construction using tropicalization and the dual cluster algebra. We then motivate asymptotic chambers using degenerate scattering diagrams.
    \item Section \ref{sec:conclusionsss}: We conclude the paper with a summary and a list of future directions. 
\end{itemize}

\noindent A short introduction to cluster algebras is provided in appendix \ref{sec:introductionclusters}. We restrict ourselves to cluster algebras whose exchange matrices are skew-symmetric, not just skew-symmetrizable. The techniques in this paper can be generalized to cluster algebras with non-skew-symmetric exchange matrices, but the formulas in this paper would require additional tweaking. \newline \newline

\noindent \textbf{Notation}: We denote the cluster variables associated with $\mathcal{A}$ and $\mathcal{X}$ as $x$ and $\hat{y}$ respectively. This notation differs from refs. \cite{Arkani-Hamed:2019rds,Henke:2019hve}, which denote cluster variables associated with $\mathcal{A}$ and $\mathcal{X}$ as $a$ and $x$ respectively. Furthermore, we denote mutations of the $k$th node as $\mu_{k}$. For example, $x_{i}=\mu_{k}x_{i}$ if $i\neq k$. Finally, we often denote cluster algebras using the notation $A_{p_{1},p_{2},\ldots,p_{n}}$, because these cluster algebras correspond to the Teichmuller space of bordered Riemann surfaces. The cluster algebra $A_{p_{1},p_{2},\ldots,p_{n}}$ corresponds to the Teichmuller space of a Riemann surface with $n$ borders and $p_{i}$ punctures on border $i$. For those unfamiliar with the connection between cluster algebra and surfaces, this notation is unimportant for our applications to $\mathcal{N}=4$ pSYM but is nice for organizational purposes. 

\section{Wall crossing, cluster algebras, and asymptotic chambers}\label{sec:wallcrossingasymptotquiver}

In this section, we develop the notion of scattering diagrams and asymptotic chambers. We begin with a short introduction to $g$-vectors before giving a relation between the scattering diagram and the $g$-vector fan of the cluster algebra. We then develop the notion of asymptotic chambers, using the $A_{1,1}$ cluster algebra as our guide. 

\subsection{Principal quivers and the \texorpdfstring{$g$}{g}-vector fan}
\label{sec:principalquiverasymptotcones}

Our goal is to find a minimal multiplicative basis of the $\hat{y}$-variables that parameterize the positive region of $\mathcal{X}$. Unfortunately, the set of $\hat{y}$-variables is very difficult to study for a cluster algebra with generic frozen variables. For example, $\hat{y}$-variables will not always be independent. To see the problem, consider the initial quiver, 
\[
\begin{tikzcd}
x_{1} \arrow[r]& x_{2} & x_{3}\arrow[l] 
\end{tikzcd}
\]
\noindent so 
\begin{equation}
\hat{y}_{1}=\frac{1}{x_{2}}, \quad \hat{y}_{2}=x_{1}x_{3}, \quad \hat{y}_{3}=\frac{1}{x_{2}} \ .  
\end{equation}
\noindent Without any frozen nodes, we trivially see that $\hat{y}_{1}=\hat{y}_{3}$. However, suppose we include the frozen node
\[
\begin{tikzcd}
y_{1}\arrow[d]& & \\
x_{1} \arrow[r]& x_{2} & x_{3}\arrow[l] \ ,
\end{tikzcd}
\]
\noindent so $\hat{y}_{1}=y_{1}/x_{2}$ and $\hat{y}_{1}\neq \hat{y}_{3}$. From this example, it is clear that the frozen nodes play a crucial role in distinguishing $\hat{y}$-variables. One approach to this problem is to simply add frozen nodes until the $\hat{y}$-variables are maximally disambiguated \cite{2006math......2259F,2018arXiv180302492S}. Only a finite, albeit large, number of frozen nodes are necessary to maximally disambiguate the $\hat{y}$-variables. 

\indent However, we are not interested in the set of all $\hat{y}$-variables but instead finding a multiplicatively independent basis. Given that any $\hat{y}$-variable can be written as a monomial of $x$-variables, we only need to maximally disambiguate $x$-variables of the cluster algebra, not the $\hat{y}$-variables. We are therefore motivated to consider a cluster algebra with a principal quiver \cite{2006math......2259F}. To construct a principal quiver,  consider an initial quiver without any frozen nodes. Then add a frozen node, $y_{i}$, to each non-frozen node, $x_{i}$, with an edge pointing from the frozen node to the mutable node. For example, the quiver 
\begin{equation}\label{A2princquiverex}
\begin{tikzcd}
y_{1} \arrow[d]& y_{2}\arrow[d] \\
x_{1} \arrow[r] & x_{2} \  
\end{tikzcd}
\end{equation}
\noindent is a principle quiver of the $A_{2}$ cluster algebra. Remarkably, the frozen nodes of a principal quiver are enough to maximally disambiguate all $x$-variables! Details of this statement are provided in appendix \ref{ref:differentiatingyvar}. We subsequently study cluster algebras with principle quivers to study the multiplicative basis of $\hat{y}$-variables of cluster algebras with arbitrary frozen nodes. Furthermore, we can choose any quiver of our cluster algebra to be the principal quiver. 

\indent We now turn to the problem of understanding the relation between $\hat{y}$-variables and $x$-variables for a cluster algebra with a principal quiver. Although we cannot write a direct map from $\hat{y}$ to $x$, attempting to do so allows us to associate a canonical vector to each $x$-variable. Suppose we start with the principal quiver. Any $x$-variable in the cluster algebra can be written as a Laurent polynomial of the $x$-variables and $y$-variables of the principle quiver. It is not generally possible to re-write this Laurent polynomial entirely in terms of $\hat{y}$-variables. However, it can be written as a polynomial of $\hat{y}$-variables of the principle quiver up to a monomial of $x$-variables of the principle quiver:
\begin{equation}\label{eq:conjecturFpolynomiagvec}
x=x^{\vec{g}}F(\hat{y}_{i}), \quad x^{\vec{g}}=\prod_{i}x_{i}^{g_{i}}  \ , 
\end{equation}
\noindent where $F(\hat{y}_{i})$ is a Laurent polynomial in $\hat{y}$-variables of the principal quiver, which we denote as $\hat{y}_{i}$. No two $x$-variables share the same $g$-vector, allowing us to associate a canonical $g$-vector to each element of the cluster algebra. As an illustrative example, again consider the $A_{2}$ cluster algebra. The $\hat{y}$-variables of (\ref{A2princquiverex}) are
\begin{equation}
\hat{y}_{1}= y_{1}x_{2}^{-1}, \quad \hat{y}_{2}=y_{2}x_{1} \ .
\end{equation}
\noindent Upon mutating $x_{1}$, we find 
\begin{equation}
\begin{split}
x_{3}&=\frac{y_{1}+x_{2}}{x_{1}}=\frac{x_{2}}{x_{1}}(1+\hat{y}_{1}) \\ &\rightarrow \ \vec{g}=(-1,1), \quad F(\hat{y}_{i})=1+\hat{y}_{1} \ .
\end{split}
\end{equation}
\noindent Mutating through all clusters yields all $F(\hat{y}_{i})$ polynomials and $g$-vectors of the $A_{2}$ cluster algebra, which are provided in table \ref{fig:A2gvectorpolynomials}. Each cluster defines a cone bounded by the $g$-vectors of the $x$-variables in the cluster. Remarkably, the cones associated with distinct clusters are nonoverlapping, which is not at all obvious from the above definition. The collection of these cones defines a (sometimes incomplete) fan.  

\begin{table}
\begin{center}
\begin{tabular}{ | m{1cm} | m{2cm}| m{1cm} | } 
\hline
$x_{i}$ & $F_{i}$ & $g_{i}$ \\ 
\hline
$x_{1}$ & $1$ & $(1,0)$ \\ 
\hline
$x_{2}$ & $1$ & $(0,1)$ \\ 
\hline
$x_{3}$ & $1+\hat{y}_{1}$ & $(-1,1)$ \\ 
\hline
$x_{4}$ & $1+\hat{y}_{1}+\hat{y}_{1}\hat{y}_{2}$ & $(-1,0)$ \\ 
\hline
$x_{5}$ & $1+\hat{y}_{2}$  & $(0,-1)$ \\ 
\hline
\end{tabular}
\end{center}
\caption{The $F(\hat{y}_{i})$ polynomials and $g$-vectors of the $A_{2}$ cluster algebra. }
\label{fig:A2gvectorpolynomials}
\end{table}
\indent In summary, we reduced the problem of finding a multiplicative basis of the $\hat{y}$-variables of a cluster algebra with generic frozen variables to finding a multiplicative basis of the $\hat{y}$-variables of a cluster algebra with a principal quiver. We then used the $\hat{y}$-variables of the principle quiver to find a map from $x$-variables to $g$-vectors.

\subsection{Scattering diagrams and wall crossing}\label{wallcrossingintroduction}

\indent In this section, we introduce the notion of scattering diagrams and wall crossing, following the review in ref. \cite{nathanreadingslides}. We then show how cluster algebras fit into the wall crossing framework, using the $A_{2}$ cluster algebra as our primary example.   

\indent A scattering diagram is defined on a lattice, $\mathbb{Z}^{N}$. We denote vectors as $\gamma$ and basis vectors as $\gamma_{i}$.\footnote{For example, if $N=3$, then $\gamma_{1}=(1,\ 0,\ 0)$, $\gamma_{2}=(0,\ 1,\ 0)$ and $\gamma_{3}=(0,\ 0,\ 1)$.} A scattering diagram requires three pieces of input data:

\begin{itemize}
    \item A collection of cones bounded by co-dimension 1 walls. Each wall in the scattering diagram is associated with a scalar function, $f(y)$. 
    \item $N$ coordinates on $\mathcal{X}$, denoted as $\hat{y}_{\gamma_{i}}$. Each coordinate corresponds to a basis vector.\footnote{We use the notation $\hat{y}_{\gamma_{i}}$, instead of $\hat{y}_{i}$, to distinguish them from $\hat{y}$-variables.}
    \item A skew-symmetric matrix, $B_{i,j}^{0}$, that defines a skew-symmetric\footnote{The scattering diagram framework can also be applied when the product is skew-symmetrizable instead of just skew-symmetric, but the following formulas requires modifications. See ref. \cite{Gaiotto:2010be}.} product for the $\gamma$,
    \begin{equation}\label{eq:skewsymprod}
    \langle \gamma_{i},\gamma_{j}\rangle=\gamma_{i}\cdot B^{0}\cdot \gamma_{j} \ .
    \end{equation}
\end{itemize}
\noindent Each cone in the fan is associated with a particular parameterization of $\mathcal{X}$ similar to how the $\hat{y}$-variables of a cluster correspond to a particular parameterization of $\mathcal{X}$. Crossing a wall between two cones corresponds to a coordinate transformation.

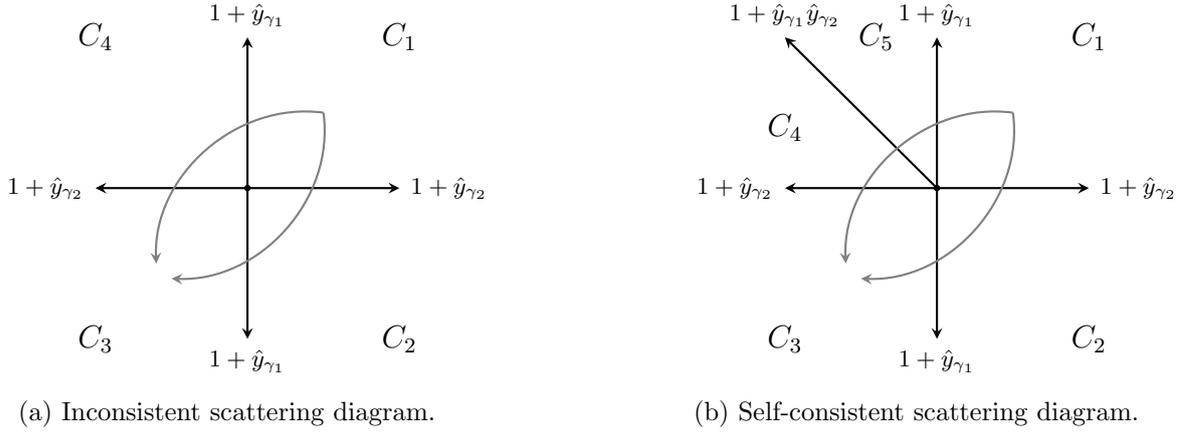
\begin{figure}
\centering
\begin{subfigure}[b]{0.4\textwidth}
    \begin{tikzpicture}[every node/.style={font=\footnotesize}]
	\draw [thick,-stealth] (0,0) -- (2,0) node[right=0.2pt]{$1+\hat{y}_{\gamma_{2}}$};
	\draw [thick,-stealth] (0,0) -- (-2,0) node[left=0.2pt]{$1+\hat{y}_{\gamma_{2}}$};
	\draw [thick,-stealth] (0,0) -- (0,2) node[above=0.2pt]{$1+\hat{y}_{\gamma_{1}}$};
	\draw [thick,-stealth] (0,0) -- (0,-2)node[below=0.2pt]{$1+\hat{y}_{\gamma_{1}}$} ;
	\filldraw (0,0) circle (1pt) ;
	\coordinate (f1) at (1,1);
    \coordinate (f2) at (-1.2,-1);
    \coordinate (f3) at (-1,-1.2);
	\draw [thick,gray,-stealth] (f1.center) to [bend right=50] (f2.center);
	\draw [thick,gray,-stealth] (f1.center) to [bend left=50] (f3.center);
	\draw (2,2)node{\large $C_{1}$}; 
	\draw (2,-2)node{\large $C_{2}$};
	\draw (-2,-2)node{\large $C_{3}$}; 
	\draw (-2,2)node{\large $C_{4}$}; 
	\end{tikzpicture}
    \caption{Inconsistent scattering diagram. }
    \label{fig:badexample2}
    \end{subfigure}
    \hfill
    \begin{subfigure}[b]{0.4\textwidth}
    \begin{tikzpicture}[every node/.style={font=\footnotesize}]
	\draw [thick,-stealth] (0,0) -- (-2,2) node[above=0.2pt]{$1+\hat{y}_{\gamma_{1}}\hat{y}_{\gamma_{2}}$};
	\draw [thick,-stealth] (0,0) -- (2,0) node[right=0.2pt]{$1+\hat{y}_{\gamma_{2}}$};
	\draw [thick,-stealth] (0,0) -- (-2,0) node[left=0.2pt]{$1+\hat{y}_{\gamma_{2}}$};
	\draw [thick,-stealth] (0,0) -- (0,2) node[above=0.2pt]{$1+\hat{y}_{\gamma_{1}}$};
	\draw [thick,-stealth] (0,0) -- (0,-2)node[below=0.2pt]{$1+\hat{y}_{\gamma_{1}}$} ;
	\filldraw (0,0) circle (1pt) ;
	\coordinate (f1) at (1,1);
    \coordinate (f2) at (-1.2,-1);
    \coordinate (f3) at (-1,-1.2);
	\draw [thick,gray,-stealth] (f1.center) to [bend right=50] (f2.center);
	\draw [thick,gray,-stealth] (f1.center) to [bend left=50] (f3.center);
	\draw (2,2)node{\large $C_{1}$}; 
	\draw (2,-2)node{\large $C_{2}$};
	\draw (-2,-2)node{\large $C_{3}$}; 
	\draw (-2,0.8)node{\large $C_{4}$};
	\draw (-0.8,2)node{\large $C_{5}$};
	\end{tikzpicture}
    \caption{Self-consistent scattering diagram. }
    \label{fig:goodexample2}
    \end{subfigure}
  \caption{Two examples of scattering diagrams. The scattering diagram on the left is inconsistent if $B_{i,j}^{0}$ equals eq. (\ref{fixedBijmatrix}). The relations between $\hat{y}_{\gamma}$ is not path independent, as shown in eq. (\ref{examplemutationpathssss}). The scattering diagram on the right is path-independent and can be identified with the $A_{2}$ cluster algebra. }
  \label{fig:4wallexameplA2}
\end{figure}

We now describe the coordinate transformation. Each co-dimension one wall is associated with a vector, $\gamma^{\perp}$, perpendicular to the wall, 
\begin{equation}\label{perpvectordefin}
\gamma^{\perp}=a_{i}\gamma_{i} \ .
\end{equation}
The sign of $\gamma^{\perp}$ is chosen so $\gamma^{\perp}$ points opposite the direction one is mutating across the wall. Furthermore, the magnitude of $\gamma^{\perp}$ is chosen so that all of its components, the $a_{i}$ in eq. (\ref{perpvectordefin}), are integers whose least common denominator is 1. Finally, we associate a unique monomial, $\hat{y}_{\gamma^{\perp}}$, to each $\gamma^{\perp}$:
\begingroup
\allowdisplaybreaks
\begin{align}
\hat{y}_{\gamma^{\perp}}&=(\prod \hat{y}_{\gamma_{i}}^{a_{i}})^{\textrm{Sign}(\gamma^{\perp}\cdot \vec{N})} \ , \\
\vec{N}&=(1,1,\ldots,1) \ . \label{definmultiplicativebasis}
\end{align}
\endgroup
\noindent For example, for a wall with the perpendicular vector $\gamma^{\perp}=( 0,\ 1,\ 1 )$, the associated monomial is 
\begin{equation}
\begin{split}
\hat{y}_{( 0,\ 1,\ 1)}= \hat{y}_{\gamma_{2}}\hat{y}_{\gamma_{3}} \ . 
\end{split}
\end{equation}
\noindent Due to the $\textrm{Sign}(\gamma^{\perp}\cdot \vec{N})$ exponent, the perpendicular vector $\gamma^{\perp}=( 0,\ -1,\ -1 )$ is associated with the same monomial,
\begin{equation}
\begin{split}
\hat{y}_{( 0,\ -1,\ -1)}=\hat{y}_{\gamma_{2}}\hat{y}_{\gamma_{3}} \ . 
\end{split}
\end{equation}
\noindent This makes sense as $( 0,\ -1,\ -1 )$ and $( 0,\ 1,\ 1 )$ correspond to the same wall and should therefore be associated with the same monomial. Although $\gamma^{\perp}$ flips sign depending on the direction you are mutating across the wall, $\hat{y}_{\gamma^{\perp}}$ is the same due to the $\textrm{Sign}(\gamma^{\perp}\cdot \vec{N})$ exponent. The mutation relation for $\hat{y}_{\gamma_{i}}$ across a wall is 
\begin{equation}\label{mutationsss}
\hat{y}_{\gamma_{i}}\rightarrow \hat{y}_{\gamma_{i}}f(\hat{y}_{\gamma^{\perp}})^{\langle \gamma_{i}, \gamma^{\perp}\rangle } \ ,
\end{equation}
\noindent which gives the $\hat{y}_{\gamma_{i}}$ of the new cone in terms of $\hat{y}_{\gamma_{i}}$ of the initial cone. To see eq. (\ref{mutationsss}) in an explicit example, suppose we are crossing from cone $C_{1}$ to cone $C_{2}$ in fig. \ref{fig:badexample2}, where we fix, 
\begin{equation}\label{fixedBijmatrix}
B^{0}_{i,j}=\begin{bmatrix}
0 & 1\\ 
-1 & 0
\end{bmatrix}  \ ,
\end{equation}
\noindent and $f(y)=1+y$ for all walls. The perpendicular vector for the relevant wall is $\gamma^{\perp}=(0,1)$, so $\hat{y}_{\gamma^{\perp}}=\hat{y}_{\gamma_{2}}$. Applying eq. (\ref{mutationsss}), the $\hat{y}_{\gamma_{i}}$ of chamber $C_{2}$ are then
\begin{equation}\label{eq:exmutationss}
\begin{split}
\hat{y}_{\gamma_{1}}&=\hat{y}_{\gamma_{1}}^{\text{I}}(1+\hat{y}_{\gamma_{2}}^{\text{I}})^{(1,0)\cdot B^{0}\cdot (0,1)}=\hat{y}_{\gamma_{1}}^{\text{I}}(1+\hat{y}_{\gamma_{2}}^{\text{I}}) \ , \\
\hat{y}_{\gamma_{2}}&=\hat{y}_{\gamma_{2}}^{\text{I}}(1+\hat{y}_{\gamma_{2}}^{\text{I}})^{(0,1)\cdot B^{0}\cdot (0,1)}=\hat{y}_{\gamma_{2}}^{\text{I}} \ , \\
\end{split}
\end{equation}
\noindent where $\hat{y}_{\gamma_{i}}^{\text{I}}$ corresponds to the $\hat{y}_{\gamma_{i}}$ of cone $C_{1}$. 

\indent For a scattering diagram to be self-consistent, the relations between the $\hat{y}_{\gamma_{i}}$ of any two cones should be path independent. To see why self-consistency is non-trivial, again consider the scattering diagram in fig. \ref{fig:badexample2} with the same $B_{i,j}^{0}$ and $f(y)$. Applying eq. (\ref{mutationsss}) to each path in fig. \ref{fig:badexample2}, we find 
\begingroup
\allowdisplaybreaks
\begin{align}
\begin{pmatrix}
\hat{y}^{\text{I}}_{\gamma_{1}}\\ 
\hat{y}^{\text{I}}_{\gamma_{2}}
\end{pmatrix}&\rightarrow \begin{pmatrix}
\hat{y}^{\text{I}}_{\gamma _1}\\ 
\frac{\hat{y}^{\text{I}}_{\gamma _2}}{\hat{y}^{\text{I}}_{\gamma _1}+1}
\end{pmatrix} \rightarrow \begin{pmatrix}
\hat{y}^{\text{I}}_{\gamma _1} \left(\frac{\hat{y}^{\text{I}}_{\gamma _2}}{\hat{y}^{\text{I}}_{\gamma
   _1}+1}+1\right)\\ 
\frac{\hat{y}^{\text{I}}_{\gamma _2}}{\hat{y}^{\text{I}}_{\gamma _1}+1}
\end{pmatrix} \ , \nonumber \\ 
\begin{pmatrix}
\hat{y}^{\text{I}}_{\gamma_{1}}\\ 
\hat{y}^{\text{I}}_{\gamma_{2}}
\end{pmatrix}&\rightarrow \begin{pmatrix}
\hat{y}^{\text{I}}_{\gamma _1} \left(\hat{y}^{\text{I}}_{\gamma _2}+1\right)\\ 
\hat{y}^{\text{I}}_{\gamma _2}
\end{pmatrix} \rightarrow \begin{pmatrix}
\hat{y}^{\text{I}}_{\gamma _1} \left(\hat{y}^{\text{I}}_{\gamma _2}+1\right)\\ 
\frac{\hat{y}^{\text{I}}_{\gamma _2}}{\hat{y}^{\text{I}}_{\gamma _1}
   \left(\hat{y}^{\text{I}}_{\gamma _2}+1\right)+1}
\end{pmatrix} \ , \label{examplemutationpathssss}
\end{align}
\endgroup
\noindent where $\hat{y}^{\text{I}}_{\gamma_{i}}$ again corresponds to the $\hat{y}_{\gamma_{i}}$ of the initial cone, $C_{1}$. The scattering diagram in fig. \ref{fig:badexample2} is inconsistent as the $\hat{y}_{\gamma_{i}}$ associated with $C_{3}$ are not path independent. To make the scattering diagram self-consistent, we must include the additional wall $\gamma^{\perp}=(1,1)$, leading to the scattering diagram in fig. \ref{fig:goodexample2}. Including this second wall, the first line in eq. (\ref{examplemutationpathssss}) becomes
\begin{equation}
\begin{split}
\begin{pmatrix}
\hat{y}^{\text{I}}_{\gamma_{1}}\\ 
\hat{y}^{\text{I}}_{\gamma_{2}}
\end{pmatrix}\rightarrow 
\begin{pmatrix}
\hat{y}^{\text{I}}_{\gamma _1}\\ 
\frac{\hat{y}^{\text{I}}_{\gamma _2}}{\hat{y}^{\text{I}}_{\gamma _1}+1}
\end{pmatrix} \rightarrow 
\begin{pmatrix}
\hat{y}^{\text{I}}_{\gamma _1} \left(\frac{\hat{y}^{\text{I}}_{\gamma _1} \hat{y}^{\text{I}}_{\gamma _2}}{\hat{y}^{\text{I}}_{\gamma _1}+1}+1\right)\\ 
\frac{\hat{y}^{\text{I}}_{\gamma _2}}{\hat{y}^{\text{I}}_{\gamma _1} \left(\hat{y}^{\text{I}}_{\gamma _2}+1\right)+1}
\end{pmatrix} \rightarrow
\begin{pmatrix}
\hat{y}^{\text{I}}_{\gamma _1} \left(\hat{y}^{\text{I}}_{\gamma _2}+1\right)\\ 
\frac{\hat{y}^{\text{I}}_{\gamma _2}}{\hat{y}^{\text{I}}_{\gamma _1} \left(\hat{y}^{\text{I}}_{\gamma _2}+1\right)+1} 
\end{pmatrix} \ ,
\end{split}    
\end{equation}
\noindent which now matches the second line of eq. (\ref{examplemutationpathssss}). 

\indent We now describe the connection between cluster algebras and scattering diagrams. The relation between scattering diagrams and cluster algebras is that the $g$-vector fan defines a scattering diagram where each cluster is dual to a cone in the scattering diagram. The $B_{i,j}^{0}$ matrix that defines the skew-symmetric product in eq. (\ref{eq:skewsymprod}) corresponds to the exchange matrix of the principal quiver. For a cluster algebra with a finite number of cones, each wall corresponds to a cluster mutation and we fix 
\begin{equation}
f(y)=1+y    
\end{equation}
\noindent for all walls. We call the walls that correspond to cluster mutations, \textit{cluster walls}. The $\hat{y}$-variables of a given cone are the $\hat{y}_{\gamma^{\perp}}$ associated with each wall that bounds the cone,
\begin{equation}\label{ygammaformulass}
\hat{y}_{j}=\prod \hat{y}_{\gamma_{i}}^{a_{i}^{j}}, \quad \gamma^{\perp}_{j}=a_{i}^{j}\gamma_{i} \ ,
\end{equation}
\noindent where $\gamma^{\perp}_{i}$ is the $\gamma^{\perp}$ associated with $\hat{y}_{i}$. Note that $\gamma^{\perp}_{j}$ points inward from the cone in this convention. Furthermore, the exchange matrix of the quiver associated with a cone is 
\begin{equation}
B_{i,j}=\langle \gamma^{\perp}_{i},\gamma^{\perp}_{j}\rangle \ .   
\end{equation}
\noindent For example, for the cone associated with the principle quiver, the principle cone, we have
\begin{equation}
\begin{split}
\gamma_{i}^{\perp}|_{\textrm{Principle Cone}}&=\gamma_{i}
\end{split}
\end{equation}
\noindent so 
\begin{equation}
\hat{y}_{\gamma_{i}}|_{\textrm{Principle Cone}}=\hat{y}_{i}, \quad B_{i,j}|_{\textrm{Principle Cone}}= \gamma_{i}\cdot B^{0}\cdot \gamma_{j}=B_{i,j}^{0} \ .
\end{equation}
\noindent The cluster mutation in eq. (\ref{yvariabels}) corresponds to both a wall crossing transform, eq. (\ref{mutationsss}), and a mutation in the $\gamma^{\perp}_{i}$. To see this, again consider the cluster algebra associated with the quiver
\[
\begin{tikzcd}
y_{1} \arrow[d]& y_{2}\arrow[d] \\
x_{1} \arrow[r] & x_{2} \ .
\end{tikzcd}
\]
\noindent The explicit computation of the $g$-vectors in table \ref{fig:A2gvectorpolynomials} reveals that it is the same as fig. \ref{fig:goodexample2}. Consider a mutation from cone $C_{1}$ to $C_{2}$. The $\hat{y}_{\gamma_{i}}$ mutation is given by eq. (\ref{eq:exmutationss}) and the $\gamma_{i}^{\perp}$ mutate as
\begin{equation}\label{eq:yvectmutateA2}
\begin{split}
\gamma^{\perp}_{1}=(1,0)&\rightarrow \gamma^{\perp}_{1}=(1,0) \ , \\
\gamma^{\perp}_{2}=(0,1)&\rightarrow \gamma^{\perp}_{2}=(0,-1) \ . 
\end{split}
\end{equation}
\noindent Combining eqs. (\ref{eq:exmutationss}) and (\ref{eq:yvectmutateA2}), the mutation relation for $\hat{y}_{i}$ is 
\begin{equation}
\begin{pmatrix}
\hat{y}_{1}\\ 
\hat{y}_{2}
\end{pmatrix}\rightarrow \begin{pmatrix}
\hat{y}_{1}(1+\hat{y}_{2})\\ 
\frac{1}{\hat{y}_{2}}
\end{pmatrix} \ ,
\end{equation}
\noindent which exactly matches the mutation relation for $\hat{y}$-variables. Again, note that it was a combination of mutating $\gamma_{i}^{\perp}$ and $\hat{y}_{\gamma_{i}}$ that gave the cluster mutation relation for the $\hat{y}$-variables. 

\indent We can also consider the scattering diagrams of more complex cluster algebras, such as the $A_{3}$ cluster algebra
\[
\begin{tikzcd}
y_{1} \arrow[d]& y_{2}\arrow[d]& y_{3}\arrow[d] \\
x_{1} \arrow[r] & x_{2}\arrow[r]& x_{3} \ ,
\end{tikzcd}
\]
\noindent which is associated with $\overline{Gr(2,6)/T}$. Since the cluster algebra is rank 3, the associated scattering diagram is 3 dimensional. From direct calculation, we found the scattering diagram given in fig. \ref{fig:exameplA2} in the Introduction. The walls are now 2 dimensional and defined by the span of two $g$-vectors. To find the wall associated with the $\hat{y}$-variable of a specific quiver, consider all the $g$-vectors bounding the dual cone except the $g$-vector of the $x$ variable associated with the same node as the $\hat{y}$-variable. The span of these two $g$-vectors defines the wall associated with the $\hat{y}$-variable.

\indent In summary, scattering diagrams are a useful framework that provide a nice way to study canonical coordinate transformations on $\mathcal{X}$. Finding self-consistent scattering diagrams is naively quite hard since you need to check that the relations between the $\hat{y}_{\gamma_{i}}$ of any two cones are path independent. The $g$-vector fans of finite cluster algebras provide a class of self-consistent scattering diagrams where cluster mutations correspond to a very specific type of wall crossing. 

\subsection{Asymptotic chambers and limiting walls}\label{sec:asymptotandlimwallss}
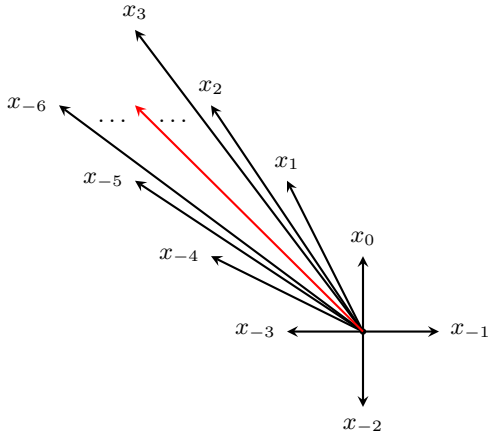
\begin{figure}
\centering
  \begin{tikzpicture}[every node/.style={font=\footnotesize}]
	%\node at (1.875,4.3) {$\mathcal{A}_5[(1,2),3,(4,5)]$};
	%\fill [fill=red!50!white] (1,0) ++(2,2) -- (1,0) -- (0,0) -- (0,1) -- ++(2,2);
	\filldraw (0,0) circle (1pt) ;
	\draw [thick,-stealth] (0,0) -- (1,0) node[right=0.2pt]{$x_{-1}$};
	\draw [thick,-stealth] (0,0) -- (-1,0) node[left=0.2pt]{$x_{-3}$};
	\draw [thick,-stealth] (0,0) -- (0,1) node[above=0.2pt]{$x_{0}$};
	\draw [thick,-stealth] (0,0) -- (0,-1)node[below=0.2pt]{$x_{-2}$} ;
	\draw (-2.5,2.8)node{$\ldots$} ;
	\draw [thick,-stealth] (0,0) -- (-1,2)node[above=0.2pt]{$x_{1}$} ;
	\draw [thick,-stealth] (0,0) -- (-2,3)node[above=0.2pt]{$x_{2}$} ;
	\draw [thick,-stealth] (0,0) -- (-3,4)node[above=0.2pt]{$x_{3}$} ;
	\draw (-3.3,2.8)node{$\ldots$} ;
	\draw [thick,-stealth] (0,0) -- (-2,1)node[left=0.2pt]{$x_{-4}$} ;
	\draw [thick,-stealth] (0,0) -- (-3,2)node[left=0.2pt]{$x_{-5}$} ;
	\draw [thick,-stealth] (0,0) -- (-4,3)node[left=0.2pt]{$x_{-6}$} ;
	
	\draw [thick,red,-stealth] (0,0) -- (-3,3) ;
	\end{tikzpicture}
  \caption{$g$-vector fan associated with the $A_{1,1}$ cluster algebra. There are an infinite number of cluster variables whose $g$-vectors approach a limiting ray, $\vec{g}_{lim}=(-1,1)$. The explicit form of the $g$-vectors is provided in eq. (\ref{eq:closedgvectosol}).}
  \label{fig:4wallexameplA1p1}
\end{figure}
We now turn to infinite cluster algebras. We will show how the scattering diagram framework provides a systematic way to study the multiplicative basis of $\hat{y}_{i}$ even when the $\hat{y}_{i}$ themselves go to infinity. Although $|\gamma_{i}^{\perp}|\rightarrow \infty$ in certain limits, so $\hat{y}_{i}\rightarrow \{ \infty,0\}$, the $\hat{y}_{\gamma}$-variables remain finite.  

\indent We first show that the $g$-vector fans of infinite cluster algebras need to include additional walls that do not correspond to cluster mutations. Furthermore, we will find the functions attached to these walls are not elements of the cluster algebra and can be identified with the mysterious cluster algebraic functions of ref. \cite{Arkani-Hamed:2019rds}. We will study the cluster algebra defined by the principal quiver
\[
\begin{tikzcd}
y_{-1} \arrow[d] & y_{0} \arrow[d] \\
x_{-1} \arrow[shift left, r]\arrow[shift right, r] & x_{0} \ ,
\end{tikzcd}
\]
\noindent as our motivating example. A review of relevant derivations and formulas for this cluster algebra are provided in appendix \ref{sec:a11cluster}. The key results are a closed form solution for $x_{i}$ with $i>0$,
\begin{equation}\label{eq:closedformexpressionforxnA1p1intext}
\begin{split}
x_{i}&=\frac{1}{2^{i+2}}[(x_{-1}+B_{+}\sqrt{\triangle})(\mathcal{P}+\sqrt{\triangle})^{i+1}+(x_{-1}-B_{+}\sqrt{\triangle})(\mathcal{P}-\sqrt{\triangle})^{i+1}] \ , \\
&\quad \mathcal{P}=\frac{y_{-1}}{x_{-1}x_{0}}+\frac{x_{0}}{x_{-1}}+\frac{x_{-1}y_{-1}y_{0}}{x_{0}} \ , \\
&\quad B_{+}=\frac{2x_{0}-x_{-1}\mathcal{P}}{\triangle} \ , \\
&\quad \triangle=\mathcal{P}^{2}-4y_{-1}y_{0}\ ,  
\end{split}
\end{equation}
\noindent and an equation for $\hat{y}_{2n-1}$ and $\hat{y}_{2n}$ in terms of $x$-variables after $2n$ mutations, 
\begin{equation}\label{eq:closedformsolution}
\hat{y}_{2n-1}=y_{0}^{n}y_{-1}^{2n+1}x_{2n}^{-2}, \quad \hat{y}_{2n}=y_{0}^{1-2n}y_{-1}^{-2n}x_{2n-1}^{2} \ .
\end{equation}
\noindent The $g$-vectors, denoted by black arrows in fig. \ref{fig:4wallexameplA1p1}, are
\begin{equation}\label{eq:closedgvectosol}
\vec{g}_{i}=\left\{\begin{matrix}
(-i,i+1) & i\geq -1 \\ 
(2+i,-i-3) & i\leq -2 
\end{matrix}\right. \ .
\end{equation}
We will now show that the self-consistency of the scattering diagram requires the existence of a new wall associated with the limiting ray that does not correspond to a standard cluster mutation. 

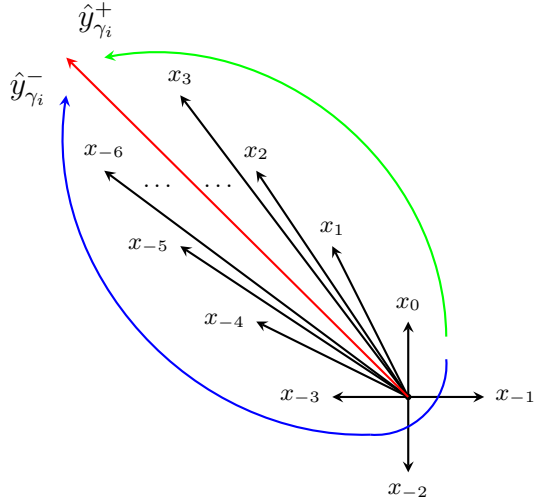
\begin{figure}
\centering
  \begin{tikzpicture}[every node/.style={font=\footnotesize}]
	%\node at (1.875,4.3) {$\mathcal{A}_5[(1,2),3,(4,5)]$};
	%\fill [fill=red!50!white] (1,0) ++(2,2) -- (1,0) -- (0,0) -- (0,1) -- ++(2,2);
	\filldraw (0,0) circle (1pt) ;
	\draw [thick,-stealth] (0,0) -- (1,0) node[right=0.2pt]{$x_{-1}$};
	\draw [thick,-stealth] (0,0) -- (-1,0) node[left=0.2pt]{$x_{-3}$};
	\draw [thick,-stealth] (0,0) -- (0,1) node[above=0.2pt]{$x_{0}$};
	\draw [thick,-stealth] (0,0) -- (0,-1)node[below=0.2pt]{$x_{-2}$} ;
	\draw (-2.5,2.8)node{$\ldots$} ;
	\draw [thick,-stealth] (0,0) -- (-1,2)node[above=0.2pt]{$x_{1}$} ;
	\draw [thick,-stealth] (0,0) -- (-2,3)node[above=0.2pt]{$x_{2}$} ;
	\draw [thick,-stealth] (0,0) -- (-3,4)node[above=0.2pt]{$x_{3}$} ;
	\draw (-3.3,2.8)node{$\ldots$} ;
	\draw [thick,-stealth] (0,0) -- (-2,1)node[left=0.2pt]{$x_{-4}$} ;
	\draw [thick,-stealth] (0,0) -- (-3,2)node[left=0.2pt]{$x_{-5}$} ;
	\draw [thick,-stealth] (0,0) -- (-4,3)node[above=0.2pt]{$x_{-6}$} ;
	\draw [thick,red,-stealth] (0,0) -- (-4.5,4.5) ;
	\coordinate (f1) at (0.5,0.8);
	\coordinate (f2) at (-4,4.5);
	\coordinate (f3) at (0.5,0.5);
	\coordinate (f4) at (-0.5,-0.5);
	\coordinate (f5) at (-4.5,4);
	\draw [thick,green,-stealth] (f1.center) to [bend right=50] (f2.center);
	\draw [thick,blue] (f3.center) to [bend left=50] (f4.center);
	\draw [thick,blue,-stealth] (f4.center) to [bend left=50] (f5.center);
	\draw (-4.1,5)node{\large $\hat{y}^{+}_{\gamma_{i}}$};
	\draw (-5,4.1)node{\large $\hat{y}^{-}_{\gamma_{i}}$};
	\end{tikzpicture}
  \caption{$g$-vector fan associated with the $A_{1,1}$ cluster algebra. There are two paths to cones asymptotically close to the limiting ray (red), which are green and blue respectively. The green path leads to the $\hat{y}^{+}_{\gamma_{i}}$ expressions while taking the blue path leads to the $\hat{y}^{-}_{\gamma_{i}}$ expressions. }
  \label{fig:4wallexameplA1p1withpaths}
\end{figure}

\indent Consider cones that are asymptotically close to the limiting ray. Importantly, the $\hat{y}_{i}$ variables go to $0$ or $\infty$ as we approach the limiting ray, which can be seen from eqs. (\ref{eq:closedformexpressionforxnA1p1intext}) and (\ref{eq:closedformsolution}). To calculate $\hat{y}_{\gamma_{i}}$ in this limit, we first express $\hat{y}_{\gamma_{i}}$ in terms of monomials of $\hat{y}$-variables. From the scattering diagram in fig. \ref{fig:4wallexameplA1p1}, the $\gamma^{\perp}_{i}$ associated with the $\hat{y}_{i}$ in eq. (\ref{eq:closedformsolution}) are 
\begin{equation}
[\gamma_{-1}^{\perp}]^{2n}=( 1+2n,2n ),  \quad   [\gamma_{0}^{\perp}]^{2n}=(-2n,1-2n ) \ ,
\end{equation}
\noindent where $[\gamma_{i}^{\perp}]^{2n}$ is the perpendicular vector to the wall associated with node $x_{i}$ after $2n$ mutations. We subsequently found that 
\begin{equation}\label{gammaA1p1formula}
\begin{split}
\gamma_{1}&=(1-2n)[\gamma_{-1}^{\perp}]^{2n}-2n[\gamma_{0}^{\perp}]^{2n} \ , \\ 
\gamma_{2}&=2n[\gamma_{-1}^{\perp}]^{2n}+(2n+1)[\gamma_{0}^{\perp}]^{2n} \ . 
\end{split}
\end{equation}
\noindent Combining eqs. (\ref{ygammaformulass}) and (\ref{gammaA1p1formula}) gives a formula for $\hat{y}_{\gamma_{i}}$ in the asymptotic limit, denoted as $\hat{y}_{\gamma_{i}}^{+}$,
\begin{equation}\label{positivelimitss}
\begin{split}
\hat{y}_{\gamma_{1}}^{+}&=\lim_{n\rightarrow \infty}(\hat{y}_{2n-1})^{1-2n}(\hat{y}_{2n})^{-2n}=16 y_{-1}(x_{-1}+B_{+}\sqrt{\triangle})^{-2}(\mathcal{P}+\sqrt{\triangle})^{-2} \ , \\
\hat{y}_{\gamma_{2}}^{+}&=\lim_{n\rightarrow \infty} (\hat{y}_{2n-1})^{2n}(\hat{y}_{2n})^{2n+1}=\frac{y_{0}}{4}(x_{-1}+B_{+}\sqrt{\triangle})^{2} \ . 
\end{split}
\end{equation}
\noindent We used the explicit formulas for $\hat{y}_{i}$ in eqs. (\ref{eq:closedformexpressionforxnA1p1intext}) and (\ref{eq:closedformsolution}) to write the final expressions for $\hat{y}_{\gamma_{1}}^{+}$ and $\hat{y}_{\gamma_{2}}^{+}$ in terms of our initial cluster variables: $y_{-1}$, $y_{0}$, $x_{-1}$ and $x_{0}$. The final expressions in eq. (\ref{positivelimitss}) are finite and provide a multiplicative basis for the $\hat{y}$-variables asymptotically close to the limiting wall if one approaches from the right. A visualization of the path is given by the green line in fig. \ref{fig:4wallexameplA1p1withpaths}. 
We then repeated the same calculation, but following the blue line in fig. \ref{fig:4wallexameplA1p1withpaths}. If one approaches from the left, the asymptotic limits of $\hat{y}_{\gamma_{i}}$, denoted as $\hat{y}_{\gamma_{i}}^{-}$, are 
\begin{equation}\label{eq:negativelimitsss}
\begin{split}
\hat{y}_{\gamma_{1}}^{-}&=\hat{y}_{\gamma_{1}}^{+}(1-\frac{\mathcal{P}-\sqrt{\mathcal{P}^{2}-4y_{-1}y_{0}}}{\mathcal{P}+\sqrt{\mathcal{P}^{2}-4y_{-1}y_{0}}})^{-4} \ ,  \\
\hat{y}_{\gamma_{2}}^{-}&=\hat{y}_{\gamma_{2}}^{+}(1-\frac{\mathcal{P}-\sqrt{\mathcal{P}^{2}-4y_{-1}y_{0}}}{\mathcal{P}+\sqrt{\mathcal{P}^{2}-4y_{-1}y_{0}}})^{4} \ . 
\end{split}
\end{equation}
\noindent The fact that eqs. (\ref{positivelimitss}) and (\ref{eq:negativelimitsss}) are not equal indicates that the scattering diagram must include another wall to be self-consistent. However, eq. (\ref{eq:negativelimitsss}) can be re-written into the suggestive form
\begin{equation}\label{genmutationrelationsss}
\hat{y}_{\gamma_{i}}^{-} = \hat{y}_{\gamma_{i}}^{+}(1-\hat{y}_{\gamma^\perp})^{-2\langle \gamma_{i},\gamma^{\perp} \rangle }, \quad  \hat{y}_{\gamma^\perp}=\hat{y}_{\gamma_{1}}^{+}\hat{y}_{\gamma_{2}}^{+}=\hat{y}_{\gamma_{1}}^{-}\hat{y}_{\gamma_{2}}^{-}, \quad \gamma^{\perp}=(1,1) \ ,    
\end{equation}
\noindent which can be matched to eq. (\ref{mutationsss}) by requiring $f(\hat{y})=(1-\hat{y})^{-2}$. Eq. (\ref{genmutationrelationsss}) shows that we must include a \textit{limiting wall} with $\gamma^{\perp}=(1,1)$ for the scattering diagram to be self-consistent. The $\hat{y}_{\gamma^{\perp}}$ associated with the limiting wall,
\begin{equation}
\begin{split}
\hat{y}_{\gamma^{\perp}}&=\hat{y}_{\gamma_{1}}^{+}\hat{y}_{\gamma_{2}}^{+} \ , \\
&=\frac{\mathcal{P}-\sqrt{\mathcal{P}^{2}-4y_{-1}y_{0}}}{\mathcal{P}+\sqrt{\mathcal{P}^{2}-4y_{-1}y_{0}}} \ ,
\end{split}
\end{equation}
\noindent takes exactly the right form for eq. (\ref{genmutationrelationsss}) to be matched with eq. (\ref{mutationsss}). The limiting wall corresponds to the red line in figs. \ref{fig:4wallexameplA1p1} and \ref{fig:4wallexameplA1p1withpaths}. Performing a mutation across this limiting wall cannot be identified with a cluster mutation in the $A_{1,1}$ scattering diagram. From the perspective of the cluster algebra, these cones are separated by an infinite number of cluster mutations. Finally, the $\hat{y}_{\gamma^{\perp}}$ of the limiting wall obeys the bound $0<\hat{y}_{\gamma^{\perp}}<1$ in contrast to normal $\hat{y}$-variables which are just positive definite. 

\indent We now briefly compare our result to previous computations in the literature. Notably, one multiplicative basis of $\hat{y}_{\gamma_{i}}^{\pm}$ is the three algebraic functions identified in ref. \cite{Drummond:2019cxm} for a given $A_{1,1}$ cluster algebra. Furthermore, due to the bound $0<\hat{y}_{\gamma^{\perp}}<1$, the cluster algebraic function attached to the limiting walls seem more like the $u$-variables identified in refs. \cite{Arkani-Hamed:2019plo, Arkani-Hamed:2020tuz}, which obey similar bounds, than standard $\hat{y}$-variables. Finally, note that the $\hat{y}_{\gamma^{\perp}}$ attached to the limiting wall is a ratio of the cluster algebraic functions defined in ref. \cite{Arkani-Hamed:2019rds}. 

\begin{figure}
\centering
  \begin{tikzpicture}[every node/.style={font=\footnotesize}]
	%\node at (1.875,4.3) {$\mathcal{A}_5[(1,2),3,(4,5)]$};
	%\fill [fill=red!50!white] (1,0) ++(2,2) -- (1,0) -- (0,0) -- (0,1) -- ++(2,2);
	\filldraw (0,0) circle (1pt) ;
    \coordinate (TL) at (-4,4);
    \coordinate (TC) at (0,4);
    \coordinate (TR) at (4,4);
    \coordinate (CL) at (-4,0);
    \coordinate (CC) at (0,0);
    \coordinate (CR) at (4,0);
    \coordinate (LL) at (-4,-4);
    \coordinate (LC) at (0,-4);
    \coordinate (LR) at (4,-4);
    \draw [thick] (TL.center) -- (LR.center);
    \draw [thick,red] (LC.center) -- (TC.center);
    \draw [thick] (CL.center) -- (CR.center);
    
    \coordinate (yB1) at (4*0.5,-4*0.5);
    \coordinate (yB2) at (4*0.25,-4*0.25);
    \coordinate (yB3) at (4*0.1666666666,-4*0.166666666);
    \coordinate (yB4) at (4*0.125,-4*0.125);
    \coordinate (yB5) at (4*0.1,-4*0.1);
    \coordinate (yB6) at (4*0.0833333,-4*0.0833333);
    \coordinate (yB7) at (4*0.0714286,-4*0.0714286);
    \coordinate (yB8) at (4*0.0625,-4*0.0625);
    \draw [blue] (TC.center) -- (yB1.center);
    \draw [blue] (TC.center) -- (yB2.center);
    \draw [blue] (TC.center) -- (yB3.center);
    \draw [blue] (TC.center) -- (yB4.center);
    \draw [blue] (TC.center) -- (yB5.center);
    \draw [blue] (TC.center) -- (yB6.center);
    \draw [blue] (TC.center) -- (yB7.center);
    \draw [blue] (TC.center) -- (yB8.center);
    
    \coordinate (yM1) at (4*0.33333,0);
    \coordinate (yM2) at (4*0.2,0);
    \coordinate (yM3) at (4*0.142857,0);
    \coordinate (yM4) at (4*0.11111111,0);
    \coordinate (yM5) at (4*0.0909091,0);
    \coordinate (yM6) at (4*0.0769231,0);
    \coordinate (yM7) at (4*0.0666667,0);
    \coordinate (yM8) at (4*0.0588235,0);
    \draw [blue] (LC.center) -- (CR.center);
    \draw [blue] (LC.center) -- (yM1.center);
    \draw [blue] (LC.center) -- (yM2.center);
    \draw [blue] (LC.center) -- (yM3.center);
    \draw [blue] (LC.center) -- (yM4.center);
    \draw [blue] (LC.center) -- (yM5.center);
    \draw [blue] (LC.center) -- (yM6.center);
    \draw [blue] (LC.center) -- (yM7.center);
    \draw [blue] (LC.center) -- (yM8.center);

    \coordinate (yT1) at (-4*0.5,4*0.5);
    \coordinate (yT2) at (-4*0.25,4*0.25);
    \coordinate (yT3) at (-4*0.1666666666,4*0.166666666);
    \coordinate (yT4) at (-4*0.125,4*0.125);
    \coordinate (yT5) at (-4*0.1,4*0.1);
    \coordinate (yT6) at (-4*0.0833333,4*0.0833333);
    \coordinate (yT7) at (-4*0.0714286,4*0.0714286);
    \coordinate (yT8) at (-4*0.0625,4*0.0625);
    \draw [blue] (LC.center) -- (yT1.center);
    \draw [blue] (LC.center) -- (yT2.center);
    \draw [blue] (LC.center) -- (yT3.center);
    \draw [blue] (LC.center) -- (yT4.center);
    \draw [blue] (LC.center) -- (yT5.center);
    \draw [blue] (LC.center) -- (yT6.center);
    \draw [blue] (LC.center) -- (yT7.center);
    \draw [blue] (LC.center) -- (yT8.center);
    
    \coordinate (yMR1) at (-4*0.33333,0);
    \coordinate (yMR2) at (-4*0.2,0);
    \coordinate (yMR3) at (-4*0.142857,0);
    \coordinate (yMR4) at (-4*0.11111111,0);
    \coordinate (yMR5) at (-4*0.0909091,0);
    \coordinate (yMR6) at (-4*0.0769231,0);
    \coordinate (yMR7) at (-4*0.0666667,0);
    \coordinate (yMR8) at (-4*0.0588235,0);
    \draw [blue] (TC.center) -- (CL.center);
    \draw [blue] (TC.center) -- (yMR1.center);
    \draw [blue] (TC.center) -- (yMR2.center);
    \draw [blue] (TC.center) -- (yMR3.center);
    \draw [blue] (TC.center) -- (yMR4.center);
    \draw [blue] (TC.center) -- (yMR5.center);
    \draw [blue] (TC.center) -- (yMR6.center);
    \draw [blue] (TC.center) -- (yMR7.center);
    \draw [blue] (TC.center) -- (yMR8.center);
    
    \node at (0,0) [circle,green,fill,inner sep=2pt]{};

    \draw (2.5,2.5)node{\large $C_{1}$} ;
    \draw (3.2,-1.7)node{\large $C_{2}$} ;
    \draw (1.7,-3.2)node{\large $C_{3}$} ;
    \draw (-2.5,-2.5)node{\large $C_{4}$} ;
    \draw (-3.2,1.7)node{\large $C_{5}$} ;
    \draw (-1.7,3.2)node{\large $C_{6}$} ;
	\end{tikzpicture}
  \caption{A schematic representation of the cone structure near the limiting ray of $A_{2,1}$, a rank 3 cluster algebra. The full scattering diagram is 3 dimensional and we are looking down on the limiting ray, which is indicated by the green dot. The red line corresponds to the limiting wall. The black lines correspond to cluster walls that intersect the limiting ray. The blue lines correspond to asymptotic walls, cluster walls that do not intersect the limiting wall and become more parallel with a limiting wall as one approaches the limiting ray. There are $6$ asymptotic chambers, each labeled by $C_{i}$. }
  \label{fig:sketchofwallA21}
\end{figure}
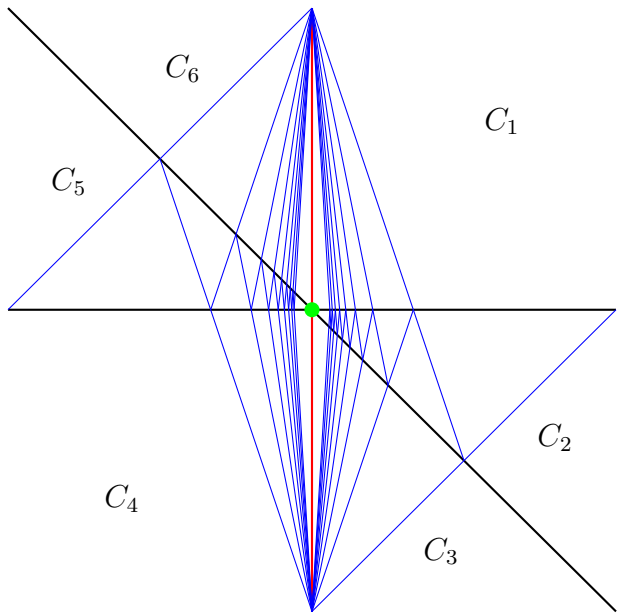

\indent Moving beyond $A_{1,1}$, we now turn to a more general discussion. We define asymptotic chambers as cones asymptotically close to the limiting ray that are separated by walls intersecting the limiting ray. For higher dimension scattering diagrams, both limiting walls and cluster walls intersect the limiting ray. Furthermore, there are always cluster walls asymptotically close to the limiting ray which do not intersect the limiting ray and become more parallel to the limiting walls as one approaches the limiting ray. These walls are \textit{asymptotic walls}. An example is sketched in fig. \ref{fig:sketchofwallA21} \cite{nimayoutubecluster}. For our definition of asymptotic chambers to be self-consistent, we must be able to ignore asymptotic walls if we are infinitesimally close to the limiting ray. If the $\hat{y}_{\gamma_{i}}$ associated with asymptotic chambers transformed non-trivially when crossing an asymptotic wall, the $\hat{y}_{\gamma_{i}}$ of asymptotic chambers would not be well defined. For example, consider the asymptotic chambers $C_{2}$ and $C_{5}$ in fig. \ref{fig:sketchofwallA21}. If $\hat{y}_{\gamma_{i}}$ transformed non-trivially across the asymptotic walls, it would be ambiguous which $\hat{y}_{\gamma_{i}}$ was associated with the asymptotic chamber. 

\indent To see whether asymptotic walls are relevant when infinitesimally close to the limiting ray, let us consider crossing one of these asymptotic walls. From the definition of asymptotic walls, the $\gamma^{\perp}$ of the asymptotic wall asymptotes to
\begin{equation}
\gamma^{\perp}\rightarrow \lim_{n\rightarrow \infty}n\times \gamma^{\perp}_{lim} \ ,
\end{equation}
\noindent where $\gamma^{\perp}_{lim}$ is the $\gamma^{\perp}$ of the limiting wall that the asymptotic wall approaches. Therefore, the wall crossing formula for the asymptotic wall reduces to
\begin{equation}\label{formualequationss}
\hat{y}_{\gamma_{i}}\rightarrow \lim_{n\rightarrow \infty}\hat{y}_{\gamma_{i}}(1+\hat{y}_{\gamma^{\perp}_{lim}}^{n})^{n\langle \gamma_{i}, \gamma^{\perp}_{lim}\rangle} \ ,
\end{equation}
\noindent which naively diverges. However, in the previous example, we found that $0<\hat{y}_{\gamma^{\perp}_{lim}}<1$ for $A_{1,1}$. If this bound holds for general $\hat{y}_{\gamma^{\perp}_{lim}}$, then eq. (\ref{formualequationss}) becomes trivial and the asymptotic walls can be ignored when asymptotically close to the limiting ray. We therefore conjecture the bound
\begin{equation}\label{boundonclusteralgebraicfunctionss}
0<\hat{y}_{\gamma^{\perp}_{lim}}<1 \ , 
\end{equation}
\noindent in $\mathcal{X}$ for all asymptotic chambers, not just those adjacent to the limiting wall. Eq. (\ref{boundonclusteralgebraicfunctionss}) is a very remarkable bound and a key conjecture of this paper. We explicitly checked that eq. (\ref{boundonclusteralgebraicfunctionss}) held for all examples studied in section \ref{sec:toymodelclusteralgebrass}. 

\indent In summary, the key insight is that the $\hat{y}_{\gamma_{i}}$ associated with asymptotic chambers are finite and can be algebraic functions of our initial coordinates. Furthermore, these $\hat{y}_{\gamma_{i}}$ obey the wall crossing formula as we mutate around the limiting ray. To find all the $\hat{y}_{\gamma_{i}}$ associated with a limiting ray, we simply need to find all the walls in the $g$-vector fan that intersect the limiting ray and then use the wall crossing formula in eq. (\ref{mutationsss}). The primary difficulty is finding all the walls that intersect the limiting ray.  

\section{Explicit calculations of asymptotic chambers}\label{sec:toymodelclusteralgebrass}

Our goal is to compute a minimal multiplicative basis for $\hat{y}_{\gamma_{i}}$ asymptotically close to limiting rays, the \textit{asymptotic symbol alphabet}. We now describe an algorithm for finding the asymptotic symbol alphabet associated with a limiting ray. A brute force search algorithm for finding asymptotic chambers is given in appendix \ref{sec:algorithlimitraytrue}. Once we found all the asymptotic chambers associated with a given limiting ray, the calculation for finding the associated symbol alphabet proceeded as follows:
\begin{enumerate}
\item Starting from the initial quiver, we performed mutations until we found a quiver with an $A_{1,1}$ subalgebra. We chose the quiver with the $A_{1,1}$ subalgebra as the principal quiver for the purposes of defining the scattering diagram and $g$-vectors.
\item Repeating the computation in section \ref{sec:asymptotandlimwallss} for the $A_{1,1}$ subalgebra, we computed the $\hat{y}_{\gamma_{i}}$ of an initial asymptotic chamber in terms of the $\hat{y}_{i}$ of the principal quiver. We denote the $\hat{y}_{\gamma_{i}}$ of the initial asymptotic chamber as $\hat{y}_{\gamma_{i}}^{0}$.
\item We computed the $\hat{y}_{\gamma_{i}}$ of all other asymptotic chambers in terms of the $\hat{y}_{\gamma_{i}}^{0}$ using wall crossing. 
\item We found a complete multiplicative basis of the asymptotic chambers' $\hat{y}_{\gamma_{i}}$ in terms of $\hat{y}_{\gamma_{i}}^{0}$.
\end{enumerate}
\noindent The multiplicative basis calculated in the final step is the asymptotic symbol alphabet associated with the limiting ray. Although each element of the multiplicative basis will be a rational function of $\hat{y}_{\gamma_{i}}^{0}$, the $\hat{y}_{\gamma_{i}}^{0}$ will themselves often be algebraic functions of $\hat{y}_{i}$.

\indent For the remainder of this section, we study a variety of cluster algebras using the above algorithm. We first dissect some lower rank cluster algebras, discussing a variety of phenomena that appear. We then move onto $\overline{Gr(4,8)/T}$, conjecturing a complete algebraic symbol alphabet for the 8-point MHV amplitude. We conclude this section by commenting on how we may need to modify the above algorithm when faced with more general types of limiting rays. 

\subsection{Lower rank cluster algebras}\label{sec:algorithlimitray}

We now consider the asymptotic chambers of some lower rank cluster algebras, finding several interesting phenomena:
\begin{itemize}
    \item $A_{2,1}$: Both cluster walls and limiting walls can intersect the limiting ray, leading to more non-trivial cluster algebraic functions.
    \item $A_{2,2}$: The scattering diagram associated with the limiting ray is not simple. A simple fan is an $N$-dimensional fan for which all cones are bound by $N$ walls.\footnote{Alternatively, a simple fan is a fan whose dual polytope is simple.}
    \item $A_{1,1,1}$: There can be multiple limiting rays and each limiting ray is associated with its own discriminant.
\end{itemize}

\subsubsection{Example: \texorpdfstring{$A_{2,1}$}{A21} }\label{sec:infiniteraysssA2p1}

\begin{figure}
\centering
  \begin{tikzpicture}[every node/.style={font=\footnotesize}]
	%\node at (1.875,4.3) {$\mathcal{A}_5[(1,2),3,(4,5)]$};
	%\fill [fill=red!50!white] (1,0) ++(2,2) -- (1,0) -- (0,0) -- (0,1) -- ++(2,2);
	\filldraw (0,0) circle (1pt) ;
    \coordinate (TL) at (-2,2);
    \coordinate (TC) at (0,2);
    \coordinate (TR) at (2,2);
    \coordinate (CL) at (-2,0);
    \coordinate (CC) at (0,0);
    \coordinate (CR) at (2,0);
    \coordinate (LL) at (-2,-2);
    \coordinate (LC) at (0,-2);
    \coordinate (LR) at (2,-2);
    \draw [thick] (TL.center) -- (LR.center);
    \draw [thick,red] (LC.center) -- (TC.center);
    \draw [thick] (CL.center) -- (CR.center);
    \draw (0,2.3)node{\large $\gamma_{c}^{\perp}$} ;
    \draw (2.3,0)node{\large $\gamma_{a}^{\perp}$} ;
    \draw (-2.3,2.3)node{\large $\gamma_{b}^{\perp}$} ;
    \draw (1.25,1.25)node{\large $C_{1}$} ;
    \draw (1.5,-0.5)node{\large $C_{2}$} ;
    \draw (0.5,-1.5)node{\large $C_{3}$} ;
    \draw (-1.25,-1.25)node{\large $C_{4}$} ;
    \draw (-1.5,0.5)node{\large $C_{5}$} ;
    \draw (-0.5,1.5)node{\large $C_{6}$} ;
    \coordinate (f1) at (-0.8,-0.8);
    \coordinate (f2) at (0.8,0.8);
    \draw [thick,gray,-stealth] (f2.center) to [bend left=70] (f1.center);
	\end{tikzpicture}
  \caption{The scattering diagram of asymptotic chambers near the limiting ray in the $A_{2,1}$ cluster algebra. We projected down onto the plane perpendicular to the limiting ray, $g_{lim}=(0,-1,0)$, and labeled the walls.}
  \label{fig:sketchofwallA21withoutasymptot}
\end{figure}
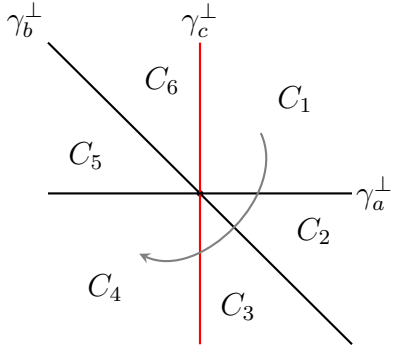

We will examine the $A_{2,1}$ cluster algebra in detail, so the algorithm is clear. The $A_{2,1}$ cluster algebra corresponds to the initial quiver, 
\[
\begin{tikzcd}
b \arrow[dr] & \\
z_{-1} \arrow[r]\arrow[u]& z_{0} \ ,
\end{tikzcd}
\]
\noindent where $b$, $z_{-1}$ and $z_{0}$ are $x$-variables and frozen variables have been suppressed. To find the limiting ray, we performed a mutation on $b$, finding the new quiver
\[
\begin{tikzcd}
x_{1}:\ b' \arrow[d] & \\
x_{2}:\ z_{-1} \arrow[shift left, r]\arrow[shift right, r]& x_{3}:\ z_{0} \arrow[ul] \ ,
\end{tikzcd}
\]
\noindent which we chose to be the principal quiver. The ``$x_{i}:$'' denotes which basis vector each $x$-variable of the principle quiver corresponds to. The corresponding exchange matrix is 
\begin{equation}\label{eq:initialswallss}
\begin{split}
B_{i,j}^{0}=\begin{bmatrix}
0 & 1 & -1\\ 
-1 & 0 & 2\\ 
1 & -2 & 0
\end{bmatrix} \ .
\end{split}    
\end{equation}
\noindent Identifying the $A_{1,1}$ subalgebra, we performed repeated mutations on the $x_{2}$ and $x_{3}$ nodes, just as in section \ref{sec:asymptotandlimwallss}, to approach the limiting ray. After repeatedly mutating the $x_{2}$ and $x_{3}$ nodes, the $g$-vectors of the $x$-variables associated with these nodes asymptotically approached
\begin{equation}
g_{lim}=(0,-1,1) \ ,    
\end{equation}
\noindent which we identified as the limiting ray. Using the algorithm in appendix \ref{sec:algorithlimitraytrue}, we found all the walls that intersect the limiting ray:
\begin{equation}\label{eq:wallss}
\begin{split}
\gamma^{\perp}_{a}&=(1,0,0) \ , \\
\gamma^{\perp}_{b}&=(1,1,1) \ , \\
\gamma^{\perp}_{c}&=(0,1,1) \ , \\
\end{split}
\end{equation}
\noindent where $\gamma_{c}$ corresponds to a limiting wall. A visualization of these walls is provided in fig. \ref{fig:sketchofwallA21withoutasymptot}, where we have taken a projection of the scattering diagram onto the plane perpendicular to the limiting ray.  This projection of the scattering diagram is the asymptotic scattering diagram. 

The $\hat{y}_{\gamma_{i}}$ of the initial asymptotic chamber, $\hat{y}_{\gamma_{i}}^{0}$, were then calculated using the same techniques as those in section \ref{sec:asymptotandlimwallss}. The $\gamma^{\perp}_{i}$ of the cone associated with the $2n$-th quiver in the sequence,
\[
\begin{tikzcd}
b' \arrow[d] & & b' \arrow[dr] & & b' \arrow[d] & & b' \arrow[dr] & &\\
z_{-1} \arrow[shift left, r]\arrow[shift right, r]& z_{0} \ , \arrow[ul] & z_{1} \arrow[u]& z_{0} \ , \arrow[shift left, l]\arrow[shift right, l] & z_{1} \arrow[shift left, r]\arrow[shift right, r]& z_{2} \ , \arrow[ul] & z_{3} \arrow[u]& z_{2} \ , \arrow[shift left, l]\arrow[shift right, l]  &  \ldots \ ,
\end{tikzcd}
\]
\noindent are 
\begin{equation}\label{eq:wallssatn}
\begin{split}
[\gamma_{1}^{\perp}]^{2n}&=(1,0,0) \ , \\
[\gamma_{2}^{\perp}]^{2n}&=(0,2n+1,2n) \ , \\
[\gamma_{3}^{\perp}]^{2n}&=(0,-2n,1-2n) \ . \\
\end{split}
\end{equation}
\noindent $[\gamma^{\perp}_{i}]^{2n}$ is the perpendicular vector of the wall associated with the node $x_{i}$ after $2n$ mutations and the limit $n\rightarrow \infty$ corresponds to our initial asymptotic chamber. Eq. (\ref{eq:wallssatn}) implies
\begin{equation}
\begin{split}
\gamma_{1}&=[\gamma_{1}^{\perp}]^{2n} \ , \\
\gamma_{2}&=(1-2n)[\gamma_{2}^{\perp}]^{2n}-2n[\gamma_{3}^{\perp}]^{2n} \ , \\
\gamma_{3}&=(2n)[\gamma_{2}^{\perp}]^{2n}+(2n+1)[\gamma_{3}^{\perp}]^{2n} \ , \\
\end{split}
\end{equation}
\noindent so
\begin{equation}\label{eq:finalequationforyhatgamma1a}
\begin{split}
\hat{y}_{\gamma_{1}}^{0}&=\lim_{n\rightarrow \infty} [\hat{y}_{1}]^{2n}=\hat{y}_{1} \frac{1}{2} \left(1+\hat{y}_{3} \hat{y}_{2}+\hat{y}_{2}+\sqrt{\triangle'}\right) \ , \\
\hat{y}_{\gamma_{2}}^{0}&=\lim_{n\rightarrow \infty}([\hat{y}_{2}]^{2n})^{1-2n}([\hat{y}_{3}]^{2n})^{-2n}=\frac{4 \hat{y}_{2} \triangle'}{\left(1-\hat{y}_{3} \hat{y}_{2}+\hat{y}_{2}+\sqrt{\triangle'}\right){}^2}\ , \\
\hat{y}_{\gamma_{3}}^{0}&=\lim_{n\rightarrow \infty} ([\hat{y}_{2}]^{2n})^{2n}([\hat{y}_{3}]^{2n})^{2n+1}=\frac{1}{4} \hat{y}_{3} \left(1+\frac{1-\hat{y}_{2} \left(\hat{y}_{3}+1\right)}{\sqrt{\triangle'}}\right){}^2\ , \\
&\quad \triangle'=\left(\hat{y}_{3}
   \hat{y}_{2}+\hat{y}_{2}+1\right){}^2-4 \hat{y}_{2} \hat{y}_{3} \ ,
\end{split}
\end{equation}
\noindent where $[\hat{y}_{i}]^{2n}$ is the $\hat{y}$-variable associated with the $x_{i}$ node after $2n$ mutations. Eq. (\ref{eq:finalequationforyhatgamma1a}) relates the $\hat{y}_{\gamma_{i}}$ of our initial asymptotic chamber to the $\hat{y}_{i}$ of our principal quiver. The formulas for $\hat{y}_{\gamma_{2}}^{0}$ and $\hat{y}_{\gamma_{3}}^{0}$ are exactly the same as eq. (\ref{positivelimitss}), except that we wrote the expression in terms of $\hat{y}$-variables of the principle quiver instead of $x$-variables. The formula for $\hat{y}_{\gamma_{1}}^{0}$ was derived using the closed form solution for the $x$-variables of the $A_{1,1}$ subalgebra in appendix \ref{sec:a11cluster} and noting that $b'$ never mutates.

\indent With the asymptotic chambers and eq. (\ref{eq:finalequationforyhatgamma1a}), we can mutate around the asymptotic scattering diagram to find all cluster algebraic functions associated with the limiting ray. For example, going along the path given in fig. \ref{fig:sketchofwallA21withoutasymptot}, the $\hat{y}_{\gamma_{i}}$ mutate as
\begin{equation}
\begin{split}
\begin{pmatrix}
\hat{y}^{0}_{\gamma_{1}} \\ 
\hat{y}^{0}_{\gamma_{2}} \\ 
\hat{y}^{0}_{\gamma_{3}}
\end{pmatrix} \rightarrow  \begin{pmatrix}
\hat{y}_{\gamma _1}^{0}  \\
\frac{\hat{y}_{\gamma _2}^{0}}{\hat{y}_{\gamma _1}^{0}+1} \\ 
\left(\hat{y}_{\gamma _1}^{0}+1\right) \hat{y}_{\gamma _3}^{0} 
\end{pmatrix} \rightarrow \begin{pmatrix}
\hat{y}_{\gamma _1}^{0}  \\
\frac{\hat{y}_{\gamma _2}^{0} \left(\hat{y}_{\gamma _1}^{0}
   \hat{y}_{\gamma _2}^{0} \hat{y}_{\gamma _3}^{0}+1\right)}{\hat{y}_{\gamma _1}^{0}+1} \\ 
\frac{\left(\hat{y}_{\gamma _1}^{0}+1\right) \hat{y}_{\gamma _3}^{0}}{\hat{y}_{\gamma
   _1}^{0} \hat{y}_{\gamma _2}^{0} \hat{y}_{\gamma _3}^{0}+1}
\end{pmatrix} \rightarrow \begin{pmatrix}
\hat{y}_{\gamma _1}^0\\
\frac{\hat{y}_{\gamma _2}^0  \left(\hat{y}_{\gamma _1}^0 \hat{y}_{\gamma _2}^0
   \hat{y}_{\gamma _3}^0+1\right)}{\left(1-\hat{y}_{\gamma _2}^0
   \hat{y}_{\gamma _3}^0\right)^{4}(\hat{y}_{\gamma _1}^0+1)}\\ \frac{\left(\hat{y}_{\gamma
   _1}^0+1\right)\left (1-\hat{y}_{\gamma _2}^0 \hat{y}_{\gamma
   _3}^0\right)^{4} \hat{y}_{\gamma _3}^0}{ \left(\hat{y}_{\gamma _1}^0 \hat{y}_{\gamma _2}^0 \hat{y}_{\gamma
   _3}^0+1\right)}
\end{pmatrix} \ . 
\end{split}    
\end{equation}
\noindent Again, the jump across the limiting wall corresponds to a generalized cluster mutation. Calculating the $\hat{y}_{\gamma_{i}}$ of all asymptotic chambers, we found the multiplicative basis
\begin{equation}\label{eq:finalalphabetsss}
\hat{y}^{0}_{\gamma_{1}}, \ \hat{y}^{0}_{\gamma_{2}}, \ \hat{y}^{0}_{\gamma_{3}}, \ (1+\hat{y}^{0}_{\gamma_{1}}), \ (1-\hat{y}^{0}_{\gamma_{2}}\hat{y}^{0}_{\gamma_{3}}), \ (1+\hat{y}^{0}_{\gamma_{1}}\hat{y}^{0}_{\gamma_{2}}\hat{y}^{0}_{\gamma_{3}}) \ .
\end{equation}
\noindent Eq. (\ref{eq:finalalphabetsss}) corresponds to all the algebraic functions associated with the limiting ray in the $A_{2,1}$ cluster algebra. Although the expressions in eq. (\ref{eq:finalalphabetsss}) look rational, remember that the $\hat{y}^{0}_{\gamma_{i}}$ are algebraic functions of $\hat{y}_{i}$. They are all algebraic in terms of $\hat{y}_{i}$ due to the presence of the quadratic root, $\sqrt{\triangle'}$.

\subsubsection{Example: \texorpdfstring{$A_{2,2}$}{A22}}\label{sec:infiniteraysssA2p2}

We now continue to the $A_{2,2}$ cluster algebra. The $A_{2,2}$ cluster algebra includes a cluster with the quiver
\[
\begin{tikzcd}
& x_{2}\arrow[dl]\arrow[dr]&\\
x_{3} \arrow[r]& x_{1}\arrow[shift left, u]\arrow[shift right, u] & x_{4} \arrow[l] \ ,
\end{tikzcd}
\]
which was chosen as our principal quiver. The asymptotic scattering diagram is slightly more complex than the $A_{2,1}$ cluster algebra, but the algorithm is the same. The limiting ray is 
\begin{figure}
\centering
  \includegraphics[scale=1]{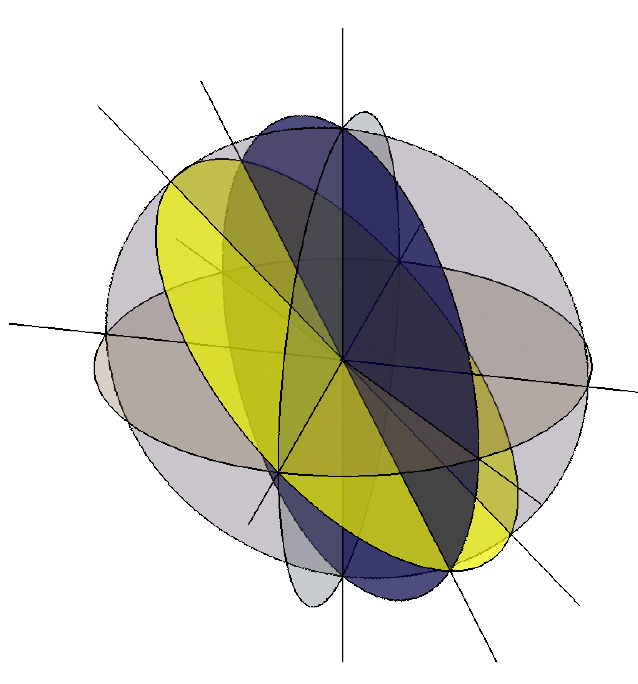} \\
  \caption{The scattering diagram associated with the asymptotic chambers of the $A_{2,2}$ cluster algebra.}
  \label{fig:A2p2asymptotcluster}
\end{figure}
\begin{equation}
g_{lim}=(-1,1,0,0)  \ .  
\end{equation}
The scattering walls are 
\begin{equation}
\begin{split}
\gamma^{\perp}\in \{ &(0,0,1,0), \ (0,0,0,1), \  (1,1,1,0), \\
&(1,1,0,1), \ (1,1,0,0) \} \ , \\
\end{split}    
\end{equation}
where the last element corresponds to the limiting wall. To visualize this scattering diagram, we project down to 3 dimensions using the basis, 
\begin{equation}
\begin{split}
\hat{e}_{1}'&=(1,1,0,0)\ , \\  
\hat{e}_{2}'&=(0,0,1,0)\ , \\  
\hat{e}_{3}'&=(0,0,0,1)\ , \\  
\end{split}    
\end{equation}
\noindent giving the asymptotic scattering diagram in fig. \ref{fig:A2p2asymptotcluster}. Note that the asymptotic scattering diagram associated with $A_{2,2}$ is not simple as there are cones bounded by 4 walls instead of 3.  

\indent We then calculated the $\hat{y}_{\gamma_{i}}^{0}$ variables of the initial asymptotic chamber in terms of the $\hat{y}$-variables of the principle quiver. The derivation is almost exactly as in section \ref{sec:infiniteraysssA2p1}, so we will not write it out here. The final result is,
\begin{equation}\label{eq:finalequationforyhatgamma1}
\begin{split}
i\in \{3,4 \}:& \ \hat{y}_{\gamma_{i}}^{0}=\frac{\hat{y}_{i} }{2} \left(1+\hat{y}_{2} \hat{y}_{1}+\hat{y}_{1}+\sqrt{\triangle'}\right) \ , \\
& \ \hat{y}_{\gamma_{1}}^{0}=\hat{y}_{1}\frac{4 \triangle'}{\left(1-\hat{y}_{2} \hat{y}_{1}+\hat{y}_{1}+\sqrt{\triangle'}\right){}^2}\ , \\
& \ \hat{y}_{\gamma_{2}}^{0}=\frac{\hat{y}_{2}}{4}  \left(1+\frac{1-\hat{y}_{1} \left(\hat{y}_{2}+1\right)}{\sqrt{\triangle'}}\right){}^2\ , \\
&\quad \triangle'=\left(\hat{y}_{2}
   \hat{y}_{1}+\hat{y}_{1}+1\right){}^2-4 \hat{y}_{1} \hat{y}_{2} \ .
\end{split}
\end{equation}
\noindent Due to the number of cones, we will not show the $\hat{y}_{\gamma_{i}}$ of each cone. A complete multiplicative basis in terms of the $\hat{y}^{0}_{\gamma_i}$ is 
\begin{equation}
\hat{y}^{0}_{\gamma _1},\ \hat{y}^{0}_{\gamma _2},\ \hat{y}^{0}_{\gamma _3},\ \hat{y}^{0}_{\gamma _4}, \ (\hat{y}^{0}_{\gamma _1} \hat{y}^{0}_{\gamma _2} \hat{y}^{0}_{\gamma
   _3}+1),\ (\hat{y}^{0}_{\gamma
   _1} \hat{y}^{0}_{\gamma _2} \hat{y}^{0}_{\gamma _4}+1),\ (\hat{y}^{0}_{\gamma _3}+1),\ (\hat{y}^{0}_{\gamma
   _4}+1), \ (1-\hat{y}^{0}_{\gamma _1} \hat{y}^{0}_{\gamma _2}) \ . 
\end{equation}
\noindent Again, $\hat{y}^{0}_{\gamma_{i}}$ are the $\hat{y}_{\gamma_{i}}$ associated with the initial asymptotic chamber approached by repeated mutations on the $x_{1}$ and $x_{2}$ nodes in the initial quiver. 

\subsubsection{Example:  \texorpdfstring{$A_{1,1,1}$}{A111}}\label{sec:notfinitescatdiag}

Our final example before moving onto $\overline{Gr(4,8)/T}$ is $A_{1,1,1}$. The $A_{1,1,1}$ cluster algebra includes a cluster with the quiver 
\[
\begin{tikzcd}
x_{2}\arrow[dr]& & & x_{6}\arrow[dl]\\
x_{1}\arrow[shift left, u]\arrow[shift right, u] & x_{3}\arrow[l]\arrow[r]& x_{4}\arrow[r]& x_{5}\arrow[shift left, u]\arrow[shift right, u]
\end{tikzcd}
\]
\noindent which was chosen as our principle quiver. Unlike the previous examples, there are actually two limiting rays:
\begin{equation}\label{eq:limitingrays}
\begin{split}
g_{lim}^{1}&=(-1,1,0,0,0,0) \ , \\   
g^{2}_{lim}&=(0,0,0,0,-1,1) \ . 
\end{split}
\end{equation}

\indent First consider the limiting ray $g^{1}_{lim}$, which is approached by performing repeated mutations on the $x_{1}$ and $x_{2}$ nodes. The formulas for $\hat{y}_{\gamma_{i}}^{0}$ are then:
\begingroup
\allowdisplaybreaks
\begin{align}
i\in \{4,5,6 \}:& \ \hat{y}_{\gamma_{i}}^{0}=\hat{y}_{i} , \nonumber \\
& \ \hat{y}_{\gamma_{3}}^{0}=\frac{\hat{y}_{3} }{2} \left(1+\hat{y}_{2} \hat{y}_{1}+\hat{y}_{1}+\sqrt{\triangle'}\right) \ ,\nonumber  \\
& \ \hat{y}_{\gamma_{1}}^{0}=\hat{y}_{1}\frac{4 \triangle'}{\left(1-\hat{y}_{2} \hat{y}_{1}+\hat{y}_{1}+\sqrt{\triangle'}\right){}^2}\ , \label{eq:finalequationforyhatgamma2} \\
& \ \hat{y}_{\gamma_{2}}^{0}=\frac{\hat{y}_{2}}{4}  \left(1+\frac{1-\hat{y}_{1} \left(\hat{y}_{2}+1\right)}{\sqrt{\triangle'}}\right){}^2\ ,  \nonumber \\
&\quad \triangle'=\left(\hat{y}_{2}
   \hat{y}_{1}+\hat{y}_{1}+1\right){}^2-4 \hat{y}_{1} \hat{y}_{2} \nonumber  \ .
\end{align}
\endgroup
\noindent However, upon applying the algorithm in appendix \ref{sec:algorithlimitraytrue}, we found that there are actually an infinite number of asymptotic chambers. Rather, an infinite number of cluster walls intersect the limiting ray. It is unsurprising that this phenomenon eventually occurs as an infinite number of walls intersect a single ray even in the $A_{2,1}$ cluster algebra. We ignored this phenomenon in section \ref{sec:infiniteraysssA2p1} because none of the rays in $A_{2,1}$ with infinitely many intersecting walls are limiting rays. 

\indent Now consider the second limiting ray in eq. (\ref{eq:limitingrays}), which we approached by performing repeated mutations on the $x_{5}$ and $x_{6}$ nodes. As we approach the second limiting ray, the limits of $\hat{y}_{\gamma_{i}}$, denoted as $\hat{y}_{\gamma_{i}}^{0'}$, are
\begingroup
\allowdisplaybreaks
\begin{align}
i\in \{1,2,3 \}:& \ \hat{y}_{\gamma_{i}}^{0'}=\hat{y}_{i} , \nonumber \\
& \ \hat{y}_{\gamma_{4}}^{0'}=\frac{\hat{y}_{4} }{2} \left(1+\hat{y}_{6} \hat{y}_{5}+\hat{y}_{5}+\sqrt{\triangle'}\right) \ , \nonumber\\
& \ \hat{y}_{\gamma_{5}}^{0'}=\hat{y}_{5}\frac{4 \triangle'}{\left(1-\hat{y}_{6} \hat{y}_{5}+\hat{y}_{5}+\sqrt{\triangle'}\right){}^2}\ , \label{eq:finalequationforyhatgamma3} \\
& \ \hat{y}_{\gamma_{6}}^{0'}=\frac{\hat{y}_{6}}{4}  \left(1+\frac{1-\hat{y}_{5} \left(\hat{y}_{6}+1\right)}{\sqrt{\triangle'}}\right){}^2\ , \nonumber \\
&\quad \triangle'=\left(\hat{y}_{6}
   \hat{y}_{5}+\hat{y}_{5}+1\right){}^2-4 \hat{y}_{5} \hat{y}_{6} \nonumber \ .
\end{align}
\endgroup
\noindent We again found an infinite number of asymptotic chambers. Note that the discriminant, $\triangle'$, of the $\hat{y}_{\gamma_{i}}^{0'}$ variables is different than that of the $\hat{y}_{\gamma_{i}}^{0}$ variables. Rather, the discriminant of the algebraic letters associated with a given set of asymptotic chambers seems to be determined by the associated limiting ray. 

\indent It is not particularly interesting for us to further study the cluster algebraic functions associated with $A_{1,1,1}$ as the asymptotic scattering diagrams contain an infinite number of asymptotic chambers. However, one could take a doubly asymptotic limit to find a 4 dimensional scattering diagram that could be finite. More concretely, one could first find the 5 dimensional asymptotic scattering diagram associated with the limiting ray $g^{1}_{lim}$ and then find the 4 dimensional asymptotic scattering diagram associated with the limiting ray of this 5 dimensional asymptotic scattering diagram. The resulting 4 dimensional asymptotic scattering diagram could be finite. We leave studying such doubly asymptotic limits to future work. 

\subsection{\texorpdfstring{$\overline{Gr(4,8)/T}$}{Gr(4,8)/T} and algebraic letters}\label{sec:gr48fullexampless} 

We now consider the algebraic letters associated with the 8-point MHV amplitude in $\mathcal{N}=4$ pSYM. Two classes of known algebraic letters are known to emerge in the $\mathcal{N}=4$ pSYM symbol alphabet at 8-point and they are related by a cyclic shift: $\langle i,j,k,l\rangle \rightarrow \langle i+1,j+1,k+1,l+1\rangle$ \cite{Zhang:2019vnm,He:2020vob}. Notably, each class of algebraic letters is associated with a unique discriminant. Since each limiting ray seems to be associated with a unique discriminant, $\triangle'$, a reasonable conjecture is that the asymptotic chambers of only two limiting rays are relevant for the 8-point MHV amplitude. Furthermore, we only need to analyze the asymptotic chambers of one of these limiting rays since we can derive the algebraic letters associated with the other limiting ray by applying a cyclic shift.\footnote{Refs. \cite{Arkani-Hamed:2019rds,Henke:2019hve,Drummond:2019cxm} have pointed out that additional types of limiting rays might be relevant for studying the symbol alphabet at higher loop. However, ref. \cite{Arkani-Hamed:2019rds} also pointed out at least some of these additional limiting rays are related by a braid group \cite{2017arXiv170200385F} to the limiting rays we study in this section. Therefore, even if the algebraic letters associated with these other limiting rays appear in the 8-point MHV symbol alphabet, it seems plausible they could be derived through braid transformations of the symbol alphabet derived in this section. } 
\begin{figure}
\[
\begin{tikzcd}
\textrm{\framebox[8ex]{$\langle 1234 \rangle$}}\arrow[dr] & & & & \\
& x_{1}:\langle 1235\rangle \arrow[r]\arrow[d]& x_{2}: \langle 1236\rangle \arrow[d]\arrow[r]& x_{3}:, \ \langle 1237\rangle\arrow[r]\arrow[d]& \textrm{\framebox[8ex]{$\langle 1235\rangle$}}  \\
& x_{4}: \langle 1245\rangle \arrow[r]\arrow[d]& x_{5}: \langle 1256\rangle \arrow[d]\arrow[ul]\arrow[r]& x_{6}: \langle 1267\rangle\arrow[ul]\arrow[r]\arrow[d]& \textrm{\framebox[8ex]{$\langle 1278 \rangle$}} \arrow[ul] \\
& x_{7}: \langle 1345\rangle \arrow[r]\arrow[d]& x_{8}: \langle 1456\rangle\arrow[ul] \arrow[d]\arrow[r]& x_{9}: \langle 1567\rangle \arrow[ul]\arrow[r]\arrow[d]& \textrm{\framebox[8ex]{$\langle 1678 \rangle$}} \arrow[ul] \\
& \textrm{\framebox[8ex]{$\langle 2345 \rangle$}} & \textrm{\framebox[8ex]{$\langle 3456 \rangle$}}\arrow[ul]  & \textrm{\framebox[8ex]{$\langle 4567 \rangle$}}\arrow[ul]  & \textrm{\framebox[8ex]{$\langle 5678 \rangle$}}\arrow[ul] \ , 
\end{tikzcd}
\]
\caption{The initial quiver for the $\overline{Gr(4,8)}$ cluster algebra \cite{scott_2006}.}
\label{initialquiverGr48}
\end{figure}
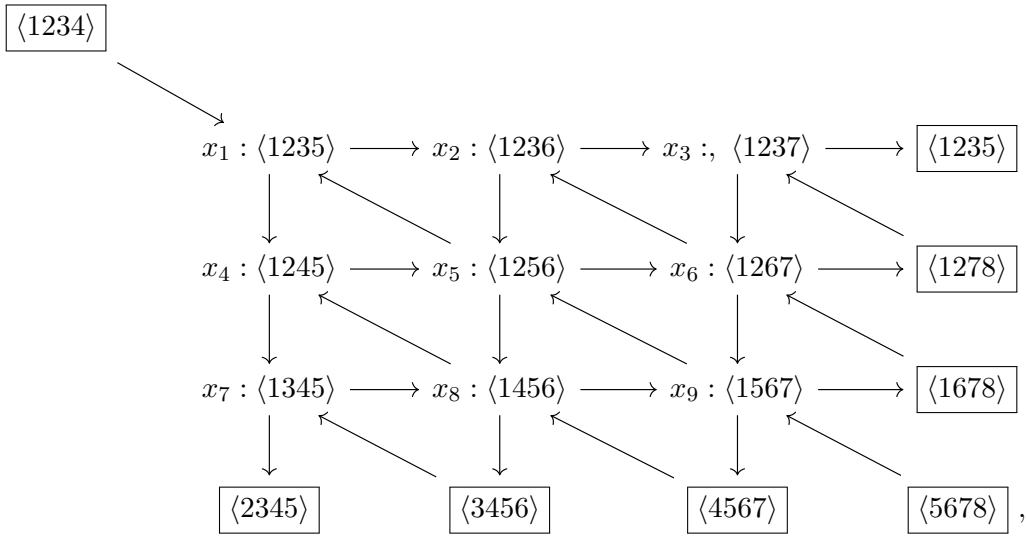

\indent We first briefly review the positive kinematic region before summarizing the computation of the algebraic letters. We parameterize kinematic space using momentum twistors,  $Z_{i}^{A}$. Due to dual conformal symmetry, we can identify $Z_{i}^{A}\in Gr(4,n)$. Furthermore, since the $Z_{i}^{A}$ are projective under a ``little group'' transform, $Z_{i}^{A}\rightarrow t_{i} Z_{i}^{A}$, we can identify $Z_{i}^{A}\in Gr(4,n)/T$. The positive kinematic region corresponds to a compactification of the positive Grassmannian, $\overline{Gr(4,n)/T}$. The cluster algebra structure of $\overline{Gr(k,n)/T}$ is well known. In particular, there is a famous initial parameterization that corresponds to the quiver in fig. \ref{initialquiverGr48} at 8-point, 
%\begin{equation}\label{initialquiverGr48}
%\begin{tikzcd}
%\textrm{\framebox[8ex]{$\langle 1234 \rangle$}}\arrow[d] & & & \\
%x_{1}:\langle 1235\rangle \arrow[r]\arrow[d]& x_{2}: \langle 1236\rangle \arrow[d]\arrow[r]& x_{3}:, \ \langle 1237\rangle\arrow[r]\arrow[d]& \textrm{\framebox[8ex]{$\langle 1235\rangle$}}  \\
%x_{4}: \langle 1245\rangle \arrow[r]\arrow[d]& x_{5}: \langle 1256\rangle \arrow[d]\arrow[ul]\arrow[r]& x_{6}: \langle 1267\rangle\arrow[ul]\arrow[r]\arrow[d]& \textrm{\framebox[8ex]{$\langle 1278 \rangle$}} \arrow[ul] \\
%x_{7}: \langle 1345\rangle \arrow[r]\arrow[d]& x_{8}: \langle 1456\rangle\arrow[ul] \arrow[d]\arrow[r]& x_{9}: \langle 1567\rangle \arrow[ul]\arrow[r]\arrow[d]& \textrm{\framebox[8ex]{$\langle 1678 \rangle$}} \arrow[ul] \\
 %\textrm{\framebox[8ex]{$\langle 2345 \rangle$}} & \textrm{\framebox[8ex]{$\langle 3456 \rangle$}}\arrow[ul]  & \textrm{\framebox[8ex]{$\langle 4567 \rangle$}}\arrow[ul]  & \textrm{\framebox[8ex]{$\langle 5678 \rangle$}}\arrow[ul] \ , 
%\end{tikzcd} 
%\end{equation}
where boxed elements in the quiver correspond to frozen variables. The $\hat{y}$-variables associated with the quiver are 
\begin{align}\label{eq:initialparameterizyhat}
&\hat{y}_{1}^{\text{I}} = \frac{\langle 1234\rangle \langle 1256 \rangle}{\langle 1236\rangle \langle 1245 \rangle}  &&\hat{y}_{2}^{\text{I}}  = \frac{\langle 1235\rangle \langle 1267 \rangle}{\langle 1237 \rangle \langle 1256 \rangle}  &&\hat{y}_{3}^{\text{I}}  = \frac{\langle 1236\rangle \langle 1278 \rangle}{\langle 1238 \rangle \langle 1267 \rangle} \notag \\
&\hat{y}_{4}^{\text{I}}  = \frac{\langle 1235\rangle \langle 1456 \rangle}{\langle 1256\rangle \langle 1345 \rangle}  &&\hat{y}_{5}^{\text{I}}  = \frac{\langle 1236 \rangle \langle 1245 \rangle \langle 1567 \rangle }{\langle 1235 \rangle \langle 1456 \rangle \langle 1267 \rangle}  &&\hat{y}_{6}^{\text{I}}  = \frac{\langle 1237 \rangle \langle 1256\rangle \langle 1678 \rangle  }{\langle 1236 \rangle  \langle 1567 \rangle \langle 1278 \rangle} \notag \\
&\hat{y}_{7}^{\text{I}} = \frac{\langle 1245\rangle \langle 3456 \rangle}{\langle 1456\rangle \langle 2345 \rangle}  &&\hat{y}_{8}^{\text{I}}  = \frac{\langle 1256 \rangle \langle 1345 \rangle \langle 4567 \rangle }{\langle 1245 \rangle \langle 3456 \rangle \langle 1567 \rangle}  &&\hat{y}_{9}^{\text{I}} = \frac{\langle 1267 \rangle \langle 1456 \rangle \langle 5678 \rangle }{\langle 1256 \rangle \langle 4567 \rangle \langle 1678 \rangle} \ ,
\end{align}
\noindent where the ``$\text{I}$'' super-script denotes how these $\hat{y}$-variables are associated with the initial quiver. Note that the quiver in fig. \ref{initialquiverGr48} was not chosen as the principal quiver for our scattering diagram. Instead, we mutated to the quiver
\begin{equation}\label{principalquiverGr48}
\begin{tikzcd}
 & & & & \hat{y}_{9}\arrow[dr]\arrow[dl]\arrow[dll]& & & \\
\hat{y}_{3} \arrow[r]& \hat{y}_{2}\arrow[r]& \hat{y}_{8}\arrow[drr]& \hat{y}_{5}\arrow[dr]& & \hat{y}_{6}\arrow[dl]& \hat{y}_{4}\arrow[l]& \hat{y}_{7}\arrow[l] \ , \\
 & & & & \hat{y}_{1}\arrow[shift left, uu]\arrow[shift right, uu] & & & 
\end{tikzcd}
\end{equation}
\noindent which was chosen to be the principal quiver, by mutating the nodes $\{1,2,4,1,6,8\}$ of the initial quiver from left to right. As argued in section \ref{sec:principalquiverasymptotcones}, no information is lost or gained by choosing different principle quivers. An explicit map from the $\hat{y}$-variables of our initial quiver to those of the chosen principal quiver is
\begingroup
\allowdisplaybreaks
\begin{align}
\hat{y}_{1}&=\frac{\left(\hat{y}^{\text{I}}_6+\hat{y}^{\text{I}}_1 \left(\hat{y}^{\text{I}}_4+1\right)
   \left(\left(\hat{y}^{\text{I}}_2+1\right) \hat{y}^{\text{I}}_6+1\right)+1\right)}{\hat{y}^{\text{I}}_1
   \hat{y}^{\text{I}}_2 \hat{y}^{\text{I}}_4} \nonumber \\ 
   &\quad \times \left(\hat{y}^{\text{I}}_8+\hat{y}^{\text{I}}_1 \left(\hat{y}^{\text{I}}_2+1\right)
   \left(\left(\hat{y}^{\text{I}}_4+1\right) \hat{y}^{\text{I}}_8+1\right)+1\right) \ , \nonumber \\
   \hat{y}_{2}&=\frac{\hat{y}^{\text{I}}_4 \hat{y}^{\text{I}}_8}{\hat{y}^{\text{I}}_8+\hat{y}^{\text{I}}_1
   \left(\hat{y}^{\text{I}}_2+1\right) \left(\left(\hat{y}^{\text{I}}_4+1\right) \hat{y}^{\text{I}}_8+1\right)+1} \ ,\nonumber \\
   \hat{y}_{3}&=\frac{\hat{y}^{\text{I}}_1 \hat{y}^{\text{I}}_2 \hat{y}^{\text{I}}_3}{\hat{y}^{\text{I}}_1
   \left(\hat{y}^{\text{I}}_2+1\right)+1} \ ,\nonumber \\
   \hat{y}_{4}&=\frac{\hat{y}^{\text{I}}_2 \hat{y}^{\text{I}}_6}{\hat{y}^{\text{I}}_6+\hat{y}^{\text{I}}_1
   \left(\hat{y}^{\text{I}}_4+1\right) \left(\left(\hat{y}^{\text{I}}_2+1\right) \hat{y}^{\text{I}}_6+1\right)+1} \ ,\nonumber \\
   \hat{y}_{5}&=\frac{\hat{y}^{\text{I}}_1 \hat{y}^{\text{I}}_2 \hat{y}^{\text{I}}_4 \hat{y}^{\text{I}}_5}{\hat{y}^{\text{I}}_1
   \left(\hat{y}^{\text{I}}_2+1\right) \left(\hat{y}^{\text{I}}_4+1\right)+1} \ , \label{eq:explciitmapfrominitialtoprinc} \\
   \hat{y}_{6}&=\frac{\hat{y}^{\text{I}}_1 \left(\hat{y}^{\text{I}}_4+1\right)+1}{\left(\hat{y}^{\text{I}}_1
   \left(\hat{y}^{\text{I}}_2+1\right) \left(\hat{y}^{\text{I}}_4+1\right)+1\right) \hat{y}^{\text{I}}_6} \ ,\nonumber \\
   \hat{y}_{7}&=\frac{\hat{y}^{\text{I}}_1 \hat{y}^{\text{I}}_4 \hat{y}^{\text{I}}_7}{\hat{y}^{\text{I}}_1
   \left(\hat{y}^{\text{I}}_4+1\right)+1} \ ,\nonumber \\
   \hat{y}_{8}&=\frac{\hat{y}^{\text{I}}_1 \left(\hat{y}^{\text{I}}_2+1\right)+1}{\left(\hat{y}^{\text{I}}_1
   \left(\hat{y}^{\text{I}}_2+1\right) \left(\hat{y}^{\text{I}}_4+1\right)+1\right) \hat{y}^{\text{I}}_8} \ , \nonumber \\
   \hat{y}_{9}&=\frac{\left(\hat{y}^{\text{I}}_1 \left(\hat{y}^{\text{I}}_2+1\right) \left(\hat{y}^{\text{I}}_4+1\right)+1\right){}^2
   \hat{y}^{\text{I}}_6 \hat{y}^{\text{I}}_8 \hat{y}^{\text{I}}_9}{\left(\hat{y}^{\text{I}}_6+\hat{y}^{\text{I}}_1
   \left(\hat{y}^{\text{I}}_4+1\right) \left(\left(\hat{y}^{\text{I}}_2+1\right) \hat{y}^{\text{I}}_6+1\right)+1\right)} \nonumber \\
   &\quad \times \frac{1}{\left(\hat{y}^{\text{I}}_8+\hat{y}^{\text{I}}_1 \left(\hat{y}^{\text{I}}_2+1\right)
   \left(\left(\hat{y}^{\text{I}}_4+1\right) \hat{y}^{\text{I}}_8+1\right)+1\right)} \ . \nonumber
\end{align}    
\endgroup
\noindent Combining eqs. (\ref{eq:initialparameterizyhat}) and (\ref{eq:explciitmapfrominitialtoprinc}) gives explicit expressions of the principal quivers' $\hat{y}$-variables in terms of external kinematic data. 

\indent We now analyze the initial asymptotic chamber using the $A_{1,1}$ subalgebra of the principal quiver. After performing an infinite number of mutations on the $x_{1}$ and $x_{9}$ nodes, the $g$-vectors of the $x_{1}$ and $x_{9}$ nodes approached
\begin{equation}\label{limitrayGR48}
g_{lim}=(-1,0,0,0,0,0,0,0,1 ) \ ,   
\end{equation}
\noindent which we identified as the limiting ray. We then found expressions for the $\hat{y}_{\gamma_{i}}^{0}$ in terms of the $\hat{y}_{i}$:
\begin{equation}\label{eq:explicitsolutiongr48princ}
\begin{split}
i\in \{2,3,4,7 \}: & \  \hat{y}_{\gamma_{i}}^{0}=\hat{y}_{i} \ , \\ 
i\in \{5,6,8 \}: &\  \hat{y}_{\gamma_{i}}^{0}=\hat{y}_{i}f(\hat{y}_{1},\hat{y}_{9}) \ , \\ 
&\ \hat{y}_{\gamma_{1}}^{0}= \frac{4 \hat{y}_{1} \triangle'}{\left(1-\hat{y}_{9} \hat{y}_{1}+\hat{y}_{1}+\sqrt{\triangle'}\right){}^2} \ , \\ 
&\ \hat{y}_{\gamma_{9}}^{0}=\frac{\hat{y}_{9}}{4} \left(1+\frac{1-\hat{y}_{1} \left(\hat{y}_{9}+1\right)}{\sqrt{\triangle'}}\right){}^2 \ , \\ 
\end{split}
\end{equation}
\noindent where
\begin{equation}
\begin{split}
f(\hat{y}_{1},\hat{y}_{9})&= \frac{1}{2} \left(1+\hat{y}_{9} \hat{y}_{1}+\hat{y}_{1}+\sqrt{\triangle'}\right) \ , \\
\triangle'&=\left(\hat{y}_{9}
   \hat{y}_{1}+\hat{y}_{1}+1\right){}^2-4 \hat{y}_{1} \hat{y}_{9}  \ .
\end{split}
\end{equation}
\noindent Unlike the $A_{2,1}$ and $A_{2,2}$ cluster algebras, not all the $\hat{y}_{\gamma_{i}}^{0}$ are algebraic function of the $\hat{y}_{i}$. Although difficult to see immediately, one can show that the discriminant, $\triangle'$, is proportional to 
\begingroup
\allowdisplaybreaks
\begin{align}
\sqrt{\triangle'}&\propto \sqrt{A^{2}-4B} \ , \nonumber \\
&\quad A=\langle 1256\rangle \langle 3478\rangle -\langle 1278\rangle \langle 3456\rangle -\langle 1234\rangle \langle 5678\rangle \ , \\
&\quad B=\langle 1234\rangle \langle 3456\rangle \langle 5678\rangle \langle 1278\rangle \ , \nonumber
\end{align}
\endgroup
\noindent which corresponds to the limiting ray $g_{1}$ in ref. \cite{Arkani-Hamed:2019rds}. The limiting ray in eq. (\ref{limitrayGR48}) looks different than the limiting ray in ref. \cite{Arkani-Hamed:2019rds} because we chose a different principal quiver to define the $g$-vector fan. While we chose the quiver in (\ref{principalquiverGr48}) as our principal quiver, the authors of ref. \cite{Arkani-Hamed:2019rds} chose the initial quiver in fig. \ref{initialquiverGr48} as their principal quiver. 

\indent We employed the algorithm in appendix \ref{sec:algorithlimitraytrue} to find all the walls that intersect the limiting ray. Performing a large number of mutations asymptotically close to the limiting ray, we eventually found 26 cluster walls and a single limiting wall:
\begingroup
\allowdisplaybreaks
\begin{align}
\gamma^{\perp}\in\{ &\text{$($0, 1, 0, 0, 0, 0, 0, 0, 0$)$}, \ 
\text{$($0, 0, 1, 0, 0, 0, 0, 0, 0$)$}, \nonumber \\ &\text{$($0, 0, 0, 1, 0, 0, 0, 0, 0$)$}, \
\text{$($0, 0, 0, 0, 1, 0, 0, 0, 0$)$}, \nonumber \\ &\text{$($0, 0, 0, 0, 0, 1, 0, 0, 0$)$}, \ \text{$($0, 0, 0, 0, 0, 0, 1, 0, 0$)$}, \nonumber \\ &\text{$($0, 1, 0, 0, 0, 0, 0, 1, 0$)$},\ \text{$($0, 1, 1, 0, 0, 0, 0, 0, 0$)$}, \nonumber \\ &\text{$($0, 0, 0, 0, 0, 0, 0, 1, 0$)$}, \ \text{$($0, 0, 0, 1, 0, 1, 0, 0, 0$)$},\nonumber \\ &\text{$($0, 0, 0, 1, 0, 0, 1, 0, 0$)$},\ \text{$($0, 1, 1, 0, 0, 0, 0, 1, 0$)$},\label{eq:wallsgr48list} \\ &\text{$($0, 0, 0, 1, 0, 1, 1, 0, 0$)$},\ \text{$($1, 0, 0, 0, 1, 0, 0, 0, 1$)$},\nonumber \\ &\text{$($1, 0, 0, 0, 0, 1, 0, 0, 1$)$},\ \text{$($1, 0, 0, 0, 0, 0, 0, 1, 1$)$},\nonumber \\ &\text{$($1, 1, 0, 0, 0, 0, 0, 1, 1$)$},\ \text{$($1, 0, 0, 1, 0, 1, 0, 0, 1$)$},\nonumber \\ &\text{$($1, 1, 1, 0, 0, 0, 0, 1, 1$)$},\ \text{$($1, 1, 0, 0, 0, 0, 0, 2, 1$)$},\nonumber \\ &\text{$($1, 0, 0, 1, 0, 1, 1, 0, 1$)$},\ \text{$($1, 0, 0, 1, 0, 2, 0, 0, 1$)$},\nonumber \\ &\text{$($1, 2, 1, 0, 0, 0, 0, 2, 1$)$},\ \text{$($1, 1, 1, 0, 0, 0, 0, 2, 1$)$},\nonumber \\ &\text{$($1, 0, 0, 1, 0, 2, 1, 0, 1$)$},\ \text{$($1, 0, 0, 2, 0, 2, 1, 0, 1$)$},\nonumber \\ &\text{$($1, 0, 0, 0, 0, 0, 0, 0, 1$)$} \} \nonumber \ ,
\end{align}
\endgroup
\noindent where the last element corresponds to the limiting wall. An extensive computer search found a complete multiplicative basis of the $\hat{y}_{\gamma_{i}}$ consists of 27 non-trivial polynomials of $\hat{y}^{0}_{\gamma_{i}}$,
\begingroup
\allowdisplaybreaks
\begin{align}
&f_{1}= \hat{y}^0_{\gamma _2}+1\ , \nonumber \\ 
&f_{2}=\hat{y}^0_{\gamma _3}+1\ , \nonumber \\ 
&f_{3}= \hat{y}^0_{\gamma_4}+1\ , \nonumber \\ 
&f_{4}= \hat{y}^0_{\gamma _5}+1\ , \nonumber \\ 
&f_{5}=\hat{y}^0_{\gamma _6}+1\ , \nonumber \\ 
&f_{6}= \hat{y}^0_{\gamma_7}+1\ , \nonumber \\ 
&f_{7}= \hat{y}^0_{\gamma _8}+1\ , \nonumber \\ 
&f_{8}=\hat{y}^0_{\gamma _2} \hat{y}^0_{\gamma _3}+\hat{y}^0_{\gamma_3}+1\ , \nonumber \\ 
&f_{9}= \hat{y}^0_{\gamma _8} \hat{y}^0_{\gamma_2}+\hat{y}^0_{\gamma _2}+1\ , \nonumber \\ 
&f_{10}= \hat{y}^0_{\gamma _6}\hat{y}^0_{\gamma _4}+\hat{y}^0_{\gamma _4}+1\ , \nonumber \\ 
&f_{11}=\hat{y}^0_{\gamma _4} \hat{y}^0_{\gamma _7}+\hat{y}^0_{\gamma_7}+1\ , \nonumber \\ 
&f_{12}= \hat{y}^0_{\gamma _2} \hat{y}^0_{\gamma_3}+\hat{y}^0_{\gamma _2} \hat{y}^0_{\gamma _8} \hat{y}^0_{\gamma_3}+\hat{y}^0_{\gamma _3}+1\ , \nonumber \\ 
&f_{13}= \hat{y}^0_{\gamma _4}\hat{y}^0_{\gamma _7}+\hat{y}^0_{\gamma _4} \hat{y}^0_{\gamma _6}\hat{y}^0_{\gamma _7}+\hat{y}^0_{\gamma _7}+1\ , \nonumber \\ 
&f_{14}=\hat{y}^0_{\gamma _1} \hat{y}^0_{\gamma _5} \hat{y}^0_{\gamma_9}+1\ , \nonumber \\ 
&f_{15}= \hat{y}^0_{\gamma _1} \hat{y}^0_{\gamma _6}\hat{y}^0_{\gamma _9}+1\ , \label{eq:explicitalphaGr48} \\ 
&f_{16}= \hat{y}^0_{\gamma _1}\hat{y}^0_{\gamma _8} \hat{y}^0_{\gamma _9}+1\ , \nonumber \\ 
&f_{17}=\hat{y}^0_{\gamma _1} \hat{y}^0_{\gamma _8} \hat{y}^0_{\gamma _9}\hat{y}^0_{\gamma _2}+\hat{y}^0_{\gamma _2}+1\ , \nonumber \\ 
&f_{18}=\hat{y}^0_{\gamma _1} \hat{y}^0_{\gamma _6} \hat{y}^0_{\gamma _9}\hat{y}^0_{\gamma _4}+\hat{y}^0_{\gamma _4}+1\ , \nonumber \\ 
&f_{19}=\hat{y}^0_{\gamma _1} \hat{y}^0_{\gamma _4} \hat{y}^0_{\gamma _9}\left(\hat{y}^0_{\gamma _6}\right)^2+\hat{y}^0_{\gamma _4} \hat{y}^0_{\gamma_6}+\hat{y}^0_{\gamma _1} \hat{y}^0_{\gamma _4} \hat{y}^0_{\gamma _9}\hat{y}^0_{\gamma _6}+\hat{y}^0_{\gamma _4}+1 , \nonumber \\ 
&f_{20}=\hat{y}^0_{\gamma _1} \hat{y}^0_{\gamma _2} \hat{y}^0_{\gamma _9}\left(\hat{y}^0_{\gamma _8}\right)^2+\hat{y}^0_{\gamma _2} \hat{y}^0_{\gamma_8}+\hat{y}^0_{\gamma _1} \hat{y}^0_{\gamma _2} \hat{y}^0_{\gamma _9}\hat{y}^0_{\gamma _8}+\hat{y}^0_{\gamma _2}+1 , \nonumber \\ 
&f_{21}=\hat{y}^0_{\gamma _2} \hat{y}^0_{\gamma _3}+\hat{y}^0_{\gamma _1}\hat{y}^0_{\gamma _2} \hat{y}^0_{\gamma _8} \hat{y}^0_{\gamma _9}\hat{y}^0_{\gamma _3}+\hat{y}^0_{\gamma _3}+1 , \nonumber \\ 
&f_{22}=\hat{y}^0_{\gamma _3} \left(\hat{y}^0_{\gamma _2}\right)^2+\hat{y}^0_{\gamma_3} \hat{y}^0_{\gamma _8} \left(\hat{y}^0_{\gamma _2}\right)^2+\hat{y}^0_{\gamma _1}\hat{y}^0_{\gamma _3} \left(\hat{y}^0_{\gamma _8}\right)^2 \hat{y}^0_{\gamma _9}\left(\hat{y}^0_{\gamma_2}\right)^2 \nonumber \\
&\quad +\hat{y}^0_{\gamma _1} \hat{y}^0_{\gamma _3}\hat{y}^0_{\gamma _8} \hat{y}^0_{\gamma _9} \left(\hat{y}^0_{\gamma _2}\right)^2+2\hat{y}^0_{\gamma _3} \hat{y}^0_{\gamma _2}+\hat{y}^0_{\gamma _3}\hat{y}^0_{\gamma _8} \hat{y}^0_{\gamma _2}+\hat{y}^0_{\gamma _1}\hat{y}^0_{\gamma _3} \hat{y}^0_{\gamma _8} \hat{y}^0_{\gamma _9}\hat{y}^0_{\gamma _2}+\hat{y}^0_{\gamma _2}+\hat{y}^0_{\gamma_3}+1 , \nonumber \\ 
&f_{23}= \hat{y}^0_{\gamma _1} \hat{y}^0_{\gamma _2}\hat{y}^0_{\gamma _3} \hat{y}^0_{\gamma _9} \left(\hat{y}^0_{\gamma_8}\right)^2+\hat{y}^0_{\gamma _2} \hat{y}^0_{\gamma _3} \hat{y}^0_{\gamma_8}+\hat{y}^0_{\gamma _1} \hat{y}^0_{\gamma _2} \hat{y}^0_{\gamma _3}\hat{y}^0_{\gamma _9} \hat{y}^0_{\gamma _8}+\hat{y}^0_{\gamma _2}\hat{y}^0_{\gamma _3}+\hat{y}^0_{\gamma _3}+1 , \nonumber \\ 
&f_{24}=\hat{y}^0_{\gamma _6} \hat{y}^0_{\gamma _7} \left(\hat{y}^0_{\gamma_4}\right)^2+\hat{y}^0_{\gamma _7} \left(\hat{y}^0_{\gamma _4}\right)^2+\hat{y}^0_{\gamma_1} \left(\hat{y}^0\right)_{\gamma _6}^2 \hat{y}^0_{\gamma _7} \hat{y}^0_{\gamma _9}\left(\hat{y}^0_{\gamma _4}\right)^2\nonumber \\
&\quad +\hat{y}^0_{\gamma _1} \hat{y}^0_{\gamma _6}\hat{y}^0_{\gamma _7} \hat{y}^0_{\gamma _9} \left(\hat{y}^0_{\gamma_4}\right)^2+\hat{y}^0_{\gamma _6} \hat{y}^0_{\gamma _7} \hat{y}^0_{\gamma _4}+2\hat{y}^0_{\gamma _7} \hat{y}^0_{\gamma _4}+\hat{y}^0_{\gamma _1}\hat{y}^0_{\gamma _6} \hat{y}^0_{\gamma _7} \hat{y}^0_{\gamma _9}\hat{y}^0_{\gamma _4}+\hat{y}^0_{\gamma _4}+\hat{y}^0_{\gamma_7}+1 , \nonumber \\ 
&f_{25}= \hat{y}^0_{\gamma _4} \hat{y}^0_{\gamma_7}+\hat{y}^0_{\gamma _1} \hat{y}^0_{\gamma _4} \hat{y}^0_{\gamma _6}\hat{y}^0_{\gamma _9} \hat{y}^0_{\gamma _7}+\hat{y}^0_{\gamma_7}+1 , \nonumber \\ 
&f_{26}= \hat{y}^0_{\gamma _1} \hat{y}^0_{\gamma _4}\hat{y}^0_{\gamma _7} \hat{y}^0_{\gamma _9} \left(\hat{y}^0_{\gamma_6}\right)^2+\hat{y}^0_{\gamma _4} \hat{y}^0_{\gamma _7} \hat{y}^0_{\gamma_6}+\hat{y}^0_{\gamma _1} \hat{y}^0_{\gamma _4} \hat{y}^0_{\gamma _7}\hat{y}^0_{\gamma _9} \hat{y}^0_{\gamma _6}+\hat{y}^0_{\gamma _4}\hat{y}^0_{\gamma _7}+\hat{y}^0_{\gamma _7}+1 \ ,  \nonumber \\
&f_{27}= 1-\hat{y}_{\gamma_{1}}^{0}\hat{y}_{\gamma_{9}}^{0} \ , \nonumber 
\end{align}  
\endgroup
\noindent and the 9 $\hat{y}^{0}_{\gamma_{i}}$, giving a symbol alphabet of 36 independent letters. When computing eqs. (\ref{eq:wallsgr48list}) and (\ref{eq:explicitalphaGr48}), we found $7348$ asymptotic chambers, in comparison to the $64$ asymptotic chambers studied in ref. \cite{Drummond:2019cxm} using slightly different methods.\footnote{Each origin cluster corresponds to two asymptotic chambers. Further discussion on the techniques in ref. \cite{Drummond:2019cxm} is given in appendix \ref{sec:comparorigincluster}.} Although we are confident we found all asymptotic chambers, we were not able to rigorously prove it as we did for $A_{2,1}$ and $A_{2,2}$. A complete search of all asymptotic chambers is very computationally challenging for $\overline{Gr(4,8)/T}$ for reasons beyond its high rank.\footnote{See appendix \ref{sec:algorithlimitraytrue} for a detailed discussion.} However, any missing asymptotic chambers should not change eqs. (\ref{eq:wallsgr48list}) or (\ref{eq:explicitalphaGr48}). Interestingly, only a subset of less than $1000$ asymptotic chambers was required to find both a complete multiplicative basis for the $\hat{y}_{\gamma_{i}}$ and all the relevant walls in the asymptotic scattering diagrams. 

\indent Combining eqs. (\ref{eq:initialparameterizyhat}), (\ref{eq:explciitmapfrominitialtoprinc}), (\ref{eq:explicitsolutiongr48princ}) and (\ref{eq:explicitalphaGr48}) gives explicit expressions for the algebraic letters in terms of momentum twistors. We have explicitly checked that the algebraic letters of ref. \cite{He:2020vob} are monomials of $\hat{y}^{0}_{\gamma_{i}}$ and $f_{i}$. Interestingly, note that many of the letters are obviously not algebraic. From eq. (\ref{eq:explicitsolutiongr48princ}), $\hat{y}^{0}_{\gamma_{2}}$, $\hat{y}^{0}_{\gamma_{3}}$, $\hat{y}^{0}_{\gamma_{4}}$, and $\hat{y}^{0}_{\gamma_{7}}$ are rational, so any $f_{i}$ that is solely a function of these variables will also be rational. From this criterion alone, the algebraic alphabet is reduced from 36 to 26 letters. Additional numerical checks show that some of these algebraic letters can further simplify to rational functions for certain momentum configurations. 

\indent These results are remarkable. There is no reason to expect that there are a finite number of asymptotic chambers associated with any limiting ray of the $\overline{Gr(4,8)/T}$ cluster algebra. In section \ref{sec:notfinitescatdiag}, we saw an explicit example of a limiting ray with an infinite number of asymptotic chambers. Furthermore, although the number of asymptotic chambers is extremely large, the multiplicative basis has rank 36 for the relevant limiting rays! We can further discard letters that are clearly not algebraic, reducing the rank of the algebraic alphabet from 72 to 52. We can now conjecture that we have found ALL algebraic letters that could appear in the $\mathcal{N}=4$ pSYM 8-point amplitude. Our $\hat{y}_{\gamma_{i}}^{0}$ coordinates show how the relations between the algebraic letters associated with the same limiting ray are inherently rational, even though the $\hat{y}_{\gamma_{i}}^{0}$ are generically algebraic functions of our initial coordinates, $\hat{y}^{\textrm{I}}_{i}$. Finally, in all examples studied in this paper, the rank of the asymptotic symbol alphabet has been equal to the number of cluster walls plus the rank of the cluster algebra. More precisely, there seems to be a correspondence between walls in the asymptotic scattering diagram, $\gamma^{\perp}$ in eq. (\ref{eq:wallsgr48list}), and polynomial letters in the asymptotic symbol alphabet, $f_{i}$ in eq. (\ref{eq:explicitalphaGr48}). At present, it is unclear to us whether this relation holds for more general cluster algebras or is a red herring.  

\subsection{Beyond \texorpdfstring{$A_{1,1}$}{A11} subalgebras}\label{sec:multiplelimitingwalls}

\indent Although this paper focuses on limiting rays associated with quadratic cluster algebraic functions, we expect that cubic cluster algebraic functions will also be relevant for studying the symbol alphabet of $\mathcal{N}=4$ pSYM beyond 8-point. Quadratic (cubic) algebraic functions are algebraic functions that are products of roots of quadratic (cubic) polynomials. To see why cubic letters should appear, note that algebraic letters can at least partially be derived from \textit{irrational} Yangian invariants, as shown in refs. \cite{Mago:2020kmp,He:2020uhb}. Using the duality between on-shell super-space variables and differentials on kinematic space,
\begin{equation}
\eta^{A}_{i}\leftrightarrow dZ^{A}_{i}    \ ,
\end{equation}
\noindent where $\eta^{A}_{i}$ are the on-shell superspace variable associated with state $i$ \cite{Elvang:2013cua,Arkani-Hamed:2017vfh}, Yangian invariants in $\mathcal{N}=4$ pSYM can be written in a manifestly dlog form:
\begin{equation}
\textrm{Yangian Invariant}\rightarrow \prod_{i} d\log(\alpha_{i}) \ , 
\end{equation}
\noindent where $\alpha_{i}$ correspond to functions of external data, $Z_{i}^{A}$, that are not necessarily rational. The $\alpha_{i}$ can be interpreted as ``letters'' of the Yangian invariant and correspond to singularities. Since we expect the branch points of N${}^{k}$MHV amplitudes to match onto branch points of MHV amplitudes, we can therefore probe the symbol alphabet of MHV amplitudes by studying the $\alpha_{i}$ that appear in Yangian invariants associated with N${}^{k}$MHV amplitudes. Starting at 11-point, we start to see irrational Yangian invariants that include \textit{cubic} algebraic letters. Therefore, we expect to find cluster algebraic functions that are cubic at 11-point. 

\indent The problem with cubic cluster algebraic functions is that it may not be possible to probe their associated asymptotic chambers using an $A_{1,1}$ subalgebra as in section \ref{sec:algorithlimitray}. If at least one asymptotic chamber of a limiting ray can be approached by repeated mutations on an $A_{1,1}$ subalgebra, then the asymptotic symbol alphabet must consist of quadratic cluster algebraic letters. To see this, note that the generating function for cluster variables in an $A_{1,1}$ subalgebra always takes the form
\begin{equation}
G_{n>0}(t)=\frac{x_{0}-x_{-1}\mathcal{F}t}{1-\mathcal{P}t+\mathcal{F}t^{2}}=\sum^{\infty}_{n=0}x_{n}t^{n} \ ,   
\end{equation}
\noindent where $\mathcal{F}$ is some product of cluster variables outside the $A_{1,1}$ subalgebra. Taking limits of $x_{i}$ generated by the above relation, such as 
\begin{equation}
\lim_{i\rightarrow \infty}x_{i}/x_{i-1} ,     
\end{equation}
\noindent will always generate a function that is either rational or quadratic, but not cubic. Since wall crossing mutations around limiting rays are always rational transformations, this means all $\hat{y}_{\gamma_{i}}$ must be either rational or quadratic.\footnote{There is a small loophole in this argument. If the asymptotic scattering diagram itself contains a limiting ray, one could take a doubly asymptotic limit as suggested at the end of section \ref{sec:notfinitescatdiag}. However, it seems unlikely to us that such doubly asymptotic limits could generate cubic algebraic letters.} Therefore, we must identify more general mutation sequences in order to approach asymptotic chambers associated with cubic algebraic functions. Such mutation sequences could correspond to generating functions with higher-order polynomials in the denominator, such that specific limits of $x_{i}$ generate cubic cluster algebraic functions. We expect the methods and results in ref. \cite{Galakhov:2013oja} may be useful for pursuing this direction. 

\section{Degenerate scattering diagrams and tropicalization}\label{sec:degscatteringtrop}

We now study speculative truncations of $\hat{y}$-variables from the perspective of scattering diagrams. We will first motivate and define the notion of a degenerate scattering diagram, commenting on the specific connection to $\mathcal{N}=4$ pSYM. Although we did not find a definite algorithm for truncating $\hat{y}$-variables, we did find that the notion of asymptotic chambers naturally emerges from degenerate scattering diagrams. 

\subsection{Scattering diagrams from tropicalization of the dual cluster algebra}\label{sec:tropicalizationprocedre}
\indent In this section, we relate the $g$-vector fan to tropicalization of the \textit{dual} cluster algebra. We then motivate degenerate fans using tropicalization arguments. 

\indent We now give a brief review of tropicalization. Since all elements of $\mathcal{O}(\mathcal{X})$ are positive Laurent polynomials where the minus operation never appears, we can consider the tropicalization of such functions. Tropicalization naturally emerges from studying the behavior of geometric spaces at small (or large) values of their coordinates. For example, given a function $f(a_{1},a_{2,},\ldots,a_{n})$, the tropical function is defined as 
\begin{equation}
\textrm{Trop}[f(a_{1},a_{2,},\ldots,a_{n})]=\lim_{\epsilon\rightarrow \infty}\frac{-1}{\epsilon}\log[f(e^{-\epsilon a_{1}},e^{-\epsilon a_{2}},\ldots, e^{-\epsilon a_{n}})]    \ .
\end{equation}
\noindent The tropicalization of a function effectively amounts to the replacements 
\begin{equation}
\begin{split}
a\times b &\rightarrow a+b \ , \\ a+b &\rightarrow \min(a,b) \ , \\ 1&\rightarrow 0 \ , \\
\end{split}    
\end{equation}
\noindent where $a$ and $b$ now take values on a semifield. For example,
\begin{equation}
\begin{split}
\textrm{Trop}[1+x]&=\min(0,x) \ , \\ 
\textrm{Trop}[1+x+xy]&=\min(0,x,x+y)  \ .  
\end{split}    
\end{equation}
\noindent Tropicalization has many applications, ranging from mirror symmetry to intersection theory. We will now review one aspect of the connection with cluster algebras. 

\indent In our tropicalization arguments, we do not consider the $\mathcal{O}(\mathcal{X})$ associated with our initial principal quiver. Instead, we consider the \textit{dual} principal quiver and the associated dual cluster algebra, $\mathcal{X}^{\vee}$. The dual principal quiver is given by the initial quiver except that we flip all arrows between mutable nodes. As an example, given the initial quiver
\[
\begin{tikzcd}
y_{1} \arrow[d]& y_{2}\arrow[d] \\
x_{1} \arrow[r] & x_{2} \ ,
\end{tikzcd}
\]
\noindent the dual quiver is
\[
\begin{tikzcd}
y_{1}^{\vee} \arrow[d]& y_{2}^{\vee}\arrow[d] \\
x_{1}^{\vee}  & x_{2}^{\vee}\arrow[l] \ .
\end{tikzcd}
\]
\noindent We now study $\mathcal{O}(\mathcal{X}^{\vee})$. For example, in the case of the $A_{2}$ cluster algebra, $\mathcal{O}(\mathcal{X}^{\vee})$ consists of
\begin{equation}\label{multiplicativsss}
\begin{split}
f_{1}&=1+\hat{y}_{1}^{\vee} \ , \\
f_{2}&=1+\hat{y}_{2}^{\vee} \ , \\
f_{3}&=1+\hat{y}_{2}^{\vee}+\hat{y}_{2}^{\vee}\hat{y}_{1}^{\vee} \ . 
\end{split}    
\end{equation}
\noindent Any $\hat{y}$-variable in $\mathcal{X}^{\vee}$ can be written as a product of functions in eq. (\ref{multiplicativsss}) and $\hat{y}_{i}^{\vee}$.  

\begin{figure}
\centering
\begin{subfigure}[b]{0.3\textwidth}
    \centering
    \begin{tikzpicture}[every node/.style={font=\footnotesize}]
	%\node at (1.875,4.3) {$\mathcal{A}_5[(1,2),3,(4,5)]$};
	%\fill [fill=red!50!white] (1,0) ++(2,2) -- (1,0) -- (0,0) -- (0,1) -- ++(2,2);
	\draw [thick,-stealth] (0,0) -- (0,2);
	\draw (1,0)node{\large $0$} ;
	\draw [thick,-stealth] (0,0) -- (0,-2);
	\draw (-1,0)node{\large $\hat{y}_{1}^{\vee}$} ;
	\end{tikzpicture}
    \caption{ $\textrm{Trop}[1+\hat{y}_{1}^{\vee}]$}
    \label{fig:tropicalizeA2b}
    \end{subfigure}
    \hfill
    \begin{subfigure}[b]{0.3\textwidth}
    \centering
    \begin{tikzpicture}[every node/.style={font=\footnotesize}]
	%\node at (1.875,4.3) {$\mathcal{A}_5[(1,2),3,(4,5)]$};
	%\fill [fill=red!50!white] (1,0) ++(2,2) -- (1,0) -- (0,0) -- (0,1) -- ++(2,2);
	\draw [thick,-stealth] (0,0) -- (2,0);
	\draw (0,1)node{\large $0$} ;
	\draw [thick,-stealth] (0,0) -- (-2,0);
	\draw (0,-1)node{\large $\hat{y}_{2}^{\vee}$} ;
	\end{tikzpicture}
    \caption{$\textrm{Trop}[1+\hat{y}_{2}^{\vee}]$}
    \label{fig:tropicalizeA2a}
    \end{subfigure}
    \hfill
    \begin{subfigure}[b]{0.3\textwidth}
    \centering
    \begin{tikzpicture}[every node/.style={font=\footnotesize}]
	%\node at (1.875,4.3) {$\mathcal{A}_5[(1,2),3,(4,5)]$};
	%\fill [fill=red!50!white] (1,0) ++(2,2) -- (1,0) -- (0,0) -- (0,1) -- ++(2,2);
	\draw [thick,-stealth] (0,0) -- (-2,2);
	\draw (1,1)node{\large $0$} ;
	\draw [thick,-stealth] (0,0) -- (0,-2);
	\draw (1,-1)node{\large $\hat{y}_{2}^{\vee}$} ;
	\draw [thick,-stealth] (0,0) -- (2,0);
	\draw (-1,-1)node{\large $\hat{y}_{\gamma_{1}}^{\vee}+\hat{y}_{2}^{\vee}$} ;
	\end{tikzpicture}
    \caption{$\textrm{Trop}[1+\hat{y}_{2}^{\vee}+\hat{y}_{1}^{\vee}\hat{y}_{2}^{\vee}]$ }
    \label{fig:tropicalizeA2c}
    \end{subfigure}
    \caption{The fan associated with the tropicalization of functions in eq. (\ref{eq:f1f2f3trop}).}
\end{figure}
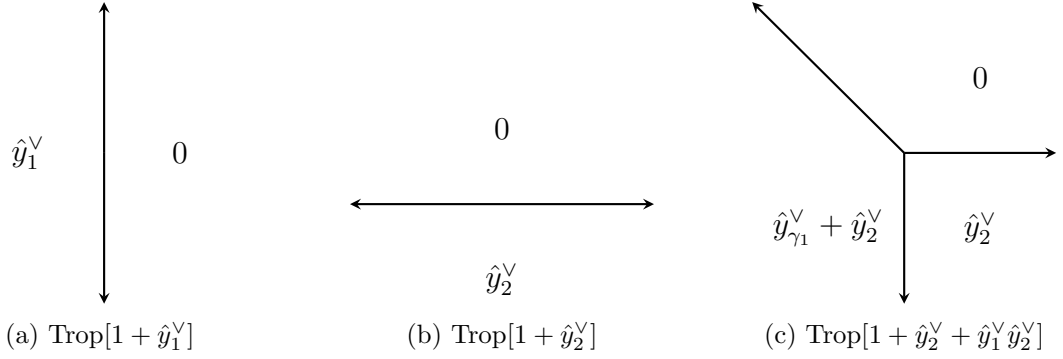
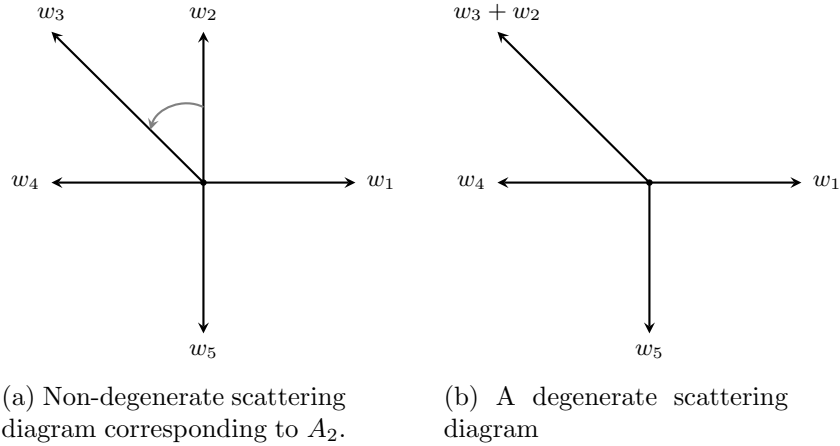
\begin{figure}
\centering
    \begin{subfigure}[b]{0.3\textwidth}
         \centering
  \begin{tikzpicture}[every node/.style={font=\footnotesize}]
	%\node at (1.875,4.3) {$\mathcal{A}_5[(1,2),3,(4,5)]$};
	%\fill [fill=red!50!white] (1,0) ++(2,2) -- (1,0) -- (0,0) -- (0,1) -- ++(2,2);
	\draw [thick,-stealth] (0,0) -- (2,0) node[right=0.2pt]{$w_{1}$};
	\draw [thick,-stealth] (0,0) -- (-2,0) node[left=0.2pt]{$w_{4}$};
	\draw [thick,-stealth] (0,0) -- (0,2) node[above=0.2pt]{$w_{2}$};
	\draw [thick,-stealth] (0,0) -- (0,-2)node[below=0.2pt]{$w_{5}$} ;
	\draw [thick,-stealth] (0,0) -- (-2,2) node[above=0.2pt]{$w_{3}$};
	\filldraw (0,0) circle (1pt) ;
	\coordinate (f1) at (0,1);
    \coordinate (f2) at (-0.7,0.7);
	\draw [thick,gray,-stealth] (f1.center) to [bend right=50] (f2.center);
	\end{tikzpicture}
    \caption{Non-degenerate scattering diagram corresponding to $A_{2}$. }
    \label{fig:degeneratediagramexamA2a}
     \end{subfigure}
     \quad \quad \quad
     \begin{subfigure}[b]{0.3\textwidth}
         \centering
    \begin{tikzpicture}[every node/.style={font=\footnotesize}]
	%\node at (1.875,4.3) {$\mathcal{A}_5[(1,2),3,(4,5)]$};
	%\fill [fill=red!50!white] (1,0) ++(2,2) -- (1,0) -- (0,0) -- (0,1) -- ++(2,2);
	\draw [thick,-stealth] (0,0) -- (2,0) node[right=0.2pt]{$w_{1}$};
	\draw [thick,-stealth] (0,0) -- (-2,0) node[left=0.2pt]{$w_{4}$};
	\draw [thick,-stealth] (0,0) -- (0,-2)node[below=0.2pt]{$w_{5}$} ;
	\draw [thick,-stealth] (0,0) -- (-2,2) node[above=0.2pt]{$w_{3}+w_{2}$};
	\filldraw (0,0) circle (1pt) ;
	\coordinate (f1) at (0,1);
    \coordinate (f2) at (-0.7,0.7);
	\end{tikzpicture}
    \caption{A degenerate scattering diagram }
    \label{fig:degeneratediagramexamA2b}
     \end{subfigure}
    \caption{A demonstration of how to derive a degenerate scattering diagram from the non-degenerate scattering diagram for the $A_{2}$ cluster algebra.}
    \label{fig:degeneratediagramexamA2}
\end{figure}

\indent The tropicalization of each $f\in \mathcal{O}(\mathcal{X}^{\vee})$ defines a fan that splits $\mathbb{Z}^{N}$ into regions where $\textrm{Trop}[f(\hat{y})]$ is constant. We simply state without proof that all such fans together give the scattering diagram in the finite case \cite{2017arXiv171206968R,2018arXiv180605094R,Arkani-Hamed:2020tuz}. For example, again consider the $A_{2}$ cluster algebra and the tropicalization of functions in eq. (\ref{multiplicativsss}):
\begin{equation}\label{eq:f1f2f3trop}
\begin{split}
f_{1}&=1+\hat{y}_{1}^{\vee} \ \ \rightarrow \ \ \textrm{Trop}[f_{1}]=\min(0,\hat{y}_{1}^{\vee}) \ , \\ 
f_{2}&=1+\hat{y}_{2}^{\vee}\ \ \rightarrow \ \  \textrm{Trop}[f_{2}]=\min(0,\hat{y}_{2}^{\vee})  \ , \\ 
f_{3}&=1+\hat{y}_{2}^{\vee}+\hat{y}_{1}^{\vee}\hat{y}_{2}^{\vee}\ \ \rightarrow \ \  \textrm{Trop}[f_{3}]=\min(0,\hat{y}_{2}^{\vee},\hat{y}_{1}^{\vee}+\hat{y}_{2}^{\vee}) \ . 
\end{split}
\end{equation}
\noindent The tropicalization of each $f_{i}$ defines a fan in $\mathbb{Z}^{2}$, which are given in fig. \ref{fig:degeneratediagramexamA2}. In this example, one can immediately see that the combination of all fans defined by tropicalization of $f_{i}\in \mathcal{O}(\mathcal{X}^{\vee})$ is equivalent to the scattering diagram for $\mathcal{X}$.\footnote{The relation between the scattering diagram of $\mathcal{X}$ and $\mathcal{O}(\mathcal{X}^{\vee})$ is easier to understand from a mirror symmetry perspective. $\mathcal{A}^{\vee}$ is dual to $\mathcal{X}$ under mirror symmetry \cite{2014arXiv1411.1394G}.} 

\indent We now motivate degenerate scattering diagrams. Suppose we do not tropicalize all regular functions in $\mathcal{O}(\mathcal{X}^{\vee})$, but only a subset. For example, suppose we only considered the tropicalization of $f_{2}$ and $f_{3}$ in eq. (\ref{eq:f1f2f3trop}). We would find only 4 walls in the scattering diagram. Naively, this does not correspond to a well-defined scattering diagram if we assume the walls are single cluster walls. However, one might conjecture that it corresponds to a \textit{degenerate scattering diagram}, where certain walls are combined so certain chambers are inaccessible. 

\subsection{Degenerate scattering diagrams}\label{sec:degeneratscatfan}

\indent We now introduce the notion of degenerate scattering diagrams to motivate this truncation. Suppose that we want to truncate some cones from the scattering diagram while keeping others. Rather we want to enforce certain conditions of the form: ``If you cross wall A, you must also cross wall B and vice-versa.'' This is a well-defined procedure if we combine walls in the scattering diagram. For example, again consider the scattering diagram associated with $A_{2}$. Suppose we consider the fan derived by only tropicalizing $f_{3}$ and $f_{2}$ in eq. (\ref{eq:f1f2f3trop}), leading to the degenerate fan in fig. \ref{fig:degeneratediagramexamA2b}. We can derive this fan from a wall combination procedure by combining walls $w_{2}$ and $w_{3}$ in the full scattering diagram in fig. \ref{fig:degeneratediagramexamA2a}. We can view this procedure as a ``wall combination procedure'' or ``cone truncation'' procedure. However, by combining walls, we lose several nice properties associated with the original scattering diagram. First, multiple functions are associated with a single degenerate wall, so wall crossing across a degenerate wall takes the form:
\begin{equation}\label{eq:multiplefunctionss}
\mu_{\gamma^{\perp}}\hat{y}_{\gamma_{i}}=\hat{y}_{\gamma_{i}}\prod_{a} f_{a}(\hat{y}_{\gamma^{\perp}_{a}})^{\langle \gamma_{i},\gamma^{\perp}_{a}\rangle }   \ . 
\end{equation}
\noindent Second, the functions $f_{a}$ in eq. (\ref{eq:multiplefunctionss}) change depending on whether you are mutating forward or backward across a degenerate wall. For instance, the functions associated with the degenerate wall in fig. \ref{fig:degeneratediagramexamA2b} take the form: 
\begin{equation}
\begin{split}
\mu^{+}_{\gamma^{\perp}=(1,1)}\hat{y}_{\gamma_{i}}&=\hat{y}_{\gamma_{i}}(1+\frac{\hat{y}_{\gamma_{1}}\hat{y}_{\gamma_{2}}}{1+\hat{y}_{\gamma_{1}}})^{\langle \gamma_{i},\gamma_{2}\rangle}(1+\hat{y}_{\gamma_{1}}+\hat{y}_{\gamma_{1}}\hat{y}_{\gamma_{2}})^{\langle \gamma_{i},\gamma_{1}\rangle} \ ,  \\
\mu^{-}_{\gamma^{\perp}=(1,1)}\hat{y}_{\gamma_{i}}&=\hat{y}_{\gamma_{i}}(1+\hat{y}_{\gamma_{1}}\hat{y}_{\gamma_{2}})^{-\langle \gamma_{i},\gamma_{2}\rangle}(1+\hat{y}_{\gamma_{1}}+\hat{y}_{\gamma_{1}}\hat{y}_{\gamma_{2}})^{-\langle \gamma_{i},\gamma_{1}\rangle} \ , \\
\end{split}
\end{equation}
\noindent where the $+$($-$) indicates if you going counter-clockwise (clockwise) around the scattering diagram. 

\indent Although the degenerate walls are useful for motivating asymptotic chambers, there is significant ambiguity in their construction. Primarily, given an arbitrary fan, we do not have a procedure for associating a unique degenerate scattering diagram to this fan. For example, again consider the fan in fig. \ref{fig:degeneratediagramexamA2}. We could construct this fan by combining wall $w_{2}$ with walls $w_{3}$ or $w_{1}$. Given only the fan, there is no canonical choice without additional input. 

\subsection{Asymptotic chambers from degenerate scattering diagrams}\label{sec:asymptotconedegersss}

\begin{figure}
\centering
  \begin{tikzpicture}[every node/.style={font=\footnotesize}]
	%\node at (1.875,4.3) {$\mathcal{A}_5[(1,2),3,(4,5)]$};
	%\fill [fill=red!50!white] (1,0) ++(2,2) -- (1,0) -- (0,0) -- (0,1) -- ++(2,2);
    \filldraw [color=blue!30] (-2,-2) -- (-2,-1) -- (-1,0) -- (0,0) -- (0,-1) -- (-1,-2) -- (-2,-2) ;
    \draw [thick] (-2,-2) -- (-2,-1) -- (-1,0) -- (0,0) -- (0,-1) -- (-1,-2) -- (-2,-2);
    \filldraw (-2,-2) circle (1pt) (-1,0) circle (1pt) (0,0) circle (1pt) (0,-1) circle (1pt) (-1,-2) circle (1pt) (-2,-1) circle (1pt);
	\end{tikzpicture}
  \caption{The dual polytope of the asymptotic scattering diagram in fig. \ref{fig:sketchofwallA21withoutasymptot}. Each asymptotic chamber corresponds to a vertex and walls between asymptotic chambers correspond to 1-dim edges.}
  \label{fig:clusterA21limfacet}
\end{figure}
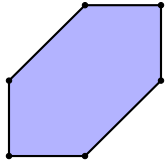

\indent We now consider the above procedure when the number of cones is infinite. We work with the degenerate cluster polytope instead of the degenerate scattering diagram.\footnote{Working with the degenerate cluster polytope is purely for visualization purposes and contains equivalent combinatorial information to the degenerate scattering diagram. A review of the map is provided in appendix \ref{sec:clusterpolytopes}.} Note that if one only tropicalizes a finite subset of $\mathcal{O}(\mathcal{X}^{\vee})$, one often finds that the associated degenerate cluster polytope includes a facet corresponding to a limiting ray.\footnote{In some sense, this facet would not appear if we tropicalized all functions in $\mathcal{O}(\mathcal{X}^{\vee})$ as the facet would be pushed to infinity.} For example, consider the following principle quiver:
\[
\begin{tikzcd}
y_{1}\arrow[r]& x_{1} \arrow[dr] & &\\
y_{2}\arrow[r]& x_{2} \arrow[r]\arrow[u]& x_{3} & y_{3}\arrow[l] \ ,
\end{tikzcd}
\]
\noindent which corresponds to the $A_{2,1}$ cluster algebra, and its dual quiver,
\[
\begin{tikzcd}
y_{1}^{\vee}\arrow[r]& x_{1}^{\vee} \arrow[d] & &\\
y_{2}^{\vee}\arrow[r]& x_{2}^{\vee} & x_{3}^{\vee}\arrow[l]\arrow[ul] & y_{3}^{\vee}\arrow[l] \ .
\end{tikzcd}
\]
\noindent Now consider the tropicalization of the following subset of regular functions of $A_{2,1}^{\vee}$:
\begin{equation}\label{eq:listlaurentpolyA21}
\begin{split}
f_{1}&=\hat{y}_{1}^{\vee}+1 \ , \\
f_{2}&=\hat{y}_{2}^{\vee}+1 \ ,\\
f_{3}&=\hat{y}_{3}^{\vee}+1\ , \\
f_{4}&=\hat{y}_{3}^{\vee}\hat{y}_{1}^{\vee}+\hat{y}_{3}^{\vee}+1 \ , \\
f_{5}&=\hat{y}_{2}^{\vee}\hat{y}_{1}^{\vee}+\hat{y}_{1}^{\vee}+1 \ , \\
f_{6}&=\hat{y}_{3}^{\vee}\hat{y}_{2}^{\vee}+\hat{y}_{3}^{\vee}+1 \ . 
\end{split}
\end{equation}
\begin{figure}
\centering
  \includegraphics[scale=0.6]{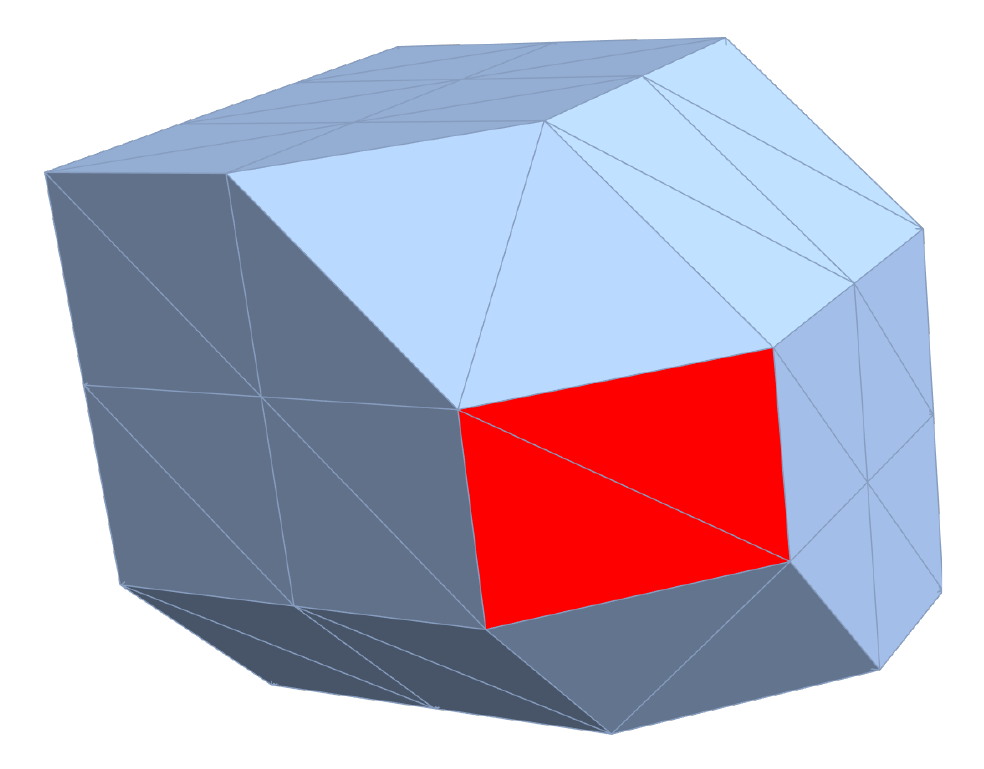}
  \caption{A degenerate cluster polytope of $A_{2,1}$ corresponding to the tropicalization of polynomials in eq. (\ref{eq:listlaurentpolyA21}). The red facet corresponds to the limiting ray.}
  \label{fig:exampleA2p1poly}
\end{figure}\noindent The corresponding polytope is given in fig. \ref{fig:exampleA2p1poly}, where the facet corresponding to the limiting ray is highlighted in red. We argue that the vertices containing this facet correspond to asymptotic chambers in the degenerate scattering diagram. Such a conjecture naturally explains the appearance of algebraic letters found at 8-point. Note that the facet associated with the limiting ray is not dual to the asymptotic scattering diagram given in section \ref{sec:infiniteraysssA2p1}, whose associated cluster polytope is given in fig. \ref{fig:clusterA21limfacet}. This is because even the asymptotic scattering diagram is degenerate.\footnote{It seems that the facet associated with the limiting ray will always be degenerate unless you include $F$-polynomials associated with points on the limiting ray. These polynomials are not elements of $\mathcal{O}(\mathcal{X}^{\vee})$ as they are not critically positive \cite{Arkani-Hamed:2019rds}. In our example, you need to include the $F$-polynomial associated with the generalization of $\mathcal{P}$ in appendix \ref{sec:a11cluster}, even though $\mathcal{P}$ is not an $x$-variable of the dual cluster algebra. } 

\indent There have been several proposals for deriving degenerate scattering diagrams. Such proposals amount to choosing finite subsets of $\mathcal{O}(\mathcal{X}^{\vee})$ to tropicalize. For example, the authors of ref. \cite{Arkani-Hamed:2019rds} proposed that the desirable subset of $\mathcal{O}(\mathcal{X}^{\vee})$ corresponds to the smallest subset of minors closed under parity: $\langle i,i+1,j,j+1\rangle$ and $\langle i,j-1,j,j+1\rangle$. These functions can be identified with some subset of $\mathcal{O}(\mathcal{X}^{\vee})$ using the ``web-variables'' originally given in ref. \cite{2003math.....12297S}. Several alternate subsets have been proposed \cite{Henke:2019hve,Arkani-Hamed:2019rds,Drummond:2019cxm,Drummond:2020kqg}. However, in contrast to our conjecture, which motivates a truncation of the clusters, these proposals argue for a truncation of the $x$-variables. In the finite case, the authors identify a subset of $x$-variables whose $g$-vectors are in bijection with facets of the degenerate cluster polytope and conjecture that this subset acts as a complete multiplicative basis for the desired $\hat{y}$-variables. In the infinite case, where facets corresponding to limiting rays appear, they conjecture the limiting rays correspond to cluster algebraic functions. It may turn out these conjectures are equivalent to our proposal. To make any definite conclusion, one would have to find a more precise procedure for isolating the correct degenerate scattering diagram, as the procedure provided here is still ambiguous. 

\section{Conclusion}\label{sec:conclusionsss}

The structure of scattering amplitudes beyond Feynman diagrams has undergone intense study in several contexts over the past 60 years. This program has been very successful at tree level, where numerous bottom-up approaches have almost completely circumvented the Lagrangian approach \cite{McGady:2013sga,Cheung:2016drk,Rodina:2016jyz,Arkani-Hamed:2016rak,Arkani-Hamed:2017jhn,Rodina:2018pcb,Elvang:2018dco,Cheung:2018oki,Elvang:2020lue}. However, a systematic understanding of how locality, causality and unitarity are precisely encoded at all orders in scattering amplitudes remains surprisingly elusive. Many approaches, ranging from topological strings on twistor space \cite{Witten:2003nn} to flat space holography \cite{Pasterski:2016qvg,Cheung:2016iub,Pasterski:2017ylz}, have given partial answers to this problem. For example, the infrared structure of scattering amplitudes is famously connected to the vacuum structure of the theory and asymptotic symmetries \cite{Kapec:2014opa,Pasterski:2015tva,Strominger:2017zoo,Arkani-Hamed:2020gyp}. Recent research suggests that the underlying structure of scattering amplitudes is deeply connected to geometric and combinatorial notions such as total positivity and motives \cite{ArkaniHamed:2012nw,Golden:2013xva,Arkani-Hamed:2017vfh,Arkani-Hamed:2017mur,Arkani-Hamed:2019mrd,Caron-Huot:2019bsq}. The amplituhedron provides a precise geometric description of integrands in $\mathcal{N}=4$ pSYM at all-loop orders. However, although the amplituhedron has led to many interesting results in the study of scattering amplitudes, it is a fundamentally perturbative description of the underlying physics. The ultimate goal of this program is a geometric description of the integrated all-loop amplitude independent of the chosen perturbation method, a ``non-perturbative geometry'' \cite{Arkani-Hamed:2019rds}. 

\indent One possible manifestation of this non-perturbative geometry is the connection between boundaries of the positive kinematic region and logarithmic branch points of integrated MHV amplitudes in $\mathcal{N}=4$ pSYM. This conjecture is more subtle than it initially appears due to ambiguities in the precise definition of the positive kinematic region, such as the chosen compactification. In this paper, we focused on studying the positive kinematic region of the MHV sector and proposed that scattering diagrams are a useful mathematical framework to study the boundary structure of the positive kinematic region. Furthermore, we developed the notion of asymptotic chambers to explain the appearance of algebraic letters in the symbol alphabet of MHV amplitudes. Interestingly, the asymptotic diagram approach provides manifestly rational relations for the asymptotic $\hat{y}_{\gamma}$-variables associated with the same limiting ray. 

\indent As a proof of concept, we used scattering diagrams to study the branch point structure of the 8-point MHV amplitude. Using the scattering diagram framework, we made a conjecture for all possible algebraic letters that could appear in the 8-point symbol alphabet. We confirmed that the algebraic letters found in explicit computations could be written as monomials of letters in our alphabet. We also developed the notion of degenerate scattering diagrams and commented on a possible truncation procedure for $\hat{y}$-variables, following the philosophy of refs. \cite{Arkani-Hamed:2019rds,Henke:2019hve,Drummond:2019cxm}. 

\indent Our results are especially interesting in the context of the Landau equations \cite{LANDAU1959181,Gurdogan:2020tip}. The Landau equations provide a direct link between the structure of the integrand and the branch points of the integrated amplitude. In particular, the branch points of amplitudes at high multiplicity and loop order have been calculated by applying the Landau equations to the amplituhedron \cite{Dennen:2016mdk,Prlina:2017azl,Prlina:2017tvx}. However, although the Landau equations provide a non-trivial probe of the integrated amplitudes' branch points, knowledge of the branch points is not enough to uniquely determine the symbol alphabet (see section 7 of ref. \cite{Prlina:2017azl}). For example, although some letters in the alphabet may take the schematic form

\begin{equation}
\phi \sim \frac{f-\sqrt{\triangle'}}{f+\sqrt{\triangle'}}    
\end{equation}

\noindent where $f$ and $\triangle'$ are rational functions of external kinematic data, the Landau equations only predict branch points of the form $\triangle'=0$. This mismatch results from how the solution to the Landau equations corresponds to the algebraic branch cut from the square-root in $\phi_{i}$ instead of the full logarithmic branch point. A related mismatch also occurs for rational branch points. Similar to how cluster algebras provide the missing link between Landau singularities and the symbol alphabet at 6-point and 7-point, asymptotic chambers provide the missing link between the algebraic symbol alphabet and specific solutions to the Landau equations at 8-point. It has been argued that the branch points of $\mathcal{N}=4$ pSYM associated with solutions to Landau equations are universal to all gauge theories. It would be interesting to understand whether the logarithmic branch points studied in this paper, which contain more information than the solutions to the Landau equations, retain any degree of universality. 

\begin{figure}
\centering
  \includegraphics[scale=0.6]{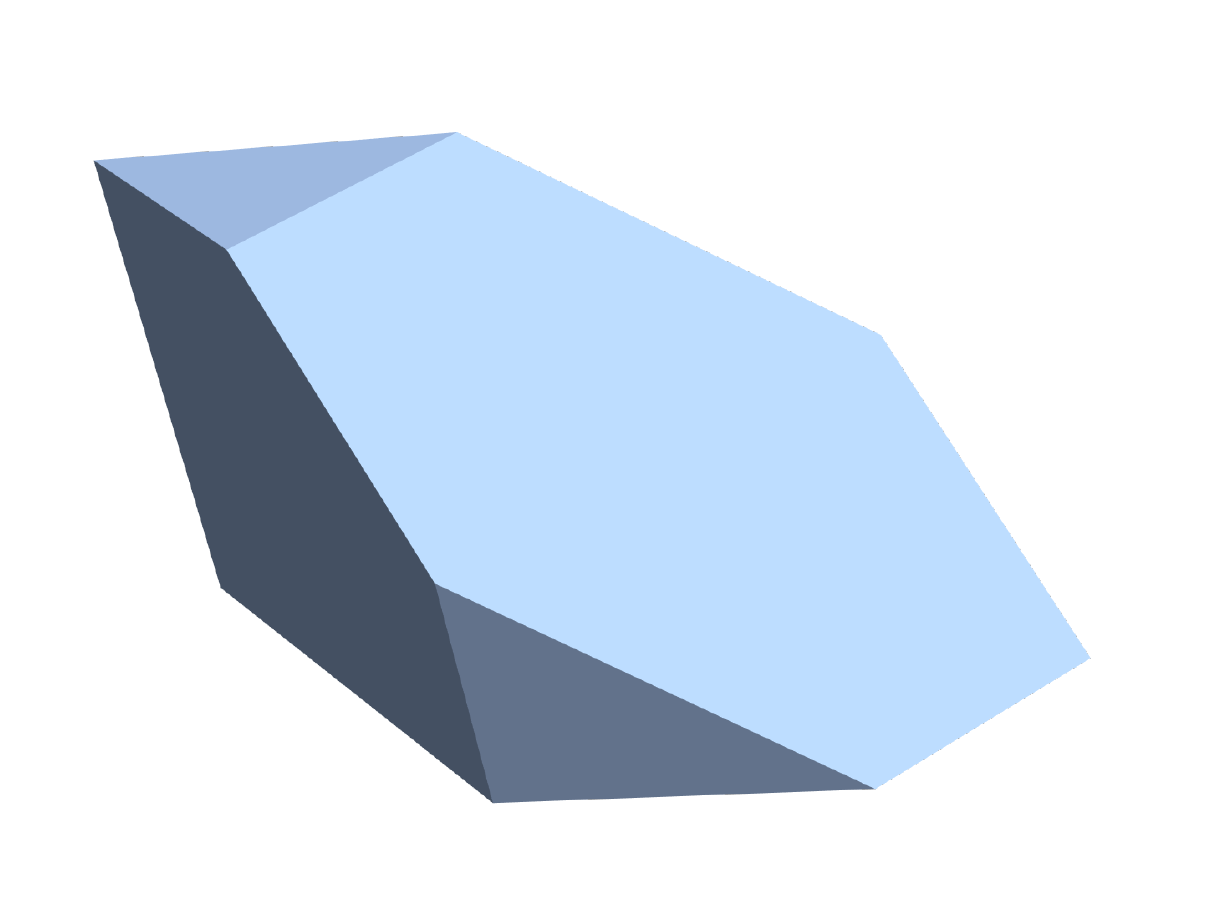}
  \caption{A degenerate cluster polytope of $A_{2,1}$ corresponding to the tropicalization of $f_{4}$ and $f_{5}$ in eq. (\ref{eq:listlaurentpolyA21}) along with eq. (\ref{eq:Fpolylimitingraysss}). }
  \label{fig:exampleA2p1polyfull}
\end{figure}

\indent The notion of degenerate scattering diagrams has applications beyond planar gauge theories, specifically higher loop integrands of $\phi^{3}$. However, it is instead the cluster polytope picture that is more interesting for studying higher loop integrands of $\phi^{3}$ \cite{Arkani-Hamed:2019vag,nimagiuliopaper} and generalized scattering amplitudes \cite{Cachazo:2019ngv,Cachazo:2019apa,Borges:2019csl,Cachazo:2019ble,Cachazo:2019xjx,Cachazo:2020uup,Cachazo:2020wgu}. Both the higher loop integrands of $\phi^{3}$ and generalized scattering amplitudes can be identified with the canonical rational function of the (degenerate) cluster polytopes discussed in section \ref{sec:degscatteringtrop} \cite{Drummond:2020kqg}. Each vertex in the cluster polytope can be mapped to a specific Feynman diagram. However, multiple vertices correspond to the same Feynman diagram, and considering all vertices in the full cluster polytope generically overcounts certain Feynman diagrams. Therefore, it is instead more natural to consider degenerate cluster polytopes, where redundant vertices have been truncated. For example, the degenerate cluster polytope associated with $A_{2,1}$, fig. \ref{fig:exampleA2p1polyfull}, is associated with the multi-trace, 1-loop 3-point integrand of $\phi^{3}$ theory \cite{nimagiuliopaper}. This degenerate polytope can be derived from the tropicalization of $f_{5}$ and $f_{6}$ in eq. (\ref{eq:listlaurentpolyA21}) along with the polynomial
\begin{equation}\label{eq:Fpolylimitingraysss}
f_{\mathcal{P}}=1+\hat{y}_{3}^{\vee}+\hat{y}_{1}^{\vee}\hat{y}_{2}^{\vee}\hat{y}_{3}^{\vee} \ ,
\end{equation}
which is the $F$-polynomial of the generalization of $\mathcal{P}$ in appendix \ref{sec:a11cluster} to $A_{2,1}$. Although the motivation for truncating the unwanted vertices is very different, the notion of truncating undesirable vertices (cones) from the degenerate cluster polytope (scattering diagram) is the same as section \ref{sec:degscatteringtrop}. 

\indent Finally, the notion of asymptotic chambers has applications outside of scattering amplitudes, such as studying coordinate systems of (higher) Teichmuller spaces. Specifically, when the cluster algebra corresponds to the (higher) Teichmuller space of a Riemann surface, one can identify cluster algebraic functions with Fenchel-Nielsen coordinates \cite{Gaiotto:2012rg}. Non-trivial relations obeyed by Fenchel-Nielsen coordinates, and their generalizations, have been studied in the context of spectral networks \cite{Hollands:2013qza,Hollands:2017ahy}. However, to our knowledge, no one has systematically studied relations between Fenchel-Nielsen coordinates in the context of scattering diagrams and cluster mutations. 

\indent Since our core result is a general framework for approaching these questions rather than individual results, there are several future directions.

\subsection{Future Directions}

\textbf{Gaining a More Systematic Understanding of Limiting Walls} \\
The most obvious direction for future work is developing a more systematic understanding of limiting walls and their connection to algebraic letters. For example, proving the bound
\begin{equation}\label{eq:conbound}
0<\hat{y}_{\gamma^{\perp}_{lim}}<1    \ ,
\end{equation}
\noindent holds in any asymptotic chamber would certainly be very interesting. However, even with eq. (\ref{eq:conbound}), it is unclear the notion of asymptotic chambers is always well defined. Alternatively, another interesting problem could be proving, or disproving, that the number of multiplicatively independent asymptotic letters associated with a limiting ray always is the number of intersecting walls plus the rank of the cluster algebra. \\

\noindent \textbf{Algorithm for Finding All Asymptotic Chambers} \\
A very practical direction for future work is the development of new algorithms for finding the $\hat{y}_{\gamma_{i}}$ of more general asymptotic chambers. For instance, if one cannot identify an $A_{1,1}$ subalgebra, one would instead need to generalize the generating function method of appendix \ref{sec:a11cluster} to more general sequences of quivers. In addition to finding explicit expressions for $\hat{y}_{\gamma_{i}}^{0}$, one also needs to find all the walls that intersect the limiting ray. The algorithm in appendix \ref{sec:algorithlimitraytrue} becomes highly inefficient for limiting rays in higher rank cluster algebras. Some possible approaches are: (1) calculating the walls using the tropicalization procedure in section \ref{sec:tropicalizationprocedre}, (2) developing some generalization of eq. (\ref{eq:mutategvectorsonly}) for asymptotic chambers, or (3) generalizing the causal diamond picture of refs. \cite{Arkani-Hamed:2019vag,nimagiuliopaper}. \\ 

\noindent \textbf{Degenerate Scattering Diagrams} \\
The discussion regarding degenerate scattering diagrams and truncated cones in section \ref{sec:asymptotconedegersss} was largely qualitative. Although the construction naturally motivated the notion of asymptotic chambers, it did not provide a concrete procedure for identifying the exact degenerate scattering diagram given the corresponding fan. A primary goal of future work would be developing the notion of canonical degenerate scattering diagrams, perhaps modifying the somewhat ad-hoc construction in section \ref{sec:degeneratscatfan}. Motivated by related results in refs. \cite{2013arXiv1309.5922G,Cachazo:2020wgu}, we note that understanding $f_{i}\in \mathcal{O}(\mathcal{X}^{\vee})$, and not just their tropicalization, could be important for developing such a notion. One path is investigating the explicit calculations in ref. \cite{He:2020ray} that connect boundary points of $\overline{Gr(k,n)/T}$ to $f_{i}$ using generalized scattering equations.  \\

\noindent \textbf{Finding Critically Positive Coordinates for $Gr_{k}(4,n)/T$} \\
We focus on the positive kinematic region defined by $Gr_{+}(4,n)/T$ due to its connection to logarithmic branch cuts of MHV amplitudes. However, the amplituhedron story naturally suggests the existence of alternate positive kinematic regions, $Gr_{k}(4,n)$, that are relevant for N${}^{k}$MHV amplitudes \cite{Arkani-Hamed:2017vfh}. These positive spaces are much more non-trivial than $Gr_{+}(4,n)$ and could be tied to the appearance of more general functions beyond the MHV sector, such as elliptic polylogarithms \cite{CaronHuot:2012ab,ArkaniHamed:2012nw,Broedel:2018iwv,Bourjaily:2018ycu,Broedel:2018qkq,Bourjaily:2020hjv}. It seems highly plausible that some notion of critically positive coordinates does generalize to these positive kinematic regions. \\

\noindent \textbf{Positive Kinematic Region of More General Theories} \\
One could also attempt to understand the significance of the positive kinematic region for more general theories. A prime target would be understanding the positive kinematic region associated with $\mathcal{N}=4$ pSYM amplitudes on the Coulomb branch \cite{Craig:2011ws,Herderschee:2019dmc}, or more general massive theories \cite{Arkani-Hamed:2017jhn,Herderschee:2019ofc}. The kinematic space of these theories does not correspond to $\overline{Gr(4,n)/T}$ since external states are massive. Alternatively, one could try to understand the positive kinematic region of more general scalar theories, such as bi-fundamental $\phi^{3}$ theory \cite{Herderschee:2019wtl,Herderschee:2020lgb} and $\phi^{p}$ theory \cite{Raman:2019utu,Kalyanapuram:2019nnf,Kalyanapuram:2020vil}. \\

\noindent \textbf{The Positive Kinematic Region and $(2,2)$ Signature} \\
Although the positive kinematic region exhibits remarkable properties and is intimately connected with the analytic structure of amplitudes, there is currently no physical explanation of its importance. However, instead of directly addressing the importance of the positive kinematic region, perhaps an easier preliminary question is the importance of $(2,2)$ signature. Using $(2,2)$ signature has often been viewed as a trick for removing ambiguities in certain computations due to the plethora of subtleties that emerge from trying to actually understand field theory dynamics in $(2,2)$ signature. However, there has been a recent revival in trying to systematically understand the behavior of theories in $(3,1)$ signature, most notably the development of light ray operators, which has led to several non-trivial results \cite{Hofman:2008ar,Hartman:2015lfa,Maldacena:2015iua,Caron-Huot:2017vep,Alday:2017vkk,Simmons-Duffin:2017nub,Kravchuk:2018htv}. It would be interesting if such an analysis in $(2,2)$ signature could lead to similarly novel results.

\acknowledgments
AH would like to thank Nima Arkani-Hamed, Will Dana, Song He, Thomas Lam, Zhenjie Li, Lecheng Ren, Chi Zhang, and Peng Zhao for stimulating discussion. In addition, AH is grateful to an anonymous referee, Niklas Henke and Georgios Papathanasiou for their constructive comments on the draft. AH would like to especially thank Henriette Elvang for her continued support and comments. AH is supported in part by the US Department of Energy under Grant No.
DESC0007859 and in part by a Leinweber Center for Theoretical Physics
Graduate Fellowship. 

\appendix

\section{Introduction to cluster algebras}\label{sec:introductionclusters}

In this appendix, we give a brief introduction to cluster algebras \cite{2001math......4151F,2003InMat.154...63F,2003math......5434B,2006math......2259F}. Thorough introductions are refs. \cite{2012arXiv1212.6263W,2016arXiv160805735F,2017arXiv170707190F,2020arXiv200809189F}, while those looking for a review that focuses on the connection with scattering amplitudes are referred to ref. \cite{Golden:2013xva}. Cluster algebras were initially motivated by the notion of total positivity. For example, one major motivating question was how much information is generically needed to prove that minors of a given matrix are positive. Due to non-linear relations between minors, this question quickly becomes very hard from a brute force approach of writing out all relations between minors and solving these polynomials directly. The advent of cluster algebras gave a different approach. 

\begin{figure}
\centering
\begin{tikzpicture}[every node/.style={font=\footnotesize,},dir/.style={decoration={markings, mark=at position \halfway with {\arrow{Latex}}},postaction={decorate}}]
\coordinate (p2) at (162:1.5);
\coordinate (p4) at (18:1.5);
\coordinate (p5) at (-54:1.5);
\node (p1) at (-126:1.5) [vertex,label={[label distance=-2pt]-126:{$1$}}] {};
\node at (p2) [vertex,label={[label distance=-2pt]162:{$2$}}] {};
\node (p3) at (0,1.5)  [vertex,label={[label distance=-1pt]90:{$3$}}] {};
\node at (p4) [vertex,label={[label distance=-2pt]18:{$4$}}] {};
\node at (p5) [vertex,label={[label distance=-2pt]-54:{$5$}}] {};
\draw [thick,black] (p1.center) -- (p3.center) (p1) -- (p4.center);
\path (p1.center) -- (p2.center) node (f1) [pos=0.5] {};
\path (p2.center) -- (p3.center) node (f2) [pos=0.5] {};
\path (p3.center) -- (p4.center) node (f3) [pos=0.5] {};
\path (p5.center) -- (p1.center) node (f5) [pos=0.5] {};
\draw [thick] (p1.center) -- (p2.center) -- (p3.center) -- (p4.center) -- (p5.center) -- cycle;
\begin{scope}[xshift=4 cm]
\coordinate (p2) at (162:1.5);
\coordinate (p4) at (18:1.5);
\coordinate (p5) at (-54:1.5);
\node (p1) at (-126:1.5) [vertex,label={[label distance=-2pt]-126:{$1$}}] {};
\node at (p2) [vertex,label={[label distance=-2pt]162:{$2$}}] {};
\node (p3) at (0,1.5)  [vertex,label={[label distance=-1pt]90:{$3$}}] {};
\node at (p4) [vertex,label={[label distance=-2pt]18:{$4$}}] {};
\node at (p5) [vertex,label={[label distance=-2pt]-54:{$5$}}] {};
\draw [thick,black] (p1.center) -- (p3.center) (p3) -- (p5.center);
\path (p1.center) -- (p2.center) node (f1) [pos=0.5] {};
\path (p2.center) -- (p3.center) node (f2) [pos=0.5] {};
\path (p3.center) -- (p4.center) node (f3) [pos=0.5] {};
\path (p5.center) -- (p1.center) node (f5) [pos=0.5] {};
\draw [thick] (p1.center) -- (p2.center) -- (p3.center) -- (p4.center) -- (p5.center) -- cycle;
\end{scope}
\begin{scope}[xshift=8 cm]
\coordinate (p2) at (162:1.5);
\coordinate (p4) at (18:1.5);
\coordinate (p5) at (-54:1.5);
\node (p1) at (-126:1.5) [vertex,label={[label distance=-2pt]-126:{$1$}}] {};
\node at (p2) [vertex,label={[label distance=-2pt]162:{$2$}}] {};
\node (p3) at (0,1.5)  [vertex,label={[label distance=-1pt]90:{$3$}}] {};
\node at (p4) [vertex,label={[label distance=-2pt]18:{$4$}}] {};
\node at (p5) [vertex,label={[label distance=-2pt]-54:{$5$}}] {};
\draw [thick,black] (p2.center) -- (p5.center) (p3) -- (p5.center);
\path (p1.center) -- (p2.center) node (f1) [pos=0.5] {};
\path (p2.center) -- (p3.center) node (f2) [pos=0.5] {};
\path (p3.center) -- (p4.center) node (f3) [pos=0.5] {};
\path (p5.center) -- (p1.center) node (f5) [pos=0.5] {};
\draw [thick] (p1.center) -- (p2.center) -- (p3.center) -- (p4.center) -- (p5.center) -- cycle;
\end{scope}
%\draw [thick] (p1.center) to [bend left=30] (p3.center) (p1.center) to [bend right=30] (p3.center);
\end{tikzpicture}
\caption{The first triangulation of the $5$-gon corresponds to the parameterization in eq. (\ref{initialminors}). Each minor in eq. (\ref{initialminors}) corresponds to an edge. The remaining triangulated $5$-gons correspond to the mutation pattern that leads to $\minorangsq{2,5}$.}
\label{fig:transformationn5}
\end{figure}
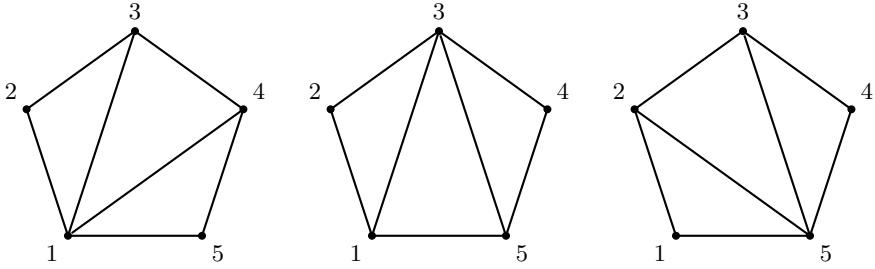

\indent Suppose you are given a $2\times n$ matrix and asked to find the minimal information needed to determine whether all $2\times 2$ minors are positive. The problem is non-trivial due to quadratic relations between minors called plucker relations:
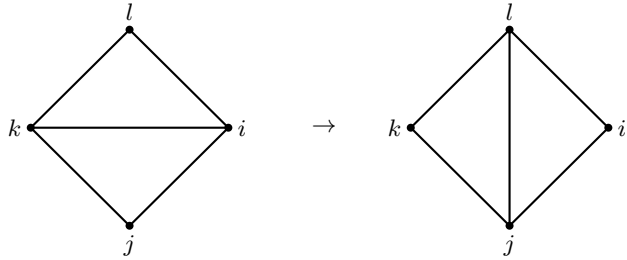
\begin{figure}
\centering
\begin{tikzpicture}[every node/.style={font=\footnotesize,},dir/.style={decoration={markings, mark=at position \halfway with {\arrow{Latex}}},postaction={decorate}}]
\coordinate (p1) at (0:1.3);
\coordinate (p2) at (-90:1.3);
\coordinate (p3) at (-180:1.3);
\coordinate (p4) at (-270:1.3);
\node (p1) at (0:1.3) [vertex,label={[label distance=-2pt]0:{$i$}}] {};
\node (p2) at (-90:1.3) [vertex,label={[label distance=-2pt]-90:{$j$}}] {};
\node (p3) at (-180:1.3) [vertex,label={[label distance=-2pt]-180:{$k$}}] {};
\node (p4) at (-270:1.3) [vertex,label={[label distance=-2pt]-270:{$l$}}] {};
\path (p1.center) -- (p2.center) node (f1) [pos=0.5] {};
\path (p2.center) -- (p3.center) node (f2) [pos=0.5] {};
\path (p3.center) -- (p4.center) node (f3) [pos=0.5] {};
\path (p4.center) -- (p1.center) node (f4) [pos=0.5] {};
\draw [thick] (p1.center) -- (p2.center) -- (p3.center) -- (p4.center) -- cycle;
\draw [thick,black] (p1.center) -- (p3.center) ;
\begin{scope}[xshift=2.2cm]
\node (p1) at (0:0) [label={[label distance=-2pt]0:{$\rightarrow$}}] {};
\end{scope}
\begin{scope}[xshift=5cm]
\coordinate (p1) at (0:1.3);
\coordinate (p2) at (-90:1.3);
\coordinate (p3) at (-180:1.3);
\coordinate (p4) at (-270:1.3);
\node (p1) at (0:1.3) [vertex,label={[label distance=-2pt]0:{$i$}}] {};
\node (p2) at (-90:1.3) [vertex,label={[label distance=-2pt]-90:{$j$}}] {};
\node (p3) at (-180:1.3) [vertex,label={[label distance=-2pt]-180:{$k$}}] {};
\node (p4) at (-270:1.3) [vertex,label={[label distance=-2pt]-270:{$l$}}] {};
\path (p1.center) -- (p2.center) node (f1) [pos=0.5] {};
\path (p2.center) -- (p3.center) node (f2) [pos=0.5] {};
\path (p3.center) -- (p4.center) node (f3) [pos=0.5] {};
\path (p4.center) -- (p1.center) node (f4) [pos=0.5] {};
\draw [thick] (p1.center) -- (p2.center) -- (p3.center) -- (p4.center) -- cycle;
\draw [thick,black] (p2.center) -- (p4.center) ;
\end{scope}
\end{tikzpicture}
\caption{A visual representation of the plucker relations for $Gr(2,n)$.}
\label{pluckerrelvis}
%Note that in the left panel, the $\mathfrak{f}$-$\mathfrak{af}$ pair $(1,2)$ and $(3,5)$ correspond to the same diagonal but counted as two different flavor lines.
\end{figure}
\begin{equation}\label{pluckerrelation}
\frac{1}{\minorangsq{i,k}}\left ( \minorangsq{i,j}\minorangsq{k,l}+\minorangsq{i,l}\minorangsq{j,k}\right )=\minorangsq{j,l} \quad i<j<k<l \ .
\end{equation}
\noindent A brute force approach would be to calculate \textit{all} quadratic relations of the form eq. (\ref{pluckerrelation}) at once and find some minimal subset directly. This computation would be problematic for even the best computers. We instead take a cluster algebra approach and find a preferred set of coordinates on the space of minors. To do so, we note that eq. (\ref{pluckerrelation}) can be visually interpreted as a mutation on the triangulation of a 4-gon with edges, $i,j,k,l$, as visualized in fig. \ref{pluckerrelvis}. Therefore, at $n=4$, a natural set of preferred minors is
\begin{equation}\label{4pointexmapless}
\minorangsq{1,2}, \ \minorangsq{2,3}, \ \minorangsq{3,4},\ \minorangsq{1,4},\  \minorangsq{1,3} \ .
\end{equation}
\noindent We can calculate the remaining coordinate, $\minorangsq{2,4}$, using eq. (\ref{pluckerrelation}), interpreting eq. (\ref{pluckerrelation}) as a mutation on the $4$-gon. Going beyond $n=4$, it is natural to start with coordinates that can be associated with the triangulation of an $n$-gon and interpret eq. (\ref{pluckerrelation}) as a mutation on this triangulated $n$-gon, just as we did for the triangulated $4$-gon. For example, consider $n=5$ and the initial coordinates:
\begin{equation}\label{initialminors}
\minorangsq{1,2}, \ \minorangsq{2,3},  \ \minorangsq{3,4},  \ \minorangsq{4,5},  \ \minorangsq{1,5},  \ \minorangsq{1,3},  \ \minorangsq{1,4}       
\end{equation}
\noindent which are associated with the first triangulation in fig. \ref{fig:transformationn5}. Suppose we want to write $\minorangsq{2,5}$ in terms of our initial coordinates. We first perform a ``mutation'' on $\minorangsq{1,4}$, finding 
\begin{equation}
\minorangsq{3,5}=\frac{1}{\minorangsq{1,4}}(\minorangsq{1,3}\minorangsq{4,5}+\minorangsq{3,4}\minorangsq{1,5})
\end{equation}
\noindent and a new triangulation where $\minorangsq{1,4}$ is replaced with $\minorangsq{3,5}$. We then perform a mutation on $\minorangsq{1,3}$, finding 
\begin{equation}
\minorangsq{2,5}=\frac{1}{\minorangsq{1,3}}(\minorangsq{3,5}\minorangsq{1,2}+\minorangsq{2,3}\minorangsq{1,5})    
\end{equation}
\noindent Therefore, assuming that all our initial minors in eq. (\ref{initialminors}) are positive, then $\minorangsq{2,5}$ must be positive as well. One can repeat the above calculation for any minor not in eq. (\ref{initialminors}), showing that all minors are positive if our initial minors in eq. (\ref{initialminors}) are positive. Note that we never mutate the edges that define the boundary of the $n$-gon. These are called frozen variables as they appear in the Plucker relations but do not themselves mutate. 

\indent The above discussion focuses on the positivity of a $Gr(2,n)$ matrix. However, we will ultimately be interested in $Gr(4,n)/T$, where $T$ acts on individual columns by a re-scaling:
\begin{equation}
Z_{i}^{A} \rightarrow t_{i} Z_{i}^{A} \ .
\end{equation} 
\noindent where $i$ and $A$ index the columns and rows respectively. Therefore, it is natural to consider the same question as above, except now for $Gr(2,n)/T$. Our minors, $\minorangsq{i,j}$, are no longer suitable coordinates as they are not invariant under $T$. Instead, we must develop a new set of coordinates, $\hat{y}$-variables, for a given triangulation that are invariant under $T$ transformations. Again consider the coordinates in eq. (\ref{initialminors}). Two natural combinations of minors invariant under $T$ are
\begin{equation} 
\hat{y}_{1}=\frac{\minorangsq{2,3}\minorangsq{1,4}}{\minorangsq{1,2}\minorangsq{3,4}}, \quad \hat{y}_{2}=\frac{\minorangsq{3,4}\minorangsq{1,5}}{\minorangsq{1,3}\minorangsq{4,5}} \ .    
\end{equation}
These variables form a natural set of coordinates on the compactified space $\overline{Gr(2,n)/T}$. To see their importance, lets interpret $\overline{Gr(2,n)/T}$ as the positive region of some manifold. Each triangulation, with its own $\hat{y}_{i}$ variables, corresponds to a different ``corner'' of $\overline{Gr(2,n)/T}$, as visualized for $n=5$ in fig. \ref{fig:exampleofcompactificationsA2}. 
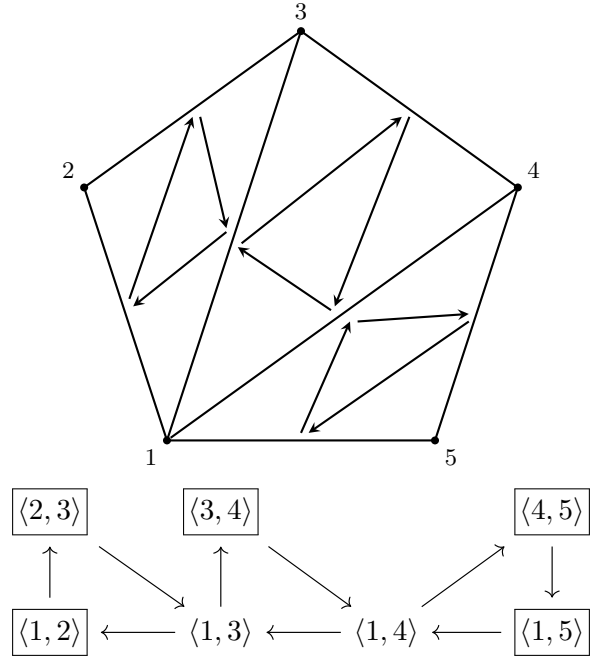
\begin{figure}[tbp]
\centering
\begin{tikzpicture}[every node/.style={font=\footnotesize,},dir/.style={decoration={markings, mark=at position \halfway with {\arrow{Latex}}},postaction={decorate}}]
\coordinate (p2) at (162:3);
\coordinate (p4) at (18:3);
\coordinate (p5) at (-54:3);
\node (p1) at (-126:3) [vertex,label={[label distance=-2pt]-126:{$1$}}] {};
\node at (p2) [vertex,label={[label distance=-2pt]162:{$2$}}] {};
\node (p3) at (0,3)  [vertex,label={[label distance=-1pt]90:{$3$}}] {};
\node at (p4) [vertex,label={[label distance=-2pt]18:{$4$}}] {};
\node at (p5) [vertex,label={[label distance=-2pt]-54:{$5$}}] {};
\draw [thick,black] (p1.center) -- (p3.center) (p1) -- (p4.center);
\path (p1.center) -- (p2.center) node (f1) [pos=0.5] {};
\path (p2.center) -- (p3.center) node (f2) [pos=0.5] {};
\path (p3.center) -- (p4.center) node (f3) [pos=0.5] {};
\path (p4.center) -- (p5.center) node (f4) [pos=0.5] {};
\path (p5.center) -- (p1.center) node (f5) [pos=0.5] {};
\path (p1.center) -- (p3.center) node (f6) [pos=0.5] {};
\path (p1.center) -- (p4.center) node (f7) [pos=0.5] {};
\draw [thick] (p1.center) -- (p2.center) -- (p3.center) -- (p4.center) -- (p5.center) -- cycle;
%\draw [thick] (p1.center) to [bend left=30] (p3.center) (p1.center) to [bend right=30] (p3.center);
\draw [thick,-stealth] ([xshift=0.1cm,yshift=-0.1cm]f2.center) -- ([xshift=-0.1cm,yshift=0.1cm]f6.center);
\draw [thick,-stealth] ([xshift=-0.1cm,yshift=0.05cm]f6.center) -- ([xshift=0.1cm,yshift=0.1cm]f1.center);
\draw [thick,-stealth] ([xshift=0.05cm,yshift=0.2cm]f1.center) -- ([xshift=-0.0 cm,yshift=-0.1cm]f2.center);
\draw [thick,-stealth] ([xshift=0.1cm,yshift=-0.1cm]f6.center) -- ([xshift=-0.1cm,yshift=-0.1cm]f3.center);
\draw [thick,-stealth] ([xshift=-0.0cm,yshift=-0.1cm]f3.center) -- ([xshift=-0.1cm,yshift=0.1cm]f7.center);
\draw [thick,-stealth] ([xshift=-0.15cm,yshift=0.05cm]f7.center) -- ([xshift=0.05cm,yshift=-0.15cm]f6.center);
\draw [thick,-stealth] ([xshift=0.2cm,yshift=-0.1cm]f7.center) -- ([xshift=-0.1cm]f4.center);
\draw [thick,-stealth] ([xshift=-0.1cm,yshift=-0.1cm]f4.center) -- ([xshift=0.1cm,yshift=0.1cm]f5.center);
\draw [thick,-stealth] ([yshift=0.1cm]f5.center) -- ([xshift=0.1cm,yshift=-0.1cm]f7.center);
\end{tikzpicture} \\
\begin{tikzcd}
\textrm{\framebox[6ex]{$\langle 2,3\rangle$}}\arrow[dr] & \textrm{\framebox[6ex]{$\langle 3,4\rangle$}} \arrow[dr] & & \textrm{\framebox[6ex]{$\langle 4,5\rangle$}} \arrow[d]\\
\textrm{\framebox[6ex]{$\langle 1,2\rangle$}} \arrow[u]& \minorangsq{1,3}\arrow[l]\arrow[u]  & \minorangsq{1,4} \arrow[l] \arrow[ur] & \textrm{\framebox[6ex]{$\langle 1,5\rangle$}} \arrow[l]
\end{tikzcd}
\caption{A triangulation of a $5$-gon and its dual quiver representation. The boxed elements in the quiver correspond to frozen nodes.}
\label{fig:quiverfromtriangulation}
\end{figure}

\indent The above strategy of finding an initial ``cluster'' of coordinates and developing a sequence of coordinate transforms turns out to be very versatile. Generalizing beyond $Gr(2,n)$, one can systematically develop the notion of a cluster algebra. Instead of a triangulation, we associate to each cluster a quiver with exchange matrix $B_{i,j}$:
\begin{equation}
B_{i,j}=\left\{\begin{matrix}
n & \textrm{if there are }n\textrm{ arrows from }i\textrm{ to }j \\ 
-n & \textrm{if there are }n\textrm{ arrows from }j\textrm{ to }i  \\ 
0 & \textrm{if there are no arrows between }j\textrm{ and }i
\end{matrix}\right. \ .    
\end{equation}
\noindent Each triangulation of an $n$-gon for $Gr(2,n)$ maps onto a triangulation in the following way

\begin{itemize}
    \item Each edge in the $n$-gon triangulation corresponds to a node in the quiver. The edges corresponding to the boundary of the $n$-gon are frozen nodes that never mutate 
    \item For each triangle in the $n$-gon triangulation, we draw a clock-wise orientated cycle in $Q$ connecting the vertices associated with the bounding edges.
\end{itemize}
\noindent For example, a visualization of the quiver associated with the first triangulation in fig. \ref{fig:transformationn5} is given in fig. \ref{fig:quiverfromtriangulation}. Given a mutation, the minors generalizes to cluster variables that mutate as 
\begin{equation}\label{ximutation}
\mu_{k}x_{i}=\left\{\begin{matrix}
\frac{1}{x_{i}}(\prod_{j\rightarrow i}x_{j}+\prod_{j\leftarrow i}x_{j}) & i=k \\ 
x_{i} & i\neq k
\end{matrix}\right.
\end{equation}
\noindent If we perform a mutation on node $k$, the quiver, and corresponding exchange matrix, mutate according to the rules
\begin{itemize}
\item Reverse all arrows going in or out of $k$,
\item For each sub-path of the form $i\rightarrow k \rightarrow j$, add the arrow $i\rightarrow j$,
\item Remove any two cycles that have formed.
\end{itemize}
\noindent One can explicitly check that eq. (\ref{ximutation}) and the preceding quiver mutation rules are a self consistent generalization of those given for $Gr(2,n)$. Finally, the $\hat{y}$ coordinates also have a natural generalization as 
\begin{equation}\label{defyvar}
\hat{y}_{i}=\prod_{j} x_{j}^{-B_{i,j}} \ ,   
\end{equation}
\noindent and mutate as 
\begin{equation}\label{yvariabels}
\mu_{j}\hat{y}_{i}=\left\{\begin{matrix}
\frac{1}{\hat{y}_{i}} & i=j\\ 
\hat{y}_{i}(1+\hat{y}_{j}^{\textrm{Sign}(B_{i,j})})^{B_{i,j}} & i\neq j
\end{matrix}\right.   \ . 
\end{equation}
\noindent We will denote the positive space parameterized by $x_{i}$ coordinates as $\mathcal{A}$ and the space parameterized by $\hat{y}_{i}$ coordinates as $\mathcal{X}$. The relation between $\mathcal{A}$ and $\mathcal{X}$ is still under active research and not completely understood. 

\indent Cluster algebras have many remarkable properties, such as the Laurent phenomenon. The cluster variable of any quiver can be written as a Laurent polynomial of $x_{i}$ of some initial cluster. For example, consider the cluster algebra associated with $Gr(2,5)$. One can show that any $\minorangsq{i',j'}$ can be written as a Laurent polynomial of $\minorangsq{i,j}$ in eq. (\ref{initialminors}). To learn about other amazing properties of cluster algebras, the reader is referred to refs. \cite{2016arXiv160805735F,2012arXiv1212.6263W}.

\section{Differentiating \texorpdfstring{$x$}{x}-variables with frozen nodes}
\label{ref:differentiatingyvar}
Like $\hat{y}$-variables, $x$-variables will also obey additional relations when there are fewer frozen nodes. To see this, again consider the $A_{3}$ cluster algebra with initial quiver:
\[
\begin{tikzcd}
x_{1} \arrow[r]& x_{2} & x_{3}\arrow[l] \ . 
\end{tikzcd}
\]
\noindent Without any frozen nodes, all $x$-variables in the cluster algebra are 
\begin{equation}\label{eq:distinctxvar}
\begin{split}
\{ &x_{1}, \quad x_{2}, \quad x_{3}  \\ 
&\frac{x_{2}+1}{x_{1}}, \quad \frac{x_{1}x_{3}+x_{2}^{2}+2x_{2}+1}{x_{1}x_{2}x_{3}}  \\
&\frac{x_{2}+1}{x_{3}}, \quad \frac{x_{1}x_{3}+x_{2}+1}{x_{1}x_{2}} \\
&\frac{x_{1}x_{3}+x_{2}+1}{x_{2}x_{3}}, \quad \frac{x_{1}x_{3}+1}{x_{2}} \} \ .
\end{split}    
\end{equation}
\noindent The minimal multiplicative basis of the $x$-variables is rank 7:
\begin{equation}
\begin{split}
\{ &x_{1}, \quad x_{2}, \quad x_{3} \\
&x_{2}+1, \quad x_{1}x_{3}+x_{2}^{2}+2x_{2}+1, \\
&x_{1}x_{3}+1, \quad x_{1}x_{3}+x_{2}+1 \} \ .
\end{split}
\end{equation}
\noindent However, suppose we include the additional frozen node, $z$, so the initial quiver is now:
\begin{equation}\label{A3examplequiver}
\begin{tikzcd}
z \arrow[d] & & \\
x_{1} \arrow[r]& x_{2} & x_{3}\arrow[l]\arrow[ull] \ . 
\end{tikzcd}
\end{equation}
\noindent All $x$-variables in the cluster algebra are now
\begin{equation}\label{eq:superdistinctxvar}
\begin{split}
\{ &z, \quad x_{1}, \quad x_{2}, \quad x_{3}  \\ 
&\frac{x_{2}+z}{x_{1}}, \quad \frac{x_{1}x_{3}z+zx_{2}^{2}+(1+z^2)x_{2}+z}{x_{1}x_{2}x_{3}}  \\
&\frac{x_{2}z+1}{x_{3}}, \quad \frac{x_{1}x_{3}z+x_{2}+z}{x_{1}x_{2}} \\
&\frac{x_{1}x_{3}+x_{2}z+1}{x_{2}x_{3}}, \quad \frac{x_{1}x_{3}+1}{x_{2}} \} \ ,
\end{split}    
\end{equation}
\noindent so the multiplicative basis for $x$-variables is now rank 10:
\begingroup
\allowdisplaybreaks
\begin{align}
\{ &z,\quad x_{1}, \quad x_{2}, \quad x_{3}, \nonumber \\
&x_{2}+z, \quad x_{1}x_{3}z+zx_{2}^{2}+(1+z^2)x_{2}+z, \label{directcomputat} \\
&x_{1}x_{3}+1, \quad x_{1}x_{3}z+x_{2}+z, \nonumber  \\
&x_{2}z+1, \quad x_{1}x_{3}+x_{2}z+1 \} \nonumber \ .
\end{align}
\endgroup
\noindent Comparing eqs. (\ref{eq:distinctxvar}) and (\ref{eq:superdistinctxvar}), one can clearly see that adding the frozen node, $z$, removes relations between the $x$-variables. Therefore, adding more frozen nodes disentangles the $x$-variables. 

\indent Remarkably, the frozen nodes of a principle quiver are enough to ensure that all the $x$-variables are maximally disentangled. To see this, note that the $x$-variables of a cluster algebra with \textit{completely arbitrary frozen nodes} can be always be written in the form,
\begin{equation}\label{eq:arbitraryformuasss}
x=x^{\vec{g}}F(\hat{y}_{i})\times (\textrm{monomial of frozen variables}) \ ,   
\end{equation}
\noindent where $\vec{g}$ and $F(\hat{y}_{i})$ are defined in eq. (\ref{eq:conjecturFpolynomiagvec}). The exact formula for computing the monomial of frozen $x$-variables is unimportant for our purposes and the reader is referred to appendix B of ref. \cite{Henke:2019hve} for details.  From eq. (\ref{eq:arbitraryformuasss}), we see that the $x$-variable of a cluster algebra with arbitrary frozen nodes is the same as the $x$-variable of cluster algebra with a principal quiver up to a monomial of frozen $x$-variables. Therefore, one multiplicative basis of the $\hat{y}$-variables of a cluster algebra with arbitrary frozen nodes is the multiplicative basis of $\hat{y}$-variables of a cluster algebra with a principal quiver in addition to all the frozen $x$-variables.  

\indent To see this result explicitly, again consider the $A_{3}$ cluster algebra. We now consider a cluster algebra with the principal quiver:
\[
\begin{tikzcd}
y_{1} \arrow[d] & y_{2} \arrow[d] & y_{3} \arrow[d] \\
x_{1} \arrow[r]& x_{2} & x_{3}\arrow[l] \ . 
\end{tikzcd}
\]
A complete basis of all $F$-polynomials is 
\begingroup
\allowdisplaybreaks
\begin{align}
f_{1}&=1+\hat{y}_{1} \nonumber \ , \\   
f_{2}&=1+\hat{y}_{2} \nonumber  \ , \\
f_{3}&=1+\hat{y}_{3} \nonumber  \ , \\
f_{4}&=1+\hat{y}_{1}+\hat{y}_{1}\hat{y}_{2} \label{F1A3bas} \ , \\
f_{5}&=1+\hat{y}_{3}+\hat{y}_{3}\hat{y}_{2} \nonumber \ , \\
f_{6}&=1+\hat{y}_{1}+\hat{y}_{3}+\hat{y}_{3}\hat{y}_{1}+\hat{y}_{3}\hat{y}_{2}\hat{y}_{1} \ . \nonumber 
\end{align}
\endgroup
We can now apply the basis in eq. (\ref{F1A3bas}) to the quiver in (\ref{A3examplequiver}), where
\begin{equation}\label{hatyvarA3}
\hat{y}_{1}=\frac{z}{x_{2}}, \quad \hat{y}_{2}=x_{1}x_{3}, \quad \hat{y}_{3}=\frac{1}{zx_{2}} \ .   
\end{equation}
Substituting eq. (\ref{hatyvarA3}) into eq. (\ref{F1A3bas}) and including the $x$-variables of the initial quiver in (\ref{A3examplequiver}) yields the multiplicative basis 
\begin{equation}\label{F1A3bas2}
\begin{split}
\{ & z,\ x_{1},\ x_{2}, \ x_{3} \ , \\
f_{1}&=\frac{x_{2}+z}{x_{2}} \ , \\   
f_{2}&=x_{1}x_{3}+1 \ , \\
f_{3}&=\frac{zx_{2}+1}{zx_{2}} \ , \\
f_{4}&=\frac{x_{1}x_{3}z+x_{2}+z}{x_{2}} \ , \\
f_{5}&=\frac{x_{1}x_{3}+x_{2}z+1}{zx_{2}}\ , \\
f_{6}&=\frac{zx_{1}x_{3}+zx_{2}^{2}+(1+z^{2})x_{2}+z}{zx_{2}^{2}} \} \ .
\end{split}    
\end{equation}
Eq. (\ref{F1A3bas2}) corresponds to a complete multiplicative basis for the $x$-variables in eq. (\ref{eq:superdistinctxvar}). A complete multiplicative basis for the $\hat{y}$-variables consists of the $f_{i}$ in eq. (\ref{F1A3bas2}), the $\hat{y}_{i}$ in eq. (\ref{hatyvarA3}) and $z$. 

\section{Review: \texorpdfstring{$A_{1,1}$}{A11} cluster algebra}
\label{sec:a11cluster}
In this appendix, we consider the cluster algebra and scattering diagram associated with the principal quiver,
\[
\begin{tikzcd}
y_{-1} \arrow[d] & y_{0} \arrow[d]\\
x_{-1} \arrow[shift left, r]\arrow[shift right, r] & x_{0} \ ,
\end{tikzcd}
\]
\noindent reviewing the results in refs. \cite{Henke:2019hve,Drummond:2019cxm,Arkani-Hamed:2019rds}. We perform repeated mutations on the nodes associated with $x_{-1}$ and $x_{0}$, starting with the $x_{-1}$ node,
\[
\begin{tikzcd}
 & y_{0} \arrow[d] & & y_{0} \arrow[d]&  & y_{0} \arrow{dl}[description]{2} &   & y_{0} \arrow{d}[description]{3} &\\
x_{-1} \arrow{r}[description]{2} & x_{0} & x_{1} \arrow[d] & x_{0} \arrow{l}[description]{2}& x_{1} \arrow{r}[description]{2} & x_{2} \arrow[u]\arrow{dl}[description]{2} & x_{3} \arrow{ur}[description]{2} \arrow{d}[description]{3} & x_{2} \arrow{l}[description]{2} &\\
y_{-1} \arrow[u] & & y_{-1} \arrow{ur}[description]{2} & & y_{-1} \arrow{u}[description]{3} & & y_{-1} \arrow{ur}[description]{4} & &\ldots \ .
\end{tikzcd}
\]
\noindent After a sequence of $2n$ mutations where $n>0$, the quiver takes the form
\[
\begin{tikzcd}
& y_{0} \arrow{dl}[description]{2n} \\
x_{2n-1} \arrow{r}[description]{2} & x_{2n} \arrow{u}[description]{2n-1}\arrow{dl}[description]{2n} \\
y_{-1} \arrow{u}[description]{2n+1}&  \ . 
\end{tikzcd}
\]
\noindent Using cluster mutations and the above representation of the quiver after $2n$ mutations, we defined a recursive solution for $x_{i}$ in this model, finding:
\begin{equation}
\begin{split}
x_{2n-1}x_{2n-3}&=(y_{-1}^{2n-1}y_{0}^{2n-2}+x_{2n-2}^{2}) \ , \\
x_{2n}x_{2n-2}&=(y_{-1}^{2n}y_{0}^{2n-1}+x_{2n-1}^{2}) \ . 
\end{split}
\end{equation}
\noindent This form of the mutation relations is still too complicated to solve analytically due to being inherently nonlinear. Instead, we identify a new variable,
\begin{equation}
\begin{split}
\mathcal{P}&=\frac{y_{-1}}{x_{-1}x_{0}}+\frac{x_{0}}{x_{-1}}+\frac{x_{-1}y_{-1}y_{0}}{x_{0}} \ ,
\end{split}
\end{equation}
\noindent such that
\begin{equation}\label{eq:Ppropertysolution}
x_{2n-1}=x_{2n-2}\mathcal{P}-x_{2n-3}\mathcal{F}, \quad \mathcal{F}=y_{-1}y_{0} \ .
\end{equation}
\noindent $\mathcal{P}$ is not an element of the cluster algebra, but a cluster-like variable associated with the limiting ray. For further discussion of $\mathcal{P}$, the reader is referred to ref. \cite{Arkani-Hamed:2019rds}. Only eq. (\ref{eq:Ppropertysolution}) is important for our purposes, which one can explicitly check. 

\indent We can solve eq. (\ref{eq:Ppropertysolution}) by first writing down the associated generating function:
\begin{equation}
G_{i>0}(t)=\frac{x_{0}-x_{-1}\mathcal{F}t}{1-\mathcal{P}t+\mathcal{F}t^{2}}=\sum_{n=0}^{\infty} x_{i}t^{i} 
\end{equation}
\noindent and then finding a closed-form expression for the derivatives of $G_{n>0}(t)$:
\begingroup
\allowdisplaybreaks
\begin{align}
x_{i}&=\frac{1}{2^{i+2}}[(x_{-1}+B_{+}\sqrt{\triangle})(\mathcal{P}+\sqrt{\triangle})^{i+1}+(x_{-1}-B_{+}\sqrt{\triangle})(\mathcal{P}-\sqrt{\triangle})^{i+1}] \ , \nonumber \\
B_{+}&=\frac{2x_{0}-x_{-1}\mathcal{P}}{\triangle} \ , \label{eq:closedformexpressionforxnA1p1} \\
\triangle&=\mathcal{P}^{2}-4\mathcal{F} \nonumber \ .  
\end{align}
\endgroup
\noindent Using our closed form expressions for $x_{i}$ in eq. (\ref{eq:closedformexpressionforxnA1p1}), it is trivial to calculate closed form expressions for $\hat{y}_{i}$ after 2n mutations:
\begin{equation}\label{yvariformula}
\hat{y}_{2n-1}=y_{0}^{2n}y_{-1}^{2n+1}x_{2n}^{-2}, \quad \hat{y}_{2n}=y_{0}^{1-2n}y_{-1}^{-2n}x_{2n-1}^{2} \ .
\end{equation}

\section{Algorithm for finding asymptotic chambers from \texorpdfstring{$A_{1,1}$}{A11} subalgebra}\label{sec:algorithlimitraytrue}
We now outline a search algorithm we used to find the asymptotic chambers associated with a limiting ray. In normal search algorithms for finite cluster algebras, one performs sequences of mutations until one finds all the cones in the fan, defining a cone by its associated $g$-vectors. This method does not work well for infinite cluster algebras where there are an infinite number of cones, even asymptotically close to the limiting ray. We partially circumvented this issue by defining a new equivalence class of cones arbitrarily close to the limiting ray called \textit{pre-asymptotic} chambers.  

\indent We first mutate the initial quiver until we find a quiver containing an $A_{1,1}$ subalgebra:
\[
\begin{tikzcd}
x_{i} \arrow[shift left, r]\arrow[shift right, r] & x_{j} \ ,
\end{tikzcd}
\]
\noindent which we define as the principal quiver. To calculate the $g$-vectors and walls of adjacent cones, we use the $g$-vector mutation formula originally derived in ref. \cite{2006math......2259F}:
\begin{equation}\label{eq:mutategvectorsonly}
\mu_{k}\vec{g}_{i}=\left\{\begin{matrix}
 \vec{g}_{i}&\textrm{if }i\neq k \\ 
 -\vec{g}_{i}+\sum_{m=1}^{N}[B_{m,k}]_{+}\vec{g}_{m}-\sum_{m=1}^{N}[B_{N+m,k}]_{+}\vec{b}_{m}& \textrm{if }i=k 
\end{matrix}\right. \ ,
\end{equation}
\noindent where $[x]_{+}=\max(x,0)$ and $\vec{b}_{m}$ is column $m$ of the initial $B_{i,j}$ matrix. This formula can be derived by combining eq. (\ref{ximutation}) in appendix \ref{sec:introductionclusters},
\begin{equation}
\mu_{k}x_{i}=\left\{\begin{matrix}
\frac{1}{x_{i}}(\prod_{j\rightarrow i}x_{j}+\prod_{j\leftarrow i}x_{j}) & i=k \\ 
x_{i} & i\neq k
\end{matrix}\right. \ ,
\end{equation}
\noindent with eq. (\ref{eq:conjecturFpolynomiagvec}) in section \ref{sec:principalquiverasymptotcones}:
\begin{equation}
x=x^{\vec{g}}F(\hat{y}_{i}), \quad x^{\vec{g}}=\prod_{i}x_{i}^{g_{i}}  \ . 
\end{equation}
\noindent This allowed us to compute the $g$-vectors of adjacent cones in the $g$-vector fan very efficiently. Using eq. (\ref{eq:mutategvectorsonly}), we calculated the $g$-vectors of cones associated with repeated mutations on nodes $x_{i}$ and $x_{j}$. The $g$-vectors of the $x_{i}$ and $x_{j}$ nodes asymptotically approached a limiting ray, $\vec{g}_{lim}$. In principle, we could now perform a brute force search, performing random mutations asymptotically close to the limiting ray until we found no new walls intersecting the limiting ray. However, this approach would be highly inefficient as there are always an infinite number of cones asymptotically close to the limiting ray. Although we did perform a brute-force search, we partially streamlined the algorithm by defining a new equivalence class of cones: pre-asymptotic chambers.

\indent Consider the schematic scattering diagram in fig. \ref{fig:sketchofwallA21}, which corresponds to the $A_{2,1}$ cluster algebra. Suppose we mutate to one of the cones in the sequence that approaches the asymptotic chamber $C_{1}$. Mutating across an asymptotic wall toward or away from the limiting ray does not give us any new information. In some sense, the sequence of cones approaching $C_{1}$ are equivalent, and therefore redundant, for the purposes of trying to find walls intersecting the limiting ray. We wish to find some criterion that allows us to avoid mutating into these redundant cones. To see what this criterion should be, we first note that the sequence of $g$-vectors along the black cluster wall generically obey the relation
\begin{equation}
\vec{g}(x_{n})-\vec{g}(x_{n-1})=\vec{g}_{lim} \ .
\end{equation}
\noindent Therefore, it is natural to consider the projection of the $g$-vectors onto the hyperplane perpendicular to the limiting ray, such that
\begin{equation}
\mathbf{P}_{\perp}(\vec{g}(x_{n})-\vec{g}(x_{n-1}))=\mathbf{P}_{\perp}\vec{g}_{lim}=\vec{0} \ .
\end{equation}
\noindent If we define equivalence classes of cones by considering the projection of their $g$-vectors, not the $g$-vectors themselves, the sequence of cones approaching $C_{1}$ correspond to the same cone under this projection. This is true for the sequences of cones approaching $C_{3}$, $C_{4}$, and $C_{6}$ as well. For $C_{2}$ and $C_{5}$, we find two classes of cone upon taking the projection. To see why, let us focus on $C_{2}$. Denoting the projection of the $g$-vectors on the two bordering cluster walls as $g_{a}$ and $g_{b}$, the two classes of cones are defined by the sets $\{g_{a},g_{a},g_{b} \}$ and $\{g_{a},g_{b},g_{b} \}$. Turning to more general cluster algebras, we define pre-asymptotic chambers as the equivalence classes of cones defined by the projection of $g$-vectors onto the hyperplane perpendicular to the limiting ray. Certain infinite sequences of cones approaching the same asymptotic chamber correspond to the same equivalence class under this projection.

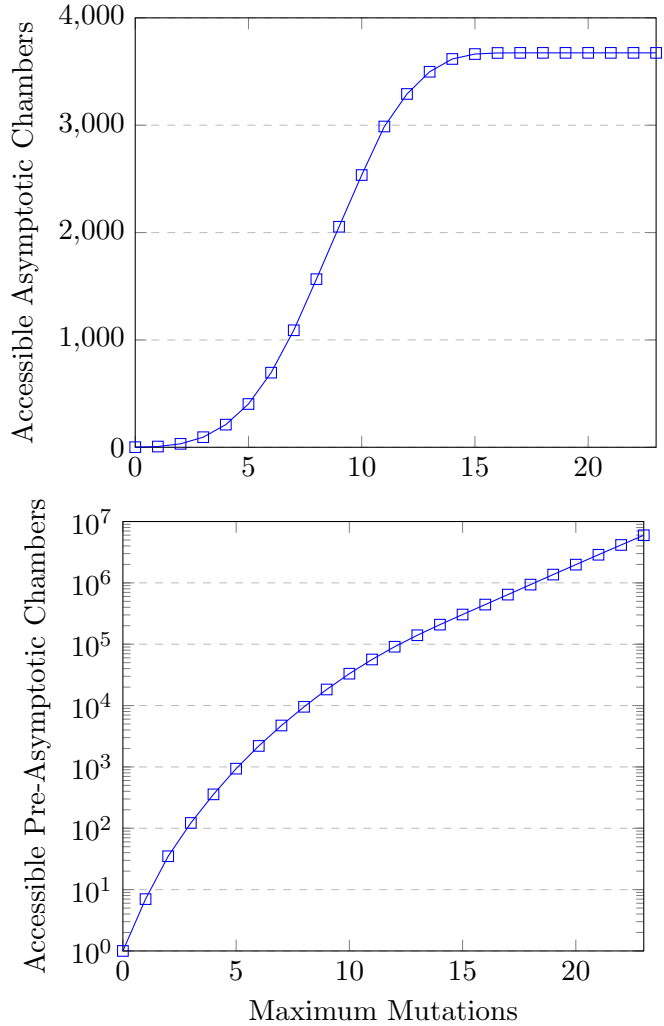
\begin{figure}[t!p]
\centering
\begin{subfigure}[b]{0.6\textwidth}
\centering
\begin{tikzpicture}
\begin{axis}[
    ylabel={Accessible Asymptotic Chambers},
    xmin=0, xmax=23,
    ymin=0, ymax=4000,
    xtick={0,5,10,15,20},
    ytick={0,1000,2000,3000,4000,6000,8000},
    legend pos=north west,
    ymajorgrids=true,
    grid style=dashed,
]

\addplot[
    color=blue,
    mark=square,
    ]
    coordinates {
    (0,1)(1,7)(2,32)(3,94)(4,211)(5,403)(6,695)(7,1090)(8,1566)(9,2054)(10,2536)(11,2988)(12,3291)(13,3498)(14,3617)(15,3663)(16,3673)(17,3674)(18,3674)(19,3674)(20,3674)(21,3674)(22,3674)(23,3674)};
    
\end{axis}
\end{tikzpicture}
\end{subfigure} \\
\begin{subfigure}[b]{0.6\textwidth}
\centering
\begin{tikzpicture}
\begin{semilogyaxis}[
    xlabel={Maximum Mutations},
    ylabel={Accessible Pre-Asymptotic Chambers},
    xmin=0, xmax=23,
    ymin=1, ymax=10000000,
    xtick={0,5,10,15,20},
    ytick={0,1,10,100,1000,10000,100000,1000000,10000000},
    legend pos=north west,
    ymajorgrids=true,
    grid style=dashed]

\addplot[
    color=blue,
    mark=square,
    ]
    coordinates {
    (0,1)(1,7)(2,35)(3,122)(4,357)(5,936)(6,2197)(7,4733)(8,9545)(9,18292)(10,33077)(11,56354)(12,91014)(13,140052)(14,208460)(15,304750)(16,442755)(17,643554)(18,937072)(19,1364726)(20,1982900)(21,2873026)(22,4149925)(23,5975433)};
    
\end{semilogyaxis}
\end{tikzpicture}
\end{subfigure}
\caption{Number of accessible asymptotic and pre-asymptotic chambers after a maximum of $X$ mutations from the initial pre-asymptotic chamber. Note that we are only considering (pre-)asymptotic chambers on one side on the limiting wall.}
\label{plotofaccesibelchambers}
\end{figure}

\indent Using pre-asymptotic chambers does not completely remove undesirable redundancies, because a single asymptotic chamber can correspond to multiple pre-asymptotic chambers. This redundancy is not a problem for lower rank cluster algebras where the number of pre-asymptotic chambers is small and finite. For the asymptotic scattering diagrams of $A_{2,1}$ and $A_{2,2}$, one can prove that the number of asymptotic chambers is finite simply by showing that the number of pre-asymptotic chambers is finite. However, for $\overline{Gr(4,8)/T}$, we found that the number of pre-asymptotic chambers is infinite, or so large it is effectively infinite, while the number of asymptotic chambers is finite. If the number of pre-asymptotic chambers is infinite, brute force mutation procedures cannot prove that you have found all asymptotic chambers. Instead, one must perform mutations on pre-asymptotic chambers until no new asymptotic chambers appear after a large number of mutations. Specifically, we found that all the discovered asymptotic chambers were within 18 mutations of our initial pre-asymptotic chamber and checked all pre-asymptotic chambers within 23 mutations of our initial pre-asymptotic chamber. Fig. \ref{plotofaccesibelchambers} is a plot of the number of asymptotic and pre-asymptotic chambers, $Y$, accessible after a maximum of $X$ mutations on the initial pre-asymptotic chamber given in section \ref{sec:gr48fullexampless}.

\indent The final subtlety to consider is that we found only a subset of the asymptotic chambers when we performed our initial search of pre-asymptotic chambers asymptotically close to the limiting ray. We could not find all pre-asymptotic chambers from a single search because we cannot use the mutation rule in eq. (\ref{eq:mutategvectorsonly}) to mutate across limiting walls. Fortunately, the asymptotic scattering diagram studied in section \ref{sec:gr48fullexampless} has only one limiting wall. We can check how many limiting walls appear by studying the asymptotic walls that appear in our brute force search. For a limiting wall to appear, asymptotic walls almost parallel the limiting wall should appear during the brute force search of cones asymptotically close to the limiting ray. However, all asymptotic walls that appeared in the $\overline{Gr(4,8)/T}$ search were asymptotically parallel to the same limiting wall, 
\begin{equation}
\gamma^{\perp}_{lim}=(1,\ 0,\ 0,\ 0,\ 0,\ 0,\ 0,\ 0,\ 1) \ .   
\end{equation}
We subsequently assumed that only one limiting wall appears in the asymptotic scattering diagram. Furthermore, we found that the limiting wall fully divides the asymptotic scattering diagram. In other words, there is no ``short-cut'' around the limiting wall. Therefore, we performed two searches of the pre-asymptotic chambers, one on each side of the limiting wall.  

\section{Comparison to the origin clusters of ref. \texorpdfstring{\cite{Drummond:2019cxm}}{[1]}}\label{sec:comparorigincluster}

In this appendix, we compare our techniques and results to those of ref. \cite{Drummond:2019cxm}. The general computation strategy of ref. \cite{Drummond:2019cxm} revolved around studying clusters with quivers of the form
\[
\begin{tikzcd}
 & & & \textrm{\framebox[3.5ex]{$f_{z}$}}\arrow[r] & z_{0}\arrow[dr]\arrow[dl]\arrow[dll]& & & \\
a_{2} & a_{1}\arrow[l]& b_{1}\arrow[l]\arrow[drr]& b_{2}\arrow[dr]& & b_{3}\arrow[r]\arrow[dl]& a_{3}\arrow[r]& a_{4}\ , \\
 & & & \textrm{\framebox[3.5ex]{$f_{w}$}} & w_{0}\arrow[l]\arrow[shift left, uu]\arrow[shift right, uu] & & & 
\end{tikzcd}
\]
where we have suppressed all frozen variables disconnected from the $A_{1,1}$ subalgebra. Such clusters were called origin clusters. The authors then used the generating function of the $A_{1,1}$ subalgebra in eq. (\ref{eq:closedformexpressionforxnA1p1}) to motivate three algebraic functions:
\begin{equation}\label{originalphabet2}
\frac{w_{0}-B_{w}\sqrt{\triangle}}{w_{0}+B_{w}\sqrt{\triangle}}\ , \quad \frac{z_{0}-B_{z}\sqrt{\triangle}}{z_{0}+B_{z}\sqrt{\triangle}}\ , \quad \frac{\mathcal{P}-\sqrt{\triangle}}{\mathcal{P}+\sqrt{\triangle}}   \ ,   
\end{equation}
where
\begingroup
\allowdisplaybreaks
\begin{align}
z_{1}&=\frac{b_{1}b_{2}b_{3}-f_{w}z_{0}^{2}}{w_{0}}\ , \quad w_{1}=\frac{b_{1}b_{2}b_{3}-f_{z}w_{0}^{2}}{z_{0}} \ , \nonumber \\ 
B_{w}&=\frac{2w_{1}-w_{0}\mathcal{P}}{\triangle}\ , \quad B_{z}=\frac{2z_{1}-z_{0}\mathcal{P}}{\triangle} \ , \\
\mathcal{P}&=\frac{f_{z}w_{0}+z_{1}}{z_{0}}\ , \quad \triangle=\mathcal{P}^{2}-4f_{w}f_{z} \ . \nonumber
\end{align}
\endgroup
The multiplicative functions in eq. (\ref{originalphabet2}) are simply an alternative multiplicative basis for the $\hat{y}_{\gamma_{i}}^{\pm}$ of the $A_{1,1}$ subalgebra found in section \ref{sec:asymptotandlimwallss}. The authors ultimately studied 32 origin clusters for each limiting ray apparently relevant for $\mathcal{N}=4$ pSYM. 

\indent Including the rational $x$-variables associated with the origin cluster, this method ultimately yields a multiplicative basis for the $\hat{y}_{\gamma_{i}}^{0}$ of the two asymptotic chambers associated with each origin cluster. Mutating $w_{0}$ ($z_{0}$) first leads to the first (second) asymptotic chamber after an infinite number of mutations on the $A_{1,1}$ subalgebra. Notably, this technique only allows one to probe the $\hat{y}_{\gamma_{i}}^{0}$ of asymptotic chambers adjacent to a limiting wall. To see this, note that the generalized mutation identified in section \ref{sec:asymptotandlimwallss} applies to the $A_{1,1}$ subalgebra when asymptotically close to the limiting ray. Since a generalized mutation corresponds to a limiting wall, the asymptotic chamber must be adjacent to a limiting wall. However, many asymptotic chambers are not adjacent to a limiting wall. For example, consider the asymptotic scattering diagram in fig. \ref{fig:sketchofwallA21withoutasymptot}, which is associated with the limiting ray of the $A_{2,1}$ cluster algebra. There are two asymptotic chambers, $C_{2}$ and $C_{5}$, that are not adjacent to the limiting wall. To probe the $\hat{y}^{0}_{\gamma_{i}}$ of such asymptotic chambers, one must either find a generating function for the sequences of cones that do not explicitly contain an $A_{1,1}$ subalgebra or use the wall-crossing techniques developed in this paper. Excluding these asymptotic chambers when calculating the multiplicative basis for $\hat{y}_{\gamma_{i}}$ leads to a truncated asymptotic symbol alphabet. For example, excluding asymptotic chambers $C_{2}$ and $C_{5}$ of $A_{2,1}$ leads to the truncated alphabet,
\begin{equation}
\hat{y}^{0}_{\gamma_{1}}, \ \hat{y}^{0}_{\gamma_{2}}, \ \hat{y}^{0}_{\gamma_{3}}, \ \frac{1+\hat{y}_{\gamma_{1}}^{0}\hat{y}_{\gamma_{2}}^{0}\hat{y}_{\gamma_{3}}^{0}}{1+\hat{y}_{\gamma_{1}}^{0}}, \ 1-\hat{y}_{\gamma_{2}}^{0}\hat{y}_{\gamma_{3}}^{0} \ ,     
\end{equation}
in comparison to the full alphabet in eq. (\ref{eq:finalalphabetsss}). 

\indent Interestingly, the $\hat{y}_{\gamma_{i}}$ of asymptotic chambers adjacent to the limiting wall are enough to derive the algebraic letters that have appeared in explicit computations. Rather, the known 18 algebraic letters that appear in the 8-point 2-loop NHMV amplitude are monomials of algebraic variables in the form of eq. (\ref{originalphabet2}). However, there is no obvious reason why additional algebraic letters could not appear at higher loop orders.

\section{Review: cluster polytopes}\label{sec:clusterpolytopes}
\begin{figure}
\centering
  \includegraphics[scale=0.6]{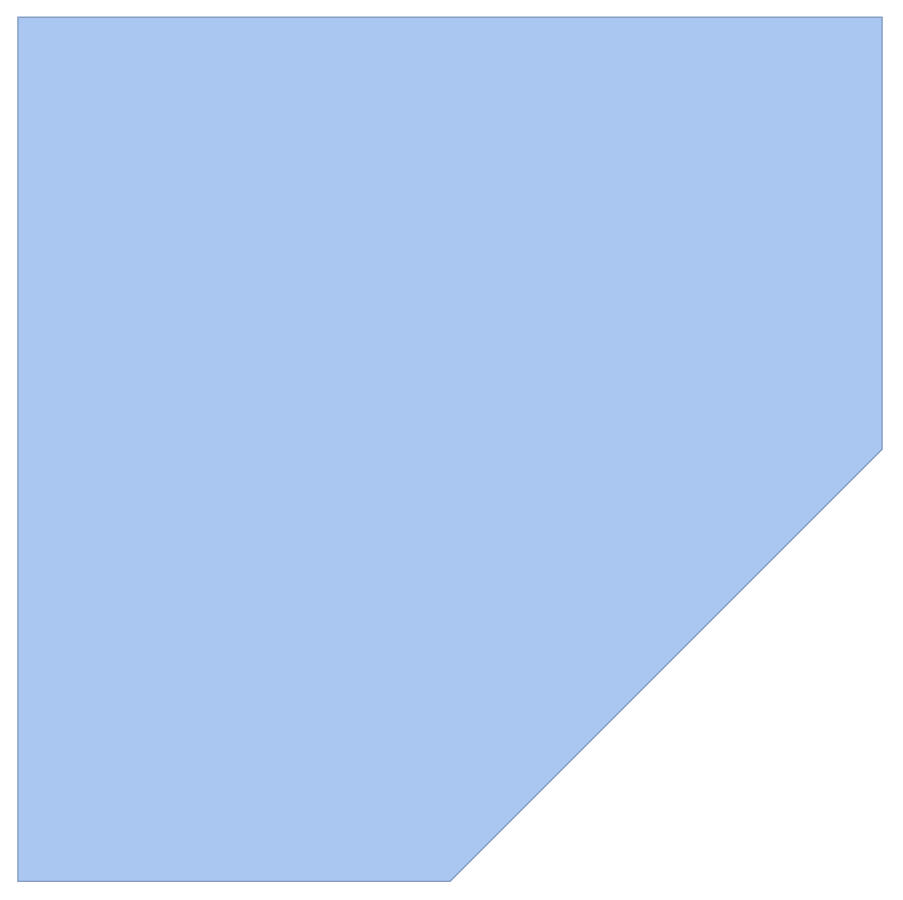} \quad \quad \quad
   \includegraphics[scale=0.6]{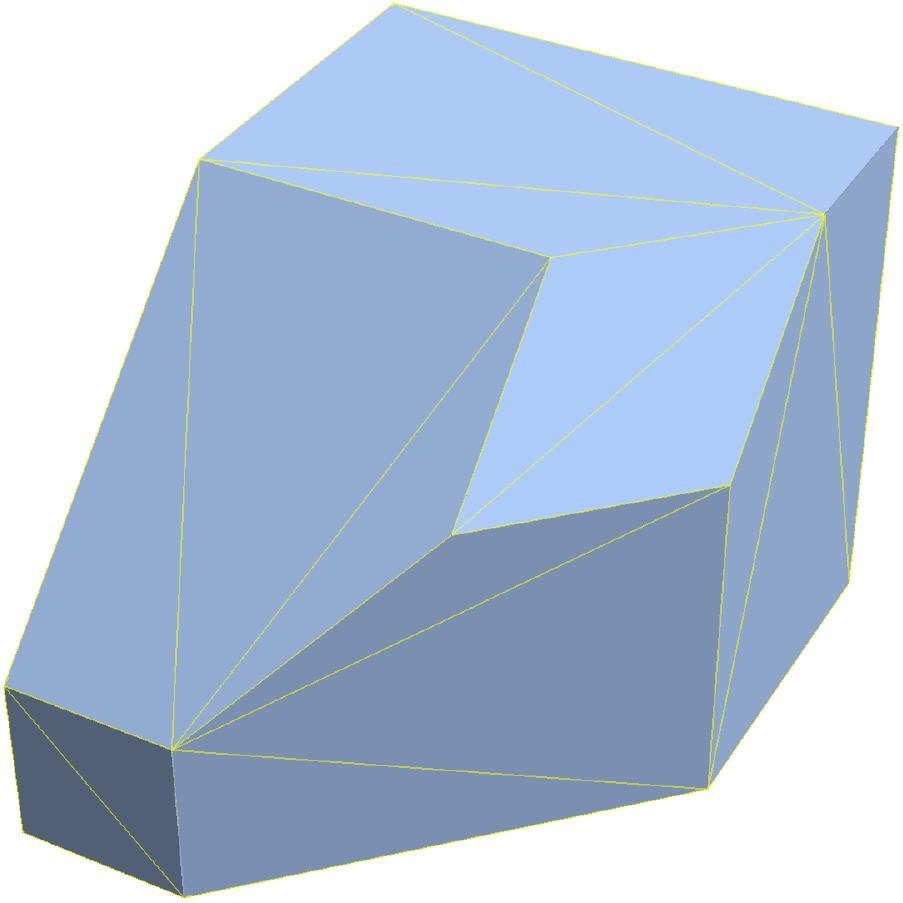} 
  \caption{Cluster polytopes corresponding to $A_{2}$ (left) and $A_{3}$ (right). }
  \label{fig:exampleA2A3cutu}
\end{figure}
Instead of investigating the scattering diagram, much research has focused on a closely related object, \textit{the cluster polytope}. Every scattering diagram is dual to a polytope where vertices correspond to cones, facets to $g$-vectors, and edges to walls. More concretely, one can define a polytope using facet vectors:
\begin{equation}
\{Y\in \mathbb{P}^{N}|Y\cdot W_{i}\geq 0 \textrm{ for all }i\} \ .
\end{equation}
\noindent For a cluster polytope, the facet vectors, $W_{i}$, match onto the $g$-vectors 
\begin{equation}\label{vectodeft}
W_{i}=(c_{i},\vec{g}_{i})    
\end{equation}
\noindent where the constants, $c_{i}$, are chosen such that the polytope has the correct vertex and facet structure. As an example, fig. \ref{fig:exampleA2A3cutu} shows the cluster polytopes associated with $A_{2}$ and $A_{3}$ \cite{Drummond:2020kqg}. The fan and polytope interpretation are equivalent and simply correspond to different visualizations of the same combinatorial data. 

\indent Given a set of tropical functions, $f_{i}$, an obvious question is how to derive the cluster polytope associated with the corresponding fan without calculating all the $c_{i}$ in eq. (\ref{vectodeft}). The answer is remarkably simple. For each $f_{i}$, we associate a corresponding polytope, $P_{i}$, by taking the convex hull of exponent vectors for each term. As an example, consider the following Laurent polynomials,
\begin{equation}
\begin{split}
f_{1}=1+x &\rightarrow \{ (0,0),\ (0,1) \} \\   
f_{2}=1+x+xy &\rightarrow \{(0,0),\ (0,1),\ (1,1)\}\ ,   
\end{split}
\end{equation}
\noindent where we have listed the vertices of the corresponding polytopes. We then consider the Minkowski sum of all such polytopes. Alternatively, we simply take the product of all such $f_{i}$ and find the associated $P$ using the same procedure
\begin{equation}
f_{1}f_{2}=(1+x)(1+x+xy)\rightarrow \{(0,0),\ (1,0),\  (1,1),\ (2,0),\ (2,1) \} \ .  
\end{equation}
\noindent A proof of this procedure is provided in ref. \cite{Arkani-Hamed:2019mrd}.

\bibliographystyle{JHEP}
\bibliography{main}

\end{document}